\documentclass[aps,prd,preprint,showpacs]{revtex4}
\newcommand{\at}{\overline{10}}

\usepackage{epsfig}
\usepackage{graphicx}

\begin{document}

\preprint{RUB-TP2-04/14}

\title{Mixing and decays of the antidecuplet in the context
of approximate SU(3) symmetry}
\author{V. Guzey}
\affiliation{Institut f{\"u}r Theoretische Physik II, Ruhr-Universit{\"a}t
Bochum, D-44780 Bochum, Germany}
\email[]{vadim.guzey@tp2.ruhr-uni-bochum.de}
\author{M.V. Polyakov}
\affiliation{Institut de Physique, B5a, Universit\'e de Li\`ege au Sart Tilman,
   B4000 Li\`ege 1, Belgium
and
Petersburg Nuclear Physics Institute, Gatchina, St. Petersburg
188300, Russia}
\email[]{maxim.polyakov@ulg.ac.be}

\begin{abstract}

We consider mixing of the antidecuplet with three 
$J^P=1/2^+$  octets (the ground-state octet, the octet containing $N(1440)$, 
$\Lambda(1600)$, $\Sigma(1660)$
and $\Xi(1690)$ and the octet containing $N(1710)$,
$\Lambda(1800)$, $\Sigma(1880)$ and $\Xi(1950)$)
in the framework of approximate flavor SU(3) symmetry.
We give general expressions for the partial decay widths of all members of 
the antidecuplet as functions of the two mixing angles. Identifying 
$N_{\at}$ with the $N(1670)$ observed by the GRAAL experiment,
we show that the considered mixing scenario can accommodate all present
 experimental and phenomenological information on the $\Theta^+$
 and $N_{\at}$ decays: 
$\Theta^+$ could be as narrow as 1 MeV; the $N_{\at} \to N\, \eta$ decay 
 is sizable, 
while the  $N_{\at} \to N\, \pi$ decay is suppressed and
the $N_{\at} \to \Lambda\, K$ decay is possibly
suppressed. Constraining the mixing angles by the $N_{\at}$ decays, we 
make definite predictions for the $\Sigma_{\at}$ decays.
We point out that $\Sigma_{\at}$ with mass near 1770 MeV
could be searched for in
 the available data on $K_S \, p$ invariant
mass spectrum, which already revealed the $\Theta^+$  peak.
It is important to experimentally verify the decay properties
 of $\Sigma(1770)$ because 
its mass and $J^P=1/2^+$ make it an attractive candidate for $\Sigma_{\at}$.
  
\end{abstract}


\maketitle

\section{Introduction}
\label{sec:intro}

Approximate flavor SU(3) symmetry of strong interactions predicts
that hadrons are grouped into certain multiplets (families)
\cite{Eightfoldway}: Singlets, octets ($\bf{8}$), decuplets
($\bf{10}$), antidecuplets ($\bf{\overline{10}}$), 27-plets,
35-plets, etc. It is a piece of textbook wisdom that all known
hadrons, which can constitute of three quarks, can be successively placed into
singlets, octets and decuplets \cite{Kokkedee,Close,Samios74} and
that other (higher) SU(3)   representations are not required. The
discoveries of the $\Theta^+$
\cite{SPRING8,DIANA,CLAS,SAPHIR,Neutrino,CLAS2,HERMES,SVD,COSYTOF,ZEUS,GRAAL,Togoo,Troyan}
 and $\Xi_{3/2}$ \cite{NA49}, if confirmed \cite{BES,ALEPH,BABAR,HERAB,CDF,FOCUS,Belle,PHENIX},
 mean the existence of the whole new exotic family -- the antidecuplet.

In QCD and in Nature, SU(3) is broken by non-equal masses of the
up, down and strange quarks. As a result, members of different
 SU(3) multiplets of the
same spin and parity can mix. If the antidecuplet has indeed
$J^P=1/2^+$ as predicted in the chiral quark-soliton approach
\cite{DPP}, then the antidecuplet can potentially mix with three
known $J^P=1/2^+$ octets and with a $J^P=1/2^+$ decuplet.
In addition to the traditional SU(3) multiplets, the antidecuplet can also mix
with a 27-plet and a 35-plet \cite{27plet}.

The degree of mixing due to SU(3) violation among SU(3) multiplets,
especially in the baryon sector,
cannot be large in order for the notion of approximate SU(3) symmetry
 to make sense. However, in the meson sector,
there are known exceptions from small mixing because of the accidental
degeneracy in mass of a singlet and an octet member.
The most celebrated example is the large (ideal) mixing between
the $\phi$ and $\omega$ vector mesons~\cite{Eightfoldway,Kokkedee}.

In our analysis, we consider 
the mixing angles (the parameters which describe the mixing)
  as small parameters.
 Because of the small width of the $\Theta^+$ 
\cite{PWA,Nussinov,Heidenbauer,Cahn,Gibbs,Sibirtsev,Sibirtsev2}, 
small mixing with the
 antidecuplet does not affect the properties of the involved
 octets. However, the decays of the ${\bf \at}$
 members dramatically depend on even very small mixing.

In this work, we examine the  scenario that the antidecuplet
mixes simultaneously with three octets --
with the ground-state octet, with
the octet containing $N(1440)$, $\Lambda(1600)$, $\Sigma(1660)$
and $\Xi(1690)$, and with the octet containing $N(1710)$,
$\Lambda(1800)$, $\Sigma(1880)$ and $\Xi(1950)$ --
and with a 27-plet and a 35-plet. The coupling constants and the mixing
angles with the ground-state octet and with  the 27-plet and 35-plet are taken
from the chiral quark soliton model \cite{27plet}.
For the other two  octets, the coupling constants are determined from
the $\chi^2$ fit to the available decay widths, and the two corresponding 
mixing angles are left as free parameters.

This strategy enables us to present the general expressions for the $\at \to B+
P$ decay widths, where $B$ is the ground-state baryon and $P$ is the
ground-state pseudoscalar meson, as functions of the two mixing angles,
the total width of the $\Theta^+$ and the pion-nucleon sigma term.
 Using scarce experimental
information on the observation, non-observation and decays of the
members of the antidecuplet, we give most probable regions of  the mixing
angles and make predictions for the yet unmeasured decay modes of
the ${\bf \at}$. This enables us to suggest the reactions most favorable 
for the excitation and identification of the members  of the antidecuplet.

\section{Antidecuplet mixing with three octets and $\at \to B+P$ decays}
\label{sec:888}
\subsection{General formalism}

We examine the scenario that the antidecuplet mixes simultaneously with
three octets. The mixing takes place through the $N$-like and
$\Sigma$-like states of the involved multiplets. 
 The physical octet $|N_1^{\rm phys} \rangle$, $|N_2^{\rm
phys}\rangle$, $|N_3^{\rm phys}\rangle$ and the antidecuplet 
$|N_{\at}^{\rm phys}\rangle$ states can be 
expressed in terms of the octet states, $|N_1\rangle$, $|N_2\rangle$ and 
$|N_3\rangle$, and the
purely antidecuplet $|N_{\at}\rangle$ state,
\begin{equation}\label{nmix}
 \left(
  \begin{array}{c}
   |N_1^{\rm phys}\rangle \\
   | N_2^{\rm phys}\rangle \\
    |N_3^{\rm phys}\rangle \\
    |N_{\at}^{\rm phys}\rangle  \\
\end{array} \right)
     = \left(
\begin{array}{cccc}
1 & 0 & 0 & \sin \theta_1 \\
0 & 1 & 0 & \sin \theta_2 \\
0 & 0 & 1 & \sin \theta_3 \\
-\sin \theta_1 & -\sin \theta_2 & -\sin \theta_3 & 1 \\
\end{array}
\right) \left(
  \begin{array}{c}
   |N_1\rangle \\
    |N_2\rangle \\
    |N_3 \rangle\\
    |N_{\at} \rangle \\
\end{array} \right) \,,
\end{equation}
assuming that the
$\theta_1$, $\theta_2$ and $\theta_3$ mixing angles are small.
A similar equation relates the physical $|\Sigma_1^{\rm phys}\rangle$,
$|\Sigma_2^{\rm phys}\rangle$, $|\Sigma_3^{\rm phys}\rangle$ and
$|\Sigma_{\at}^{\rm phys}\rangle$ states to the octet 
$|\Sigma_1\rangle$,
$|\Sigma_2\rangle$, $|\Sigma_3\rangle$ states and the purely antidecuplet
$|\Sigma_{\at}\rangle$ state after the replacement $\theta_i \to
\theta_i^{\Sigma}$. 

In our analysis, we assume that the $\theta_i$ mixing angles are small,
$\sin \theta_i={\cal O}(\epsilon)$.
The small parameter $\epsilon$ describes not only the violation of flavor
SU(3) symmetry due to the non-zero mass of the strange quark, but also 
a possible additional dynamical suppression of the mixing between exotic and 
non-exotic states. In addition, the smallness of $\theta_i$ rests on the 
previous phenomenological analyses of non-exotic baryon multiplets, which
observed only small mixing angles, see e.g. \cite{Samios74}.

In our analysis, we systematically neglect ${\cal O}(\epsilon^2)$ terms.
Therefore, the mixing matrix in Eq.~(\ref{nmix}) is unitary up to
${\cal O}(\epsilon^2)$ corrections.

It is important to note that, in general, the $|N_1\rangle$, 
$|N_2\rangle$ and $|N_3\rangle $ states can mix among 
themselves. Indeed, our analysis \cite{GPreview} shows that the  
$|N_2\rangle $ and $|N_3\rangle $ states are slightly 
mixed and that the  $|N_1\rangle$ state can be considered as a pure state
 (it does not have components from other SU(3) multiplets). 
Therefore, in practice the problem of mixing of the antidecuplet with the 
three octets is solved in two steps. First, we determine the SU(3) coupling
 constants and the mixing angles for the three octets. Because of the smallness
of $\Gamma_{\Theta^+}$ and $\theta_i$, it is legitimate to neglect the 
antidecuplet admixture at this stage.
Details of this analysis are presented in Appendix~\ref{sec:34}.
Second, the resulting octet states are mixed with the antidecuplet states.
 Naturally, since the possible mixing among the octets was already been taken
 into
 account in step one, it is sufficient to consider only the mixing of
 each individual $|N_i^{\rm phys} \rangle$ ($i=1,2,3$) with $|N_{\at}^{\rm phys} \rangle$ -- see Eq.~(\ref{nmix}).
The same procedure applies to the considered octet and antidecuplet 
$\Sigma$ states.

The assumption of small mixing angles with the antidecuplet is in stark
contrast
with the quark model calculations~\cite{JW,Stancu}, which automatically
 lead to almost ideal~\footnote{In quark models, the mixing with the
antidecuplet is exactly ideal in the SU(3) limit.} mixing~\cite{Faiman,Cohen}.
Indeed, eigenstates of a generic quark model Hamiltonian have separately 
almost well-defined number of strange quarks and strange antiquarks.
In the language of approximate flavor SU(3) symmetry, this can be realized 
only by almost ideal mixing of SU(3) states, which 
contain both non-strange and hidden strange components, such that the states 
resulting after the mixing contain mostly either non-strange quarks or
strange quarks.

Of course, both small mixing and large mixing scenarios are assumptions which
must be confronted with the data. A straightforward $\chi^2$ analysis
of the available $N(1440)$ and  $N(1710)$ partial decays widths shows 
that the popular  scenario of Jaffe and Wilczek \cite{JW}, which 
assumes nearly ideal 
 mixing of  $N(1440)$ (mostly octet state) with $N(1710)$ 
(mostly antidecuplet state),
is inconsistent with the  experimental data on the $N(1440)$ and $N(1710)$
decays \cite{PDG}: It is impossible to simultaneously accommodate 
$\Gamma_{\Theta^+} \leq 10$ MeV and a large $\Gamma_{N(1440)\to N \, \pi}$. 
The same conclusion, but  without the $\chi^2$ fit, was obtained in 
\cite{DP2004,Cohen}.

The physical states are eigenstates of the mass operator $\hat{M}$. 
The corresponding physical masses $N_i^{\rm phys}$ are 
\begin{equation}
N_i^{\rm phys} \equiv \langle N_i^{\rm phys}|\hat{M}|N_i^{\rm phys} \rangle=
N_i+\sin^2 \theta_i N_{\at}=N_i+{\cal O}(\epsilon^2) \,,
\end{equation} 
where $N_i$ is the mass of the unmixed $|N_i \rangle$ state and $i=1,2,3,\at$.
Thus, to the leading order in the SU(3)-violation effect, the physical masses 
are equal to the corresponding masses of the unmixed states.
 This means that the
Gell-Mann--Okubo mass formulas are not sensitive to the small mixing, 
if it is treated consistently.
Keeping only terms linear in the mass of the strange quark, it is not 
legitimate to estimate the mixing angles from the Gell-Mann--Okubo mass
splitting formula, as was done for instance in \cite{DP2004}.
Instead, as will be shown below, one has to consider decays since, in the
presence of multiplet mixing, the decay widths contain  a first power of the
mixing angles.

Since the mechanism of the mixing of
the $N$-states and $\Sigma$-states is the same, the mixing
angles $\theta_i$ and  $\theta_i^{\Sigma}$ are related \cite{DP2004}
\begin{equation}\label{thetasigma}
\sin \theta_i \left(N_i^{\rm phys}-N_{\at}^{\rm phys} \right)=\sin
\theta_i^{\Sigma} \left(\Sigma_i^{\rm phys}-\Sigma_{\at}^{\rm
phys} \right) \,.
\end{equation}
However, since $N_i^{\rm phys}-N_{\at}^{\rm phys}=\Sigma_i^{\rm phys}-\Sigma_{\at}^{\rm phys}+{\cal O}(\epsilon)$ and $\theta_i,\theta_i^{\Sigma} \propto {\cal O}(\epsilon)$, 
\begin{equation}\label{thetasigma2}
\theta_i =\theta_i^{\Sigma} + {\cal O}(\epsilon)=\theta_i^{\Sigma} \,,
\end{equation}
when we consistently neglect ${\cal O}(\epsilon^2)$ terms.

In our analysis we assume that SU(3) symmetry is
violated only by the non-equal masses of hadrons inside a given multiplet
and by the multiplet mixing
and that it is preserved~\footnote{An example of analysis assuming violation
of SU(3) in decay vertices can be found in \cite{Dobson78}.} 
 in the decays. The success of this assumption 
was proven in \cite{Samios74,GPreview}.
This allows one to make definite predictions for the $\at \to B + P$ 
transitions in terms of the antidecuplet and octet universal coupling 
constants.
The general SU(3) formula for the $\bf{\at} \to \bf{8} +\bf{8}$
coupling constants reads
\begin{equation}
g_{B_1 B_2 P }=-G_{\at}\ \frac{1}{\sqrt{5}} \left(
\begin{array}{cc}
8 & 8 \\
Y_2 T_2 & Y_\phi T_\phi
\end{array}\right|\left.\begin{array}{c}
          \at \\Y_1 T_1
          \end{array}\right) \,,
\label{eq:anti1088}
\end{equation}
where $G_{\at}$ is the antidecuplet universal coupling constant; 
the factors in parenthesis are SU(3) isoscalar factors, which are known for
any SU(3) multiplets \cite{deSwart}; $Y_{1,2}$ and $T_{1,2}$ are hypercharges
and isospins of the baryons $B_{1,2}$; $Y_{\phi}$ and $T_{\phi}$
are the hypercharge and isospin of the pseudoscalar meson.

Because of the mixing with octets, we also need the coupling constants
for the  $\bf{\at} \to \bf{\at} +{\bf 8}$,
$\bf{8} \to \bf{8} + \bf{8}$ and $\bf{8} \to \bf{10} + \bf{8}$
transitions. The former is defined as
\begin{equation}
g_{B_1 B_2 P }=H_{\at} \frac{1}{2 \sqrt{2}} \left(
\begin{array}{cc}
\at & 8 \\
Y_2 T_2 & Y_\phi T_\phi
\end{array}\right|\left.\begin{array}{c}
          \at \\Y_1 T_1
          \end{array}\right) \,.
\label{eq:anti10108}
\end{equation}

In the SU(3) symmetric limit, the coupling constants
 for the $\bf{8} \to \bf{8} + \bf{8}$ transition 
 are parametrized in
terms of the universal octet coupling constant $G_8$, the ratio
$\alpha=F/D$ (we choose our notations in such a way that
$\alpha=2/3$ for the ground-state octet in the naive quark
model) and SU(3) isoscalar factors
\begin{equation}
g_{B_1 B_2 P}=G_8\ \frac{3}{\sqrt{20}} \left(
\begin{array}{cc}
8 & 8 \\
Y_2 T_2 & Y_\phi T_\phi
\end{array}\right|\left.\begin{array}{c}
          8_S\\Y_1 T_1
          \end{array}\right)
\left[1+\alpha\ \frac{3}{\sqrt 5}\ \frac{ \left(
\begin{array}{cc}
8 & 8 \\
Y_2 T_2 & Y_\phi T_\phi
\end{array}\right|\left.\begin{array}{c}
          8_A\\Y_1 T_1
          \end{array}\right)
}{\left(
\begin{array}{cc}
8 & 8 \\
Y_2 T_2 & Y_\phi T_\phi
\end{array}\right|\left.\begin{array}{c}
          8_S\\Y_1 T_1
          \end{array}\right)
} \right] \,. 
\label{eq:888}
\end{equation}

Finally, the $\bf{8} \to \bf{10} + \bf{8}$ coupling constant is defined in 
terms of the universal coupling constant $G_{10}$ 
\begin{equation}
g_{B_1 B_2 P}=G_{10} \left(
\begin{array}{cc}
10 & 8 \\
Y_2 T_2 & Y_\phi T_\phi
\end{array}\right|\left.\begin{array}{c}
          8 \\Y_1 T_1
          \end{array}\right) \,.
\label{eq:8108}
\end{equation}

Equations~(\ref{eq:anti1088})-(\ref{eq:8108}) are written in the SU(3)
symmetric limit. The mixing between the antidecuplet and the octets
results in the mixing of the coupling constants which is 
controlled by the mixing
matrix of Eq.~(\ref{nmix}). The coupling constants for the decays 
of the $\bf{\at}$
members are summarized by Eqs.~(\ref{eq:decays:theta})-(\ref{eq:decays:xi}). 
For the only decay mode of the $\Theta^+$, one has \cite{DPP}
\begin{equation}
g_{\Theta^+ N\,K}=\frac{1}{\sqrt{5}} \left(G_{\at}+\sin \theta_1 H_{\at} \frac{\sqrt{5}}{4} \right) \,.
\label{eq:decays:theta}
\end{equation}

The coupling constants for the $N_{\at}^{\rm phys}$ decays read
\begin{eqnarray}
&& g_{N_{\at}  N\, \pi}= \frac{1}{2 \sqrt{5}}\left(G_{\at}+\sin \theta_1 
\left(H_{\at} \frac{\sqrt{5}}{4}-G_8 \frac{7}{\sqrt{5}} \right)
-\sum_{i=2,3} \sin \theta_i \,g_{N_i  N\, \pi}
 \right) \,, 
 \nonumber\\
&& g_{N_{\at}  N\, \eta}= \frac{1}{2 \sqrt{5}}\left(-G_{\at}+\sin \theta_1 
\left(H_{\at} \frac{\sqrt{5}}{4}-G_8 \frac{1}{\sqrt{5}} \right)
+\sum_{i=2,3} \sin \theta_i \,g_{N_i  N\, \eta} \right)\,,  \nonumber\\
&& g_{N_{\at}  \Lambda\, K}= \frac{1}{2 \sqrt{5}}\left(G_{\at}+\sin \theta_1 
G_8 \frac{4}{\sqrt{5}} 
+ \sum_{i=2,3} \sin \theta_i \, g_{N_i  \Lambda\, K}\right)\,,  \nonumber\\
&& g_{N_{\at}  \Sigma\, K}= \frac{1}{2 \sqrt{5}}\left(G_{\at}
+\sin \theta_1^{\Sigma} H_{\at} \frac{\sqrt{5}}{2}+
\sin \theta_1 
G_8 \frac{2}{\sqrt{5}}+ \sum_{i=2,3} \sin \theta_i \, g_{N_i  \Sigma\, K} \right) \,,\nonumber\\
&& g_{N_{\at}  \Delta\, \pi}= \frac{2}{\sqrt{5}}
\left(\sin \theta_1 G_8 + \sum_{i=2,3}
\sin \theta_i \, g_{N_{i}  \Delta\, \pi}\right) \,.
\label{eq:decays:n}
\end{eqnarray}
In Eqs.~(\ref{eq:decays:n}), $G_8$ refers to the ground-state baryon octet;
$g_{N_2  BP}$ refer to the transition between the octet
containing the Roper $N(1440)$ and ground-state octet; 
$g_{N_3 BP}$ refer to the transition between the octet containing the $N(1710)$ and ground-state octet;  $g_{N_{i}  \Delta\, \pi}$
 are the universal couplings for the transition
between the octets and the ground-state decuplet.
 The $g_{N_2  BP}$,  $g_{N_3  BP}$ and $g_{N_{i}  \Delta\, \pi}$
 parameters can be 
determined by considering two-body hadronic decays of the two octets, see 
\cite{GPreview} and Appendix~\ref{sec:34}.
Note that the $N_{\at} \to   \Delta\, \pi$ decay is possible only due to the
mixing. 

Turning to the $\Sigma_{\at}$, its coupling constants read
\begin{eqnarray}
&& g_{\Sigma_{\at}  \Lambda\, \pi}= \frac{1}{2 \sqrt{5}}\left(G_{\at}
-\sin \theta_1^{\Sigma} 
G_8 \frac{3}{\sqrt{5}}- \sum_{i=2,3}
\sin \theta_i ^{\Sigma}\, g_{\Sigma_{i}  \Lambda\, \pi} \right) \,, \nonumber\\
&& g_{\Sigma_{\at}  \Sigma\, \eta}= -\frac{1}{2 \sqrt{5}}\left(G_{\at}
+\sin \theta_1^{\Sigma} 
G_8 \frac{3}{\sqrt{5}}+ \sum_{i=2,3}
\sin \theta_i ^{\Sigma} \,g_{\Sigma_{i}  \Sigma\, \eta} \right) \,, \nonumber\\
&& g_{\Sigma_{\at}  \Sigma\, \pi}= \frac{1}{\sqrt{30}}\left(G_{\at}+
\sin \theta_1 \left(H_{\at} \frac{\sqrt{5}}{2}-G_8 \sqrt{5} \right)
- \sum_{i=2,3}
\sin \theta_i ^{\Sigma}\, g_{\Sigma_{i}  \Sigma\, \pi} \right) \,, \nonumber\\
&& g_{\Sigma_{\at}  \Xi\, K}= \frac{1}{\sqrt{30}}\left(G_{\at}+
\sin \theta_1^{\Sigma} G_8 \frac{14}{\sqrt{20}}
+ \sum_{i=2,3} \sin \theta_i^{\Sigma}\, g_{\Sigma_{i}  \Xi\, K} \right) \,, \nonumber \\
&& g_{\Sigma_{\at}  N\, \overline{K}}= \frac{1}{\sqrt{30}}\left(-G_{\at}+
\sin  \theta_1 H_{\at} \frac{\sqrt{5}}{2}+
\sin \theta_1^{\Sigma} G_8 \frac{4}{\sqrt{20}}
+ \sum_{i=2,3}\sin \theta_i ^{\Sigma} \, g_{\Sigma_{i}  N\, \overline{K}}
\right) \,, \nonumber \\
&& g_{\Sigma_{\at}  \Sigma_{10}\, \pi}= \frac{\sqrt{30}}{15}\left(G_{8} \sin \theta_1 + \sum_{i=2,3}  \sin \theta_i ^{\Sigma}\,g_{\Sigma_{i}  \Sigma_{10}\, \pi}  \right) \,.
\label{eq:decays:s}
\end{eqnarray}
Like in the case of the $N_{\at} \to \Delta \, \pi$ decay, the
$\Sigma_{\at} \to  \Sigma_{10}(1385)\, \pi$ decay is only possible because
 of the mixing.

Finally, the coupling constants for the $\Xi_{\at}$ decays are
\begin{eqnarray}
&&g_{\Xi_{\at} \Sigma \, \overline{K}}=-\frac{1}{\sqrt{10}} \left(G_{\at}-
\sin \theta_1^{\Sigma} H_{\at} \frac{\sqrt{5}}{4} \right) \,, \nonumber \\
&&g_{\Xi_{\at} \Xi \, \pi}=\frac{1}{\sqrt{10}} G_{\at} \,.
\label{eq:decays:xi}
\end{eqnarray}
One sees that the $\Xi_{\at} \to \Xi \, \pi$ decay is unique in that 
it completely determines the $G_{\at}$ coupling constant.

Equations~(\ref{eq:decays:theta})-(\ref{eq:decays:xi}) generalize the 
corresponding expressions of \cite{Arndt2004} because in addition to the 
mixing with the ground-state octet, we consider simultaneous mixing with two
additional octets. 
%

 Using Eqs.~(\ref{eq:decays:theta})-(\ref{eq:decays:xi}),
one can readily calculate the $\bf{\at} \to \bf{8}+\bf{8}$ partial
 decay widths
 \cite{DPP,Arndt2004},
\begin{equation}
\Gamma(B_{1} \to B_2+P)=3 |g_{B1B2 P}|^2 \frac{|\vec{p}|^3}{2 \pi (M_1+M_2)^2} \frac{M_2}{M_1} \,,
\label{eq:ps}
\end{equation}
where $|\vec{p}|$ is the center-of-mass momentum in the final state; $M_1$ and
$M_2$ are the masses of the initial and final baryon, respectively.

It is important to note that there is no universal prescription for the choice
of the phase space factor in Eq.~(\ref{eq:ps}): Different choices of the 
phase space factor
 correspond to different mechanisms of SU(3) violation, which is out of
 theoretical control. Therefore, it is purely a phenomenological issue
which phase space factor to use in the calculation of the partial decay width.
For instance, it has been known since the 70's that Eq.~(\ref{eq:ps})
 works poorly for the decays of the ground-state decuplet
and that it should be modified \cite{Samios74}. 
This issue was recently hotly debated \cite{phasefactor} in relation to
 the prediction of
the total width of the $\Theta^+$ in the chiral quark soliton model.
We emphasize that the heart of the problem lies not in a 
particular dynamical model for strong interactions (be it a chiral quark 
soliton model or any generic quark model) but in a phenomenological, i.e. 
rather general, 
observation that approximate flavor SU(3) symmetry cannot simultaneously
 describe
 all four 
decays of the ground-state decuplet.
One possible modification, which helps to remedy the problem,
leads to the following expression for the $\bf{\at} \to \bf{8}+\bf{8}$
 partial decay widths
\cite{DPP}
\begin{equation}
\Gamma(B_{1} \to B_2+P)=3 |g_{B1B2 P}|^2 \frac{|\vec{p}|^3}{2 \pi (M_1+M_2)^2} \frac{M_2}{M_1}\left(\frac{M_1}{M_2}\right)^2=3 |g_{B1B2 \phi}|^2 \frac{|\vec{p}|^3}{2 \pi (M_1+M_2)^2} \frac{M_1}{M_2} \,.
\label{eq:ps2}
\end{equation}

\subsection{Input parameters}

In order to use Eqs.~(\ref{eq:decays:theta})-(\ref{eq:decays:xi})
in practice, one has to have an input for the coupling constants and 
mixing angles. In the present work, we used the chiral 
quark soliton model results for the $G_8$ and 
$H_{\at}$ coupling constants, $2 G_{\at}- H_{\at} \approx G_8  \approx 18$ \cite{DPP,Arndt2004}, and for the mixing angle with the ground-state octet \cite{27plet}, 
\begin{equation}
\sin \theta_1=\sqrt{5} c_{\at}=-I_2 \frac{\sqrt{5}}{15} \left(\alpha+\frac{1}{2}\gamma \right) \,,
\label{eq:theta1}
\end{equation}
where the parameters $I_2$, $\alpha$ and $\gamma$ depend on
the pion-nucleon sigma term,  $\Sigma_{\pi \, N}$,
\begin{eqnarray}
&&\frac{1}{I_2}=608.7 - 2.9 \, \Sigma_{\pi \, N} \, \nonumber\\
&&\alpha=336.4-12.9 \, \Sigma_{\pi \, N} \, \nonumber\\
&&\gamma=-475.94+8.6 \, \Sigma_{\pi \, N} \,.
\label{eq:mom_inertia}
\end{eqnarray}
As follows from Eq.~(\ref{eq:decays:theta}), when one fixes the total width 
of the $\Theta^+$, there are two solutions for $G_{\at}$ and $H_{\at}$ because 
one essentially has to solve a quadratic equation in order to find them.

The octet-octet transition coupling constants
$g_{N_{2} BP}$ and  $g_{N_{3} BP}$ and the octet-decuplet 
coupling constants $g_{N_{i} \Delta \,\pi}$ are determined by performing
a $\chi^2$ fit to the experimentally measured two-body hadronic decays, see
\cite{GPreview}, where this approach
 was applied for the 
systematization of all SU(3) multiplets. 
Also, in Appendix~\ref{sec:34} we summarize the derivation of
Eq.~(\ref{eq:newcoupling:ns}), see below. 
In order to have the same notations
as in  \cite{GPreview} and also for brevity, we call the octet containing
$N(1440)$, $\Lambda(1600)$, $\Sigma(1660)$ and $\Xi(1690)$ octet 3, 
and the octet containing
$N(1710)$, $\Lambda(1810)$, $\Sigma(1880)$ and $\Xi(1950)$ -- octet 4.
 The key observation  that the mixing with the 
antidecuplet does not influence the decays of octets 3 and 4 enables us to
completely determine $g_{N_{2} BP}$,  $g_{N_{3} BP}$ and
 $g_{N_{i} \Delta \,\pi}$
 from the $\chi^2$ fit to the available experimental
 data on partial decay widths of the octets \cite{PDG}.

An analysis of \cite{GPreview} shows that it is impossible to describe
the decays of octet 4 without mixing it with some other SU(3) multiplet
because SU(3) predicts incorrectly the sign of the 
$\Sigma(1880) \to \Sigma \, \pi$ amplitude. If the data on the
two-star $\Sigma(1880)$ are taken seriously, this presents a serious challenge
to our approach based on approximate SU(3) symmetry.
 A possible solution, which remedies
the problem and produces an  acceptably low value of $\chi^2$ per degree
of freedom, is to mix octets 4 with octet 3.
The resulting values of the coupling constants for the
$N(1440)$, $N(1710)$, $\Sigma(1660)$ and $\Sigma(1880)$, which should be used 
in Eqs.~(\ref{eq:decays:n}) and (\ref{eq:decays:s}),
 are summarized in
Eq.~(\ref{eq:newcoupling:ns})
\begin{eqnarray}
&&g_{N_2 N \, \pi}=34.9  \qquad \qquad  g_{N_3 N \, \pi}=7.69
\nonumber\\
&&g_{N_2 N \, \eta}=-0.992 \qquad \qquad  g_{N_3 N \, \eta}=-2.60
\nonumber\\
&&g_{N_2 \Lambda \, K}=18.0 \qquad \qquad  g_{N_3 \Lambda \, K}=5.14 
 \nonumber\\
&&g_{N_2 \Sigma \, K}=16.0 \qquad \qquad g_{N_3 \Sigma \, K}=-0.051 
\nonumber\\
&&g_{N_2 \Delta \, \pi}=16.7 \qquad \qquad   g_{N_3 \Delta \, \pi}=13.6 \nonumber\\
&&g_{\Sigma_2 \Lambda \, \pi} =17.1 \qquad \qquad g_{\Sigma_3 \Lambda \, \pi}=-1.49 \nonumber\\
&&g_{\Sigma_2 \Sigma \, \eta}=g_{\Sigma_2 \Lambda \, \pi} \qquad \qquad   g_{\Sigma_3 \Sigma \, \eta} =g_{\Sigma_3 \Lambda \, \pi}\nonumber\\
&&g_{\Sigma_2 \Sigma \, \pi}= 20.2 \qquad  \qquad g_{\Sigma_3 \Sigma \, \pi}=3.09 \nonumber\\
&&g_{\Sigma_2 \Xi \, K}=35.8 \qquad   \qquad g_{\Sigma_3 \Xi \, K}=-0.694 \nonumber\\
&&g_{\Sigma_2 N \, \overline{K}}=15.5 \qquad   \qquad g_{\Sigma_3 N \, \overline{K}}=-3.79 \nonumber\\
&&g_{\Sigma_2 \Sigma_{10} \, \pi}=19.4 \qquad   \qquad 
g_{\Sigma_3 \Sigma_{10} \, \pi}=9.31 \,.
\label{eq:newcoupling:ns}
\end{eqnarray}

The expressions for the coupling constants, Eqs.~(\ref{eq:decays:theta})-(\ref{eq:decays:xi}) and (\ref{eq:newcoupling:ns}), combined with the phase
 space factors (\ref{eq:ps}) and (\ref{eq:ps2}) enable one to make 
predictions for all
$\bf{\at} \to \bf{8} +\bf{8}$ and $\bf{\at} \to \bf{10} +\bf{8}$ decay
 modes as functions of the two
 mixing angles $\theta_2$ and $\theta_3$, the pion-nucleon sigma term
$\Sigma_{\pi \, N}$ (which determines the $\theta_1$ mixing angle, see 
Eqs.~(\ref{eq:theta1}) and \ref{eq:mom_inertia}))
 and the total width of the $\Theta^+$.

\section{Predictions for antidecuplet decays}
\label{sec:main}

While different experiments give slightly different masses of the $\Theta^+$,
nevertheless the mass of the $\Theta^+$ can be considered established:
We use $m_{\Theta^+}=1540$ MeV.
There is only one experiment, the CERN NA49 experiment \cite{NA49}, which 
reports the $\Xi_{\at}$ signal with $m_{\Xi_{\at}}=1862 \pm 2$ MeV:
 This is the value we use in our analysis.
Since we do not use the Gell-Mann--Okubo mass formula for the antidecuplet
to determine the mixing angles,
the value of $m_{\Xi_{\at}}$ affects only the  $\Xi_{\at}$ decays.
Note that the validity of the NA49 analysis was challenged in 
\cite{NA49:critique} and that there is a number of null results on the
$\Xi^{--}$ search \cite{ALEPH,BABAR,HERAB,CDF,FOCUS}.
The masses of $N_{\at}$ and, especially, of $\Sigma_{\at}$ states
are not known. In this work, we  identify the peak
around 1670 MeV seen in the $\gamma \, n \to n\, \eta$ reaction by the GRAAL
 experiment with $N_{\at}$ and, thus, use $m_{N_{\at}}=1670$ MeV.
It was predicted that the mass of $N_{\at}$ should be in this range in
\cite{DP2004,Arndt2004}.
 Finally,
we simply assume that $m_{\Sigma_{\at}}=1765$ MeV, i.e. that $\Sigma_{\at}$
is equally-spaced between $N_{\at}$ and $\Xi_{\at}$. 

\subsection{Decays of $\Theta^+$ and the value of $G_{\at}$}

\begin{figure}
\includegraphics[width=12cm,height=12cm]{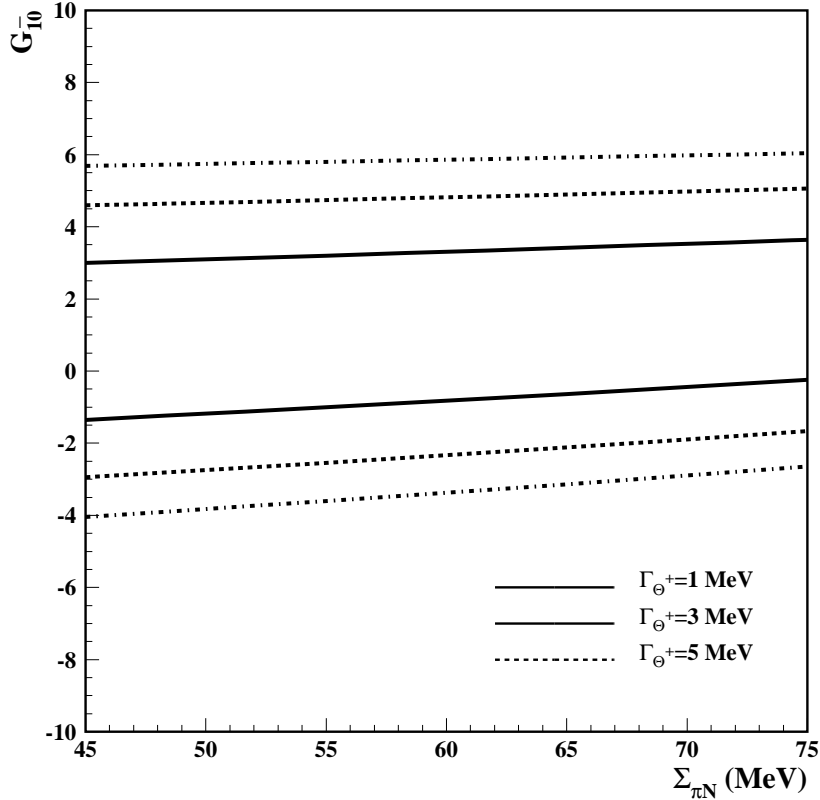}
\caption{The $G_{\at}$ coupling constant as a function of
$\Sigma_{\pi \, N}$ and at different $\Gamma_{\Theta^+}$. 
At each $\Sigma_{\pi \, N}$ and $\Gamma_{\Theta^+}$, there two values of 
$G_{\at}$: Positive and negative.
}
\label{fig:g10}
\end{figure}

In our analysis, we take $\Gamma_{\Theta^+}$  and $\Sigma_{\pi \, N}$ as
external parameters which are varied in the following intervals:
$1 \leq \Gamma_{\Theta^+} \leq 5$ MeV; $45 \leq \Sigma_{\pi \, N} \leq 75$ MeV.
We vary $\Sigma_{\pi \, N}$ in a wide interval between the values
obtained by \cite{kh,Gasser} and \cite{Pavan}.
Thus, at given $\Gamma_{\Theta^+}$  and $\Sigma_{\pi \, N}$ (the latter
fully determines $\theta_1$), the $G_{\at}$ coupling constant is found
from Eq.~(\ref{eq:decays:theta}) by solving a quadratic equation.
The two solutions are presented in Fig.~\ref{fig:g10}.
As seen in Fig.~\ref{fig:g10}, $|G_{\at}| \ll G_8 \approx 18$, which justifies one of 
our key assumptions that mixing with the antidecuplet affects the
 decay properties of the considered octets only little and, hence, can
be neglected.

\subsection{Decays of $\Xi_{\at}$}

\begin{figure}
\includegraphics[width=12cm,height=12cm]{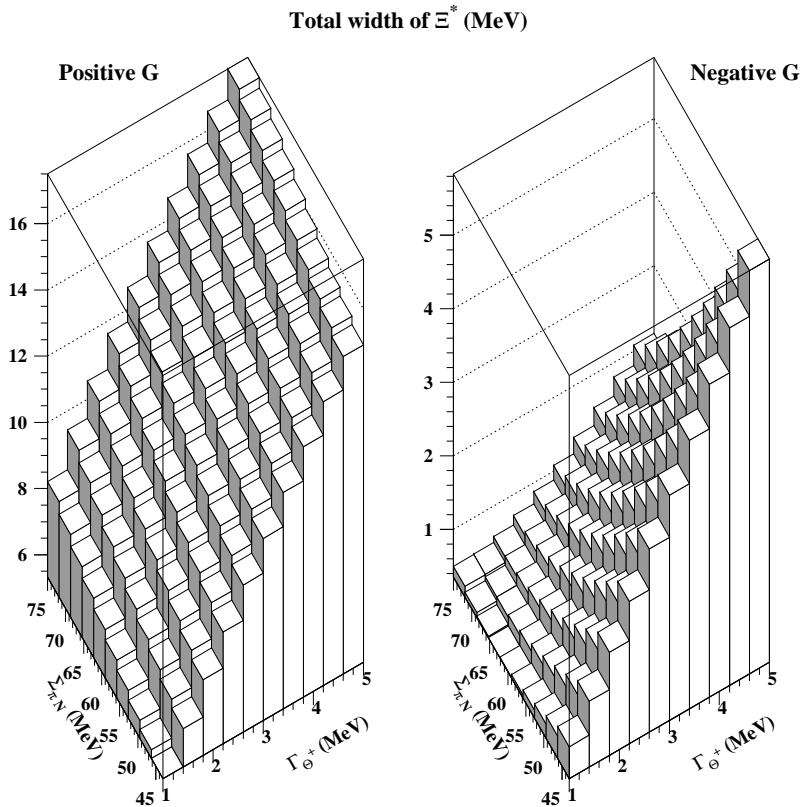}
\caption{The total width of $\Xi_{\at}$ as a function of $\Gamma_{\Theta^+}$ 
 and $\Sigma_{\pi \, N}$. The two plots correspond to the positive and
 negative solutions for 
$G_{\at}$.
}
\label{fig:xi}
\end{figure}

The $\Xi_{\at}$ partial decay widths
 depend only on $\Gamma_{\Theta^+}$  and $\Sigma_{\pi \, N}$, 
see Eq.~(\ref{eq:decays:xi}).
The total width of $\Xi_{\at}$ as a function of these two parameters is 
presented in Fig.~\ref{fig:xi}. The two plots correspond to the 
two solutions for the $G_{\at}$
coupling constant at given $\Gamma_{\Theta^+}$  and $\Sigma_{\pi \, N}$ (labeled as ``Positive G'' and ``Negative G'' in the plot).
The NA49 experiment gives the upper limit of the total width of the
$\Xi_{\at}$: $\Gamma_{\Xi_{\at}} < 18$ MeV  \cite{NA49}. Therefore, the both
solutions are compatible with the experimental upper limit. However, a slight
 increase in the mass of $\Xi_{\at}$ rather significantly increases 
$\Gamma_{\Xi_{\at}}$. Therefore, the scenario with 
a positive $G_{\at}$ at 
large $\Gamma_{\Theta^+}$ and $\Sigma_{\pi \, N}$  might lead to too large 
$\Gamma_{\Xi_{\at}}$. 

\subsection{Decays of $N_{\at}$}

\begin{figure}
\includegraphics[width=12cm,height=12cm]{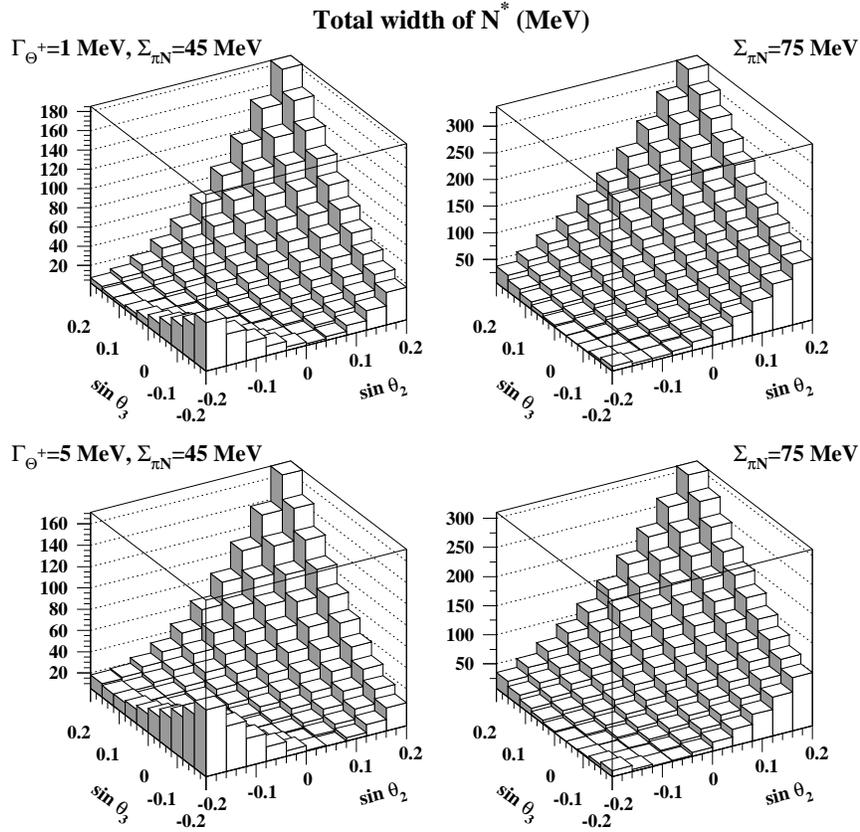}
\caption{The total width of $N_{\at}$ as a function of $\theta_2$
and $\theta_3$ at $\Gamma_{\Theta^+}=1$ and 5 MeV and 
$\Sigma_{\pi \, N}=45$ and 75 MeV.}
\label{fig:ntot}
\end{figure}

Next we turn to the decays of $N_{\at}$. An examination shows that the 
total width of $N_{\at}$
 only weakly depends on $\Gamma_{\Theta^+}$ and
 $G_{\at}$ and that  the dependence on  $\Sigma_{\pi \, N}$ is more important
because the $\theta_1$ mixing angle crucially depends on $\Sigma_{\pi \, N}$,
see Eq.~(\ref{eq:theta1}).
As an example of this trend, in Fig.~\ref{fig:ntot} we present $\Gamma_{N_{\at}}$ as a function of the $\theta_2$
and $\theta_3$ mixing angles for $\Gamma_{\Theta^+}=1$ MeV (upper row) and
 5 MeV (lower row)
and for $\Sigma_{\pi \, N}=45$ and 75 MeV. All plots correspond to the positive
$G_{\at}$ solution. 
One sees from Fig.~\ref{fig:ntot} that the total width of $N_{\at}$
depends very dramatically on all mixing angles and, in general, can be
very large.

\begin{figure}
\includegraphics[width=11cm,height=11cm]{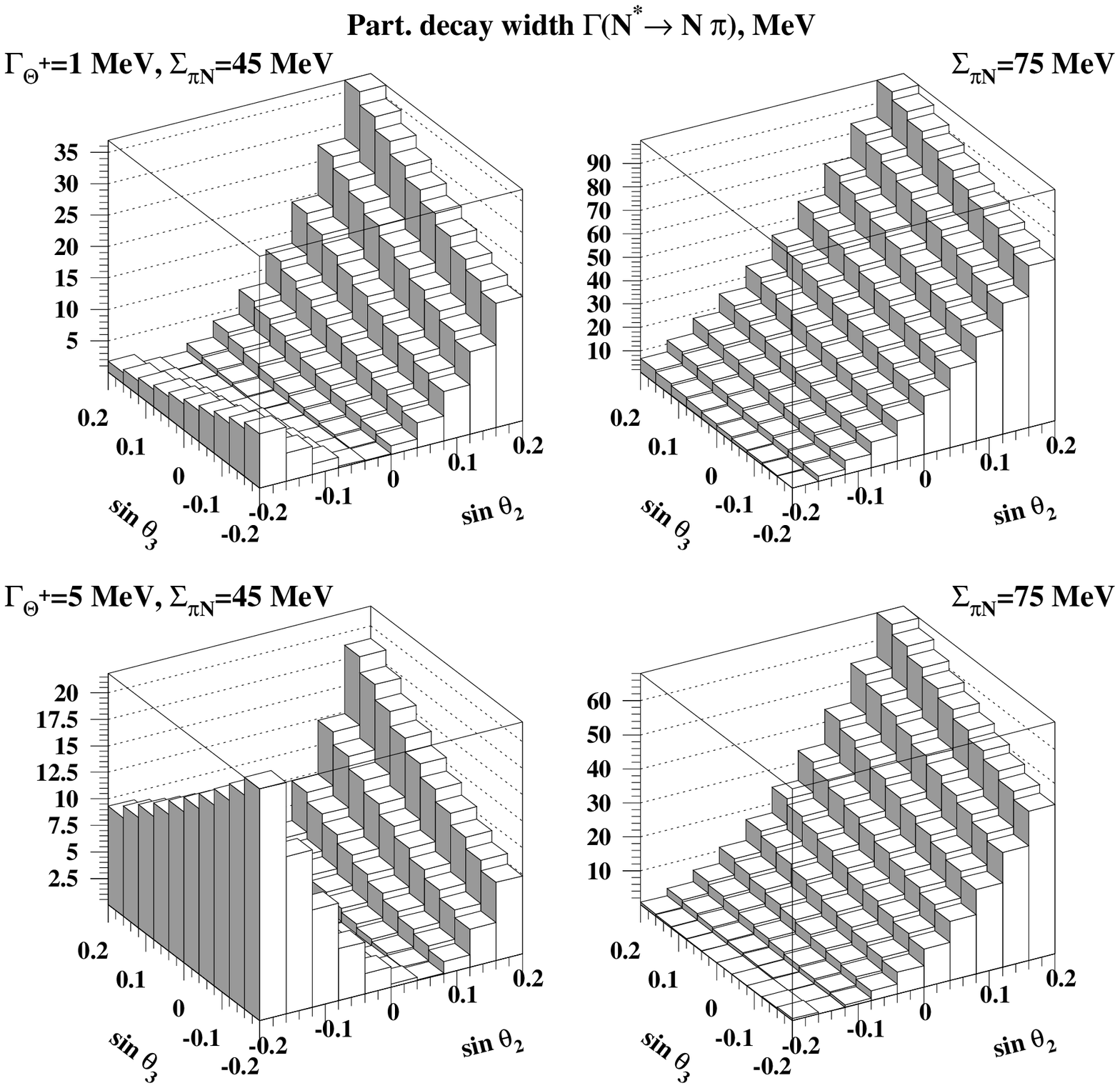}
\includegraphics[width=11cm,height=11cm]{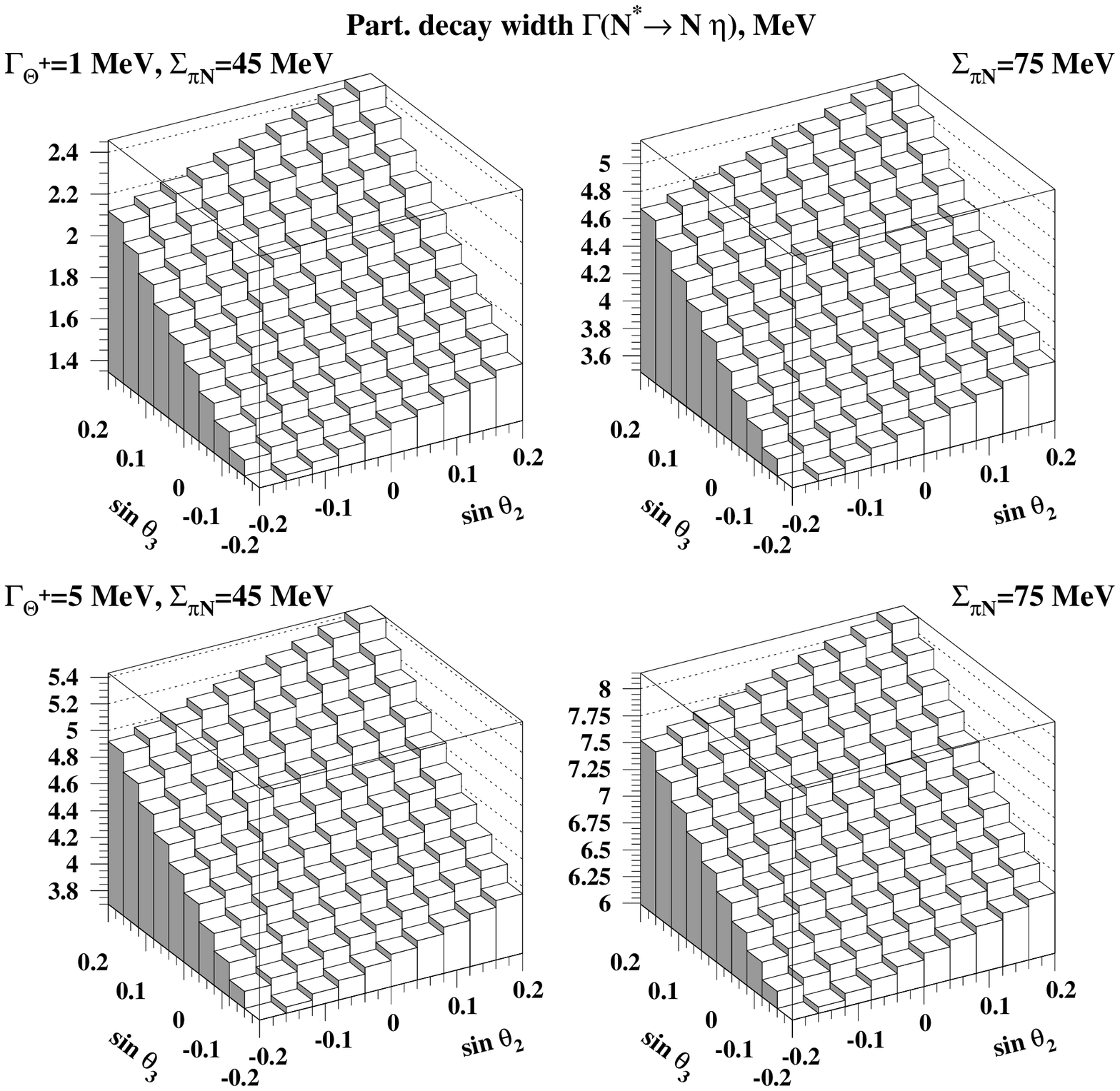}
\caption{$\Gamma_{N_{\at} \to N \, \pi}$ and $\Gamma_{N_{\at} \to N \, \eta}$
 as functions of $\theta_2$
and $\theta_3$ for $\Gamma_{\Theta^+}=1$ and 5 MeV and 
$\Sigma_{\pi \, N}=45$ and 75 MeV.}
\label{fig:nneta}
\end{figure}

\begin{figure}
\includegraphics[width=11cm,height=11cm]{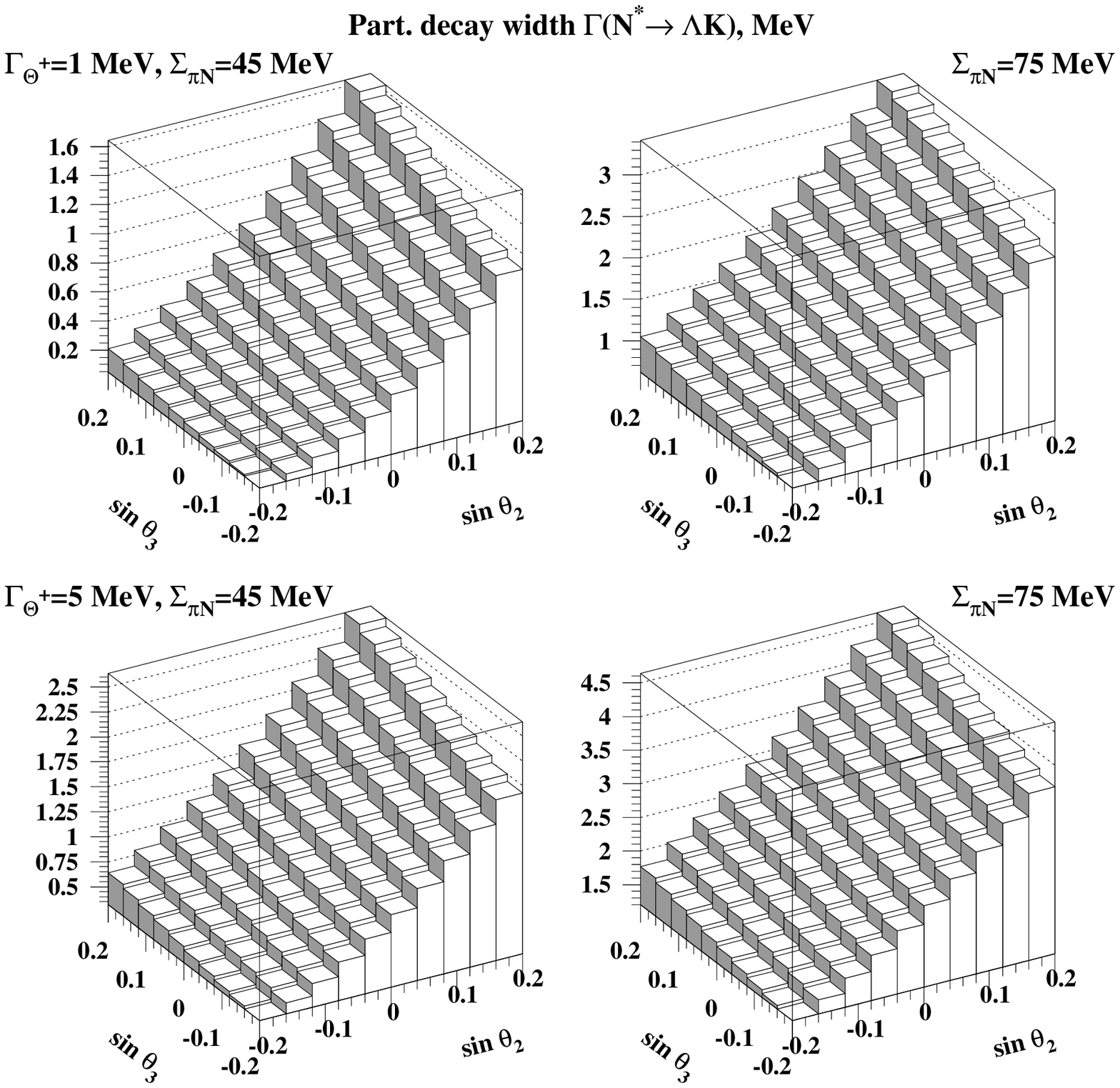}
\includegraphics[width=11cm,height=11cm]{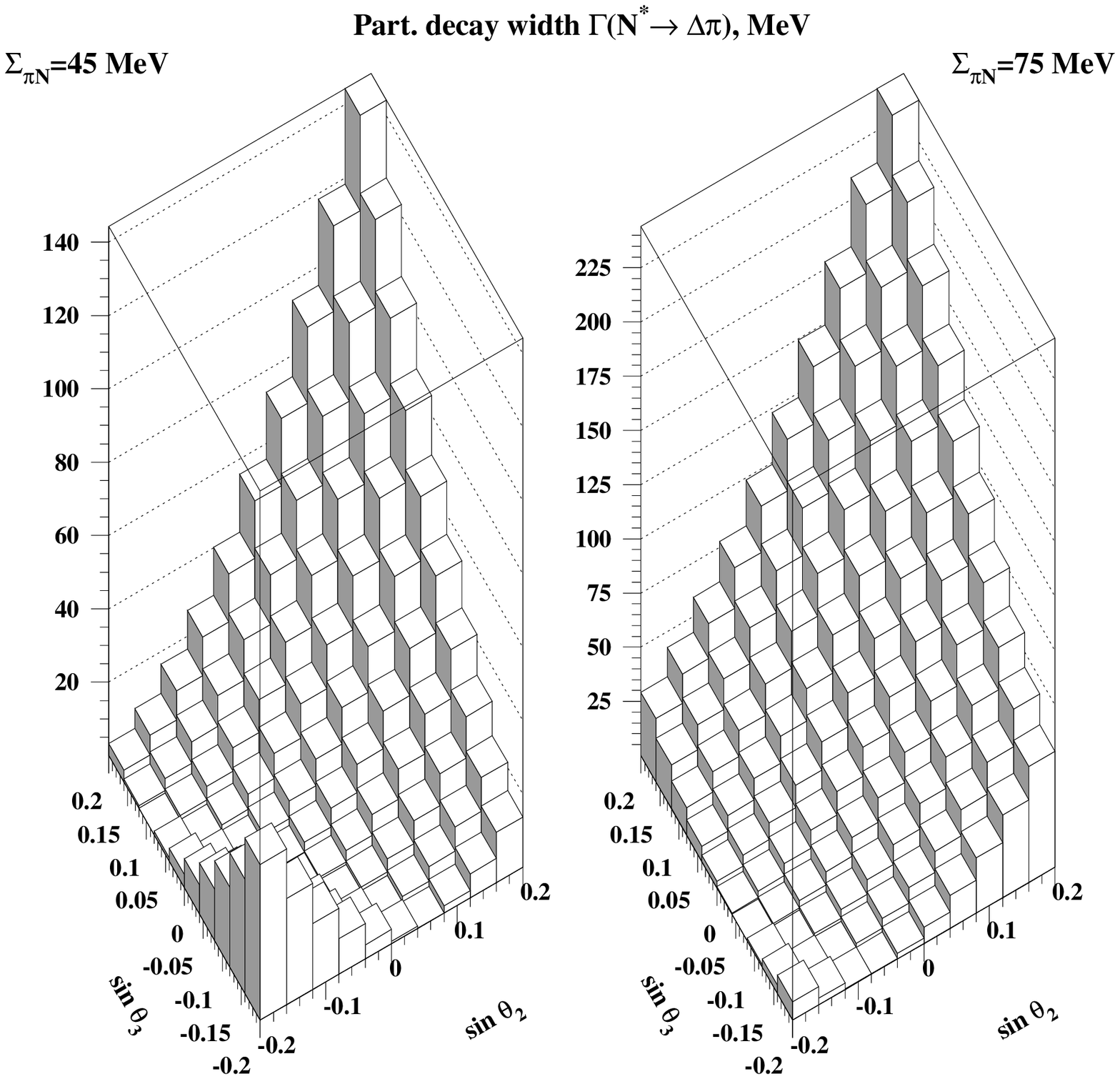}
\caption{$\Gamma_{N_{\at} \to \Lambda \, K}$ and $\Gamma_{N_{\at} \to \Delta \, \pi}$ as functions of $\theta_2$
and $\theta_3$ for  $\Gamma_{\Theta^+}=1$ and 5 MeV and 
$\Sigma_{\pi \, N}=45$ and 75 MeV. $\Gamma_{N_{\at} \to \Delta \, \pi}$
does not depend on $\Gamma_{\Theta^+}$. }
\label{fig:ndpi}
\end{figure}

It is instructive to separately examine different $N_{\at}$ partial decay
widths. As an example of such an analysis, we plot $\Gamma_{N_{\at} \to N \, \pi}$, $\Gamma_{N_{\at} \to N \, \eta}$, $\Gamma_{N_{\at} \to \Lambda \, K}$
and $\Gamma_{N_{\at} \to \Delta \, \pi}$ at $\Gamma_{\Theta^+}=1$ and 5 MeV 
and $\Sigma_{\pi \, N}=45$ and 75 MeV as functions of 
$\theta_2$ and $\theta_3$ in Figs.~\ref{fig:nneta} and 
\ref{fig:ndpi}.
Again, we present the results
with the positive $G_{\at}$ solution because the results with the negative 
$G_{\at}$ solution give too large $\Gamma_{N_{\at} \to N \, \pi}$ which seems 
to be ruled out by the PWA of \cite{Arndt2004}.
Note that the $N_{\at} \to \Sigma \, K$ decay is kinematically impossible for
the used $N_{\at}$ mass.

Figures~\ref{fig:nneta} and \ref{fig:ndpi}
 reveal the following approximate correlation
 between the partial decays widths.
Small $\Gamma_{N_{\at} \to N \, \pi}$ is correlated with small $\Gamma_{N_{\at} \to \Lambda \, K}$ and $\Gamma_{N_{\at} \to \Delta \, \pi}$. At the same time,
$\Gamma_{N_{\at} \to N \, \eta}$ does not have to be small. 
The $\Gamma_{N_{\at} \to \Delta \, \pi}$ peaks at large positive values of the
 mixing angles because the decay is possible only due to the mixing.
The variation of $\Gamma_{\Theta^+}$ or $\Sigma_{\pi \, N}$ in the considered
ranges does not 
significantly change this
trend (except for $\Gamma_{N_{\at} \to N \, \pi}$ at $\Gamma_{\Theta^+}=5$ MeV
 and  $\Sigma_{\pi \, N}=45$ MeV and for $\Gamma_{N_{\at} \to \Delta \, \pi}$
at $\Sigma_{\pi \, N}=45$ MeV in the region $\sin \theta_{2,3} \approx -0.2$)
 but merely affects the absolute values of the partial decay widths.

The trend of the correlation among the partial decay widths of $N_{\at}$
presented in Figs.~\ref{fig:nneta} and \ref{fig:ndpi} 
seems to be in a broad agreement with the present experimental situation. 
First, the PWA analysis of \cite{Arndt2004}
indicates that the candidate $N_{\at}$ state with mass near 1680 MeV
 should have a small partial decay width for the decay into the $N \, \pi$ final state,  
$\Gamma_{N_{\at} \to  N\, \pi} \leq 0.5$ MeV. We find that such small 
$\Gamma(N_{\at} \to  N\, \pi)$ solutions do exist (see the following 
discussion).
 Second, the GRAAL experiment indicates the existence of a narrow 
nucleon resonance near 1670 MeV in the reaction $\gamma \, n \to n \,
 \eta$ \cite{GRAAL2}. Therefore, $\Gamma_{N_{\at} \to  N\, \eta}$ should not 
be too small. Third, the STAR collaboration 
observes a narrow peak at $1734 \pm 0.5 \pm 5$ MeV 
and only a weak indication of a narrow peak at $1693 \pm 0.5$ MeV
in the $\Lambda \, K_{S}$ invariant
mass \cite{STAR}.
The former peak is interpreted as a candidate for $N_{\at}$; the latter is
hypothesized to be a candidate for the $\Xi(1690)$ state.
This does not fit well our picture of the $N_{\at}$ decays. Therefore, 
until the STAR results and conclusions are proven by other groups,
in the present analysis we ignore the peak at $1734$ MeV and assume that
the peak at $1693$ MeV corresponds to the $N_{\at}$. In this case, 
we interpret the STAR results as an indication that 
 the $\Gamma_{N_{\at} \to  \Lambda \, K}$ is not larger than 1-2 MeV, i.e.
the decay is possibly suppressed.

An examination of the general expressions for the $N_{\at}$ decay widths in
Eq.~(\ref{eq:decays:n}) shows that the picture of the $N_{\at}$ decays, which
emerges from the present experimental information, can be qualitatively
justified. Indeed, because of the minus sign in front of the positive $G_{\at}$
and $G_8$ coupling constants
and negative values of $g_{N_i N\, \eta}$
 in the expression for  $g_{N_{\at}  N\, \eta}$, the $g_{N_{\at}  N\, \eta}$
coupling constant can be enhanced compared to the $g_{N_{\at}  N\, \pi}$
coupling constant where the terms proportional to $G_8$ and 
$g_{N_{i}  N\, \pi}$  partially cancel
the $G_{\at}$ contribution. Note that this logic works only if $G_{\at}$
is positive. Therefore, unless specified, we always give our predictions
 for the positive $G_{\at}$, see Fig.~\ref{fig:g10}. 
 As to the $N_{\at} \to \Lambda \, K$ decay,
its partial width is suppressed in any case by the phase space factor.

In order to quantitatively examine how well the above mentioned constraints
 on the partial decay widths of  $N_{\at}$ are satisfied and at which mixing
angles, in Figs.~\ref{fig:nneta_cut_paper} and \ref{fig:ndpi_cut_paper}
 we show the partial decay widths of
 Figs.~\ref{fig:nneta} and \ref{fig:ndpi}
 only at  those $\theta_2$ and $\theta_3$ which correspond
 to $\Gamma_{N_{\at} \to  N\, \pi} \leq 1$ MeV. If this criterion is not met,
 the partial decay widths are not shown (they are formally set to zero).

\begin{figure}
\includegraphics[width=11cm,height=11cm]{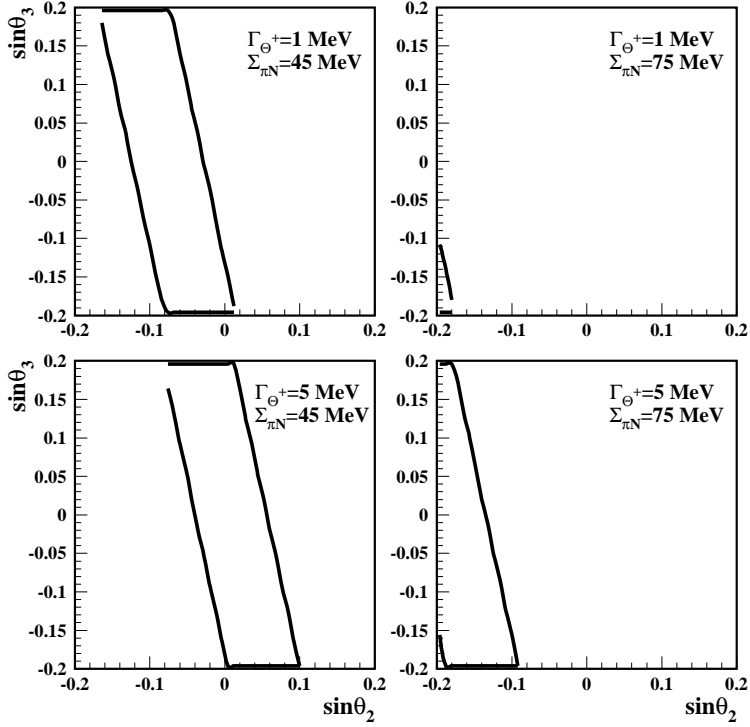}
\caption{The regions of the $\theta_2$ and $\theta_2$
mixing angles allowed by the  
$\Gamma_{N_{\at} \to  N\, \pi} \leq 1$ MeV condition.}
\label{fig:ugli}
\end{figure}

\begin{figure}
\includegraphics[width=11cm,height=11cm]{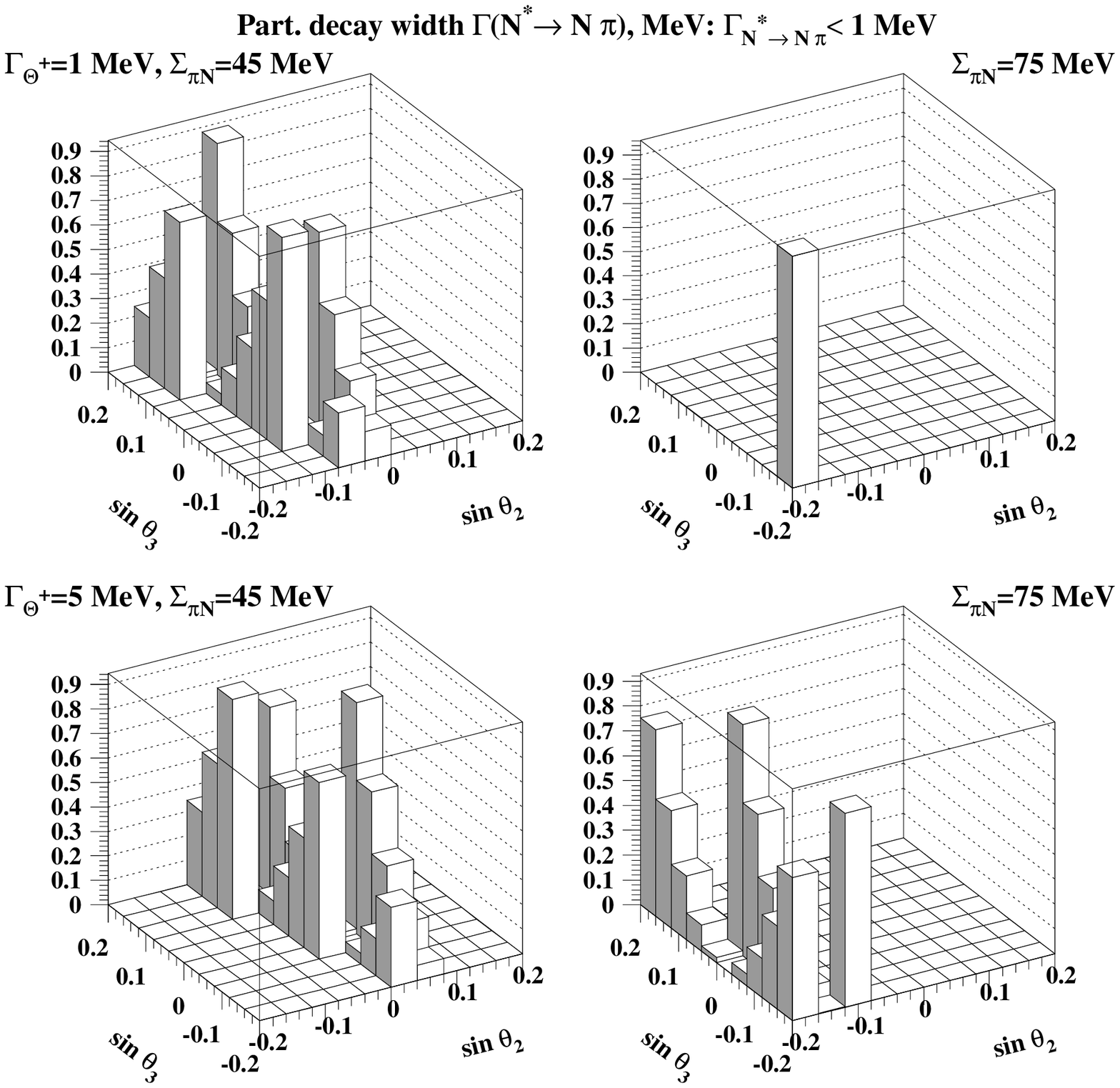}
\includegraphics[width=11cm,height=11cm]{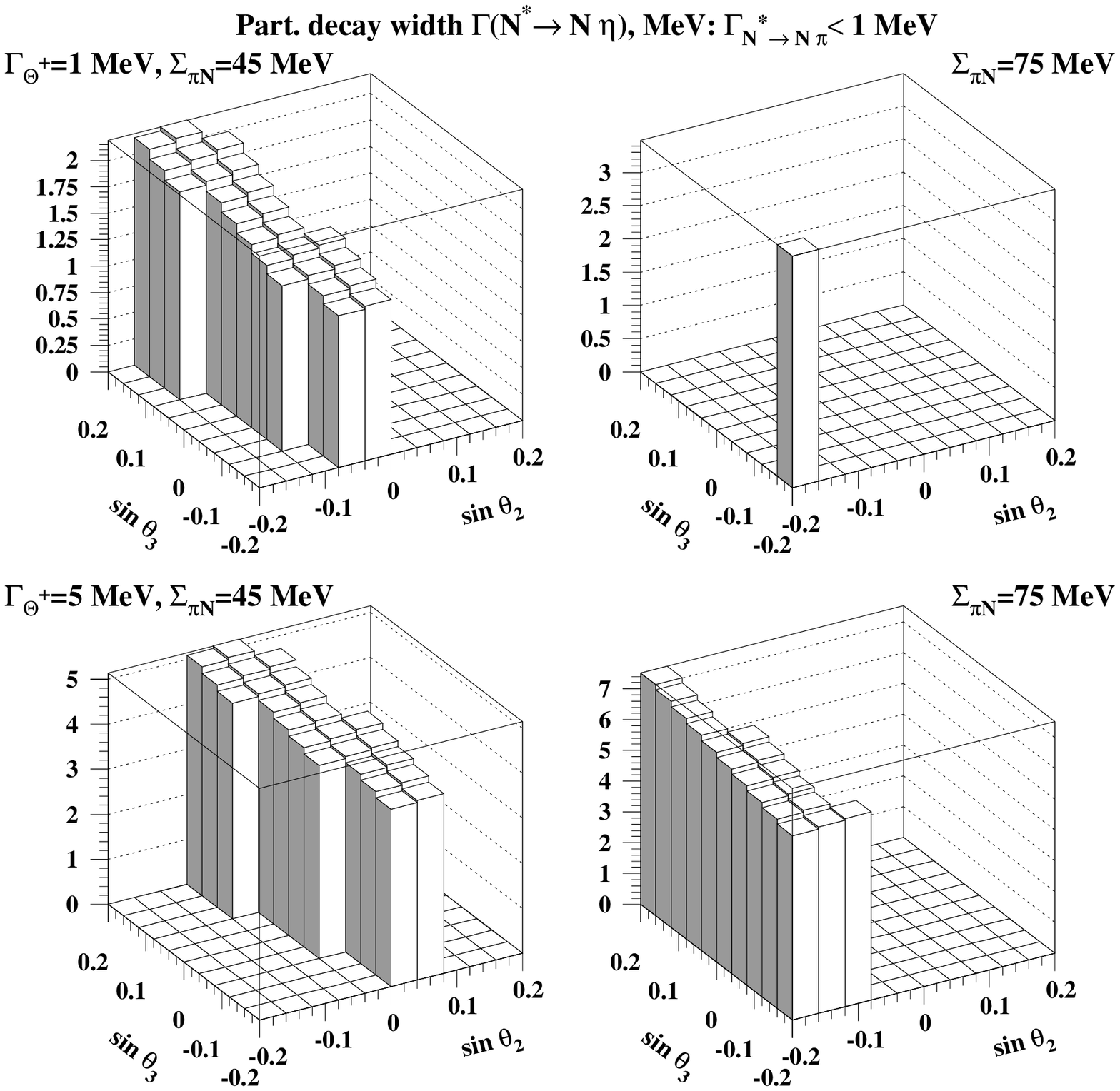}
\caption{$\Gamma_{N_{\at} \to N \, \pi}$ and $\Gamma_{N_{\at} \to N \, \eta}$ 
as functions of $\theta_2$
and $\theta_3$.  The decay widths are shown only where 
 $\Gamma_{N_{\at} \to  N\, \pi} \leq 1$ MeV.}
\label{fig:nneta_cut_paper}
\end{figure}

\begin{figure}
\includegraphics[width=11cm,height=11cm]{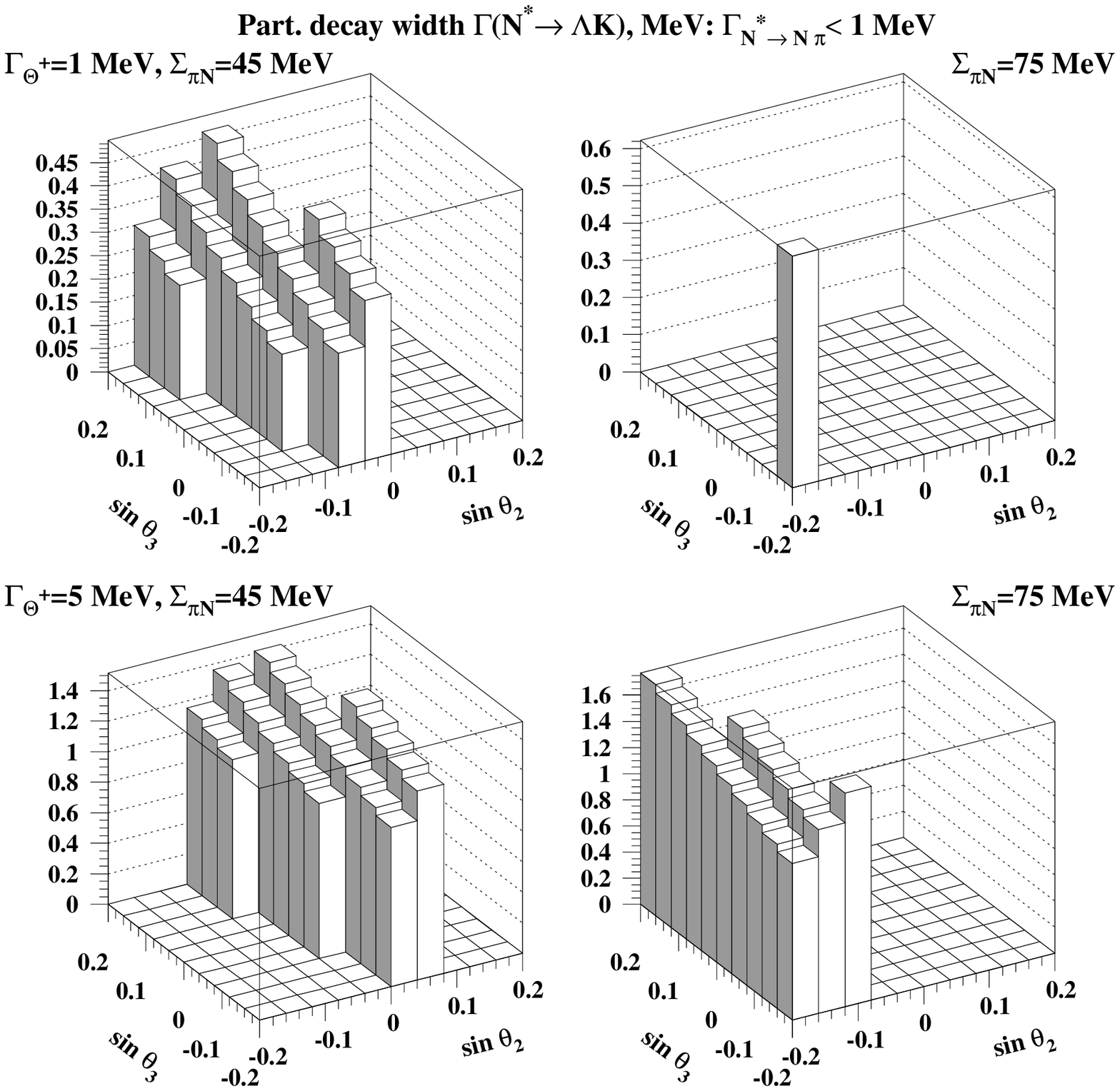}
\includegraphics[width=11cm,height=11cm]{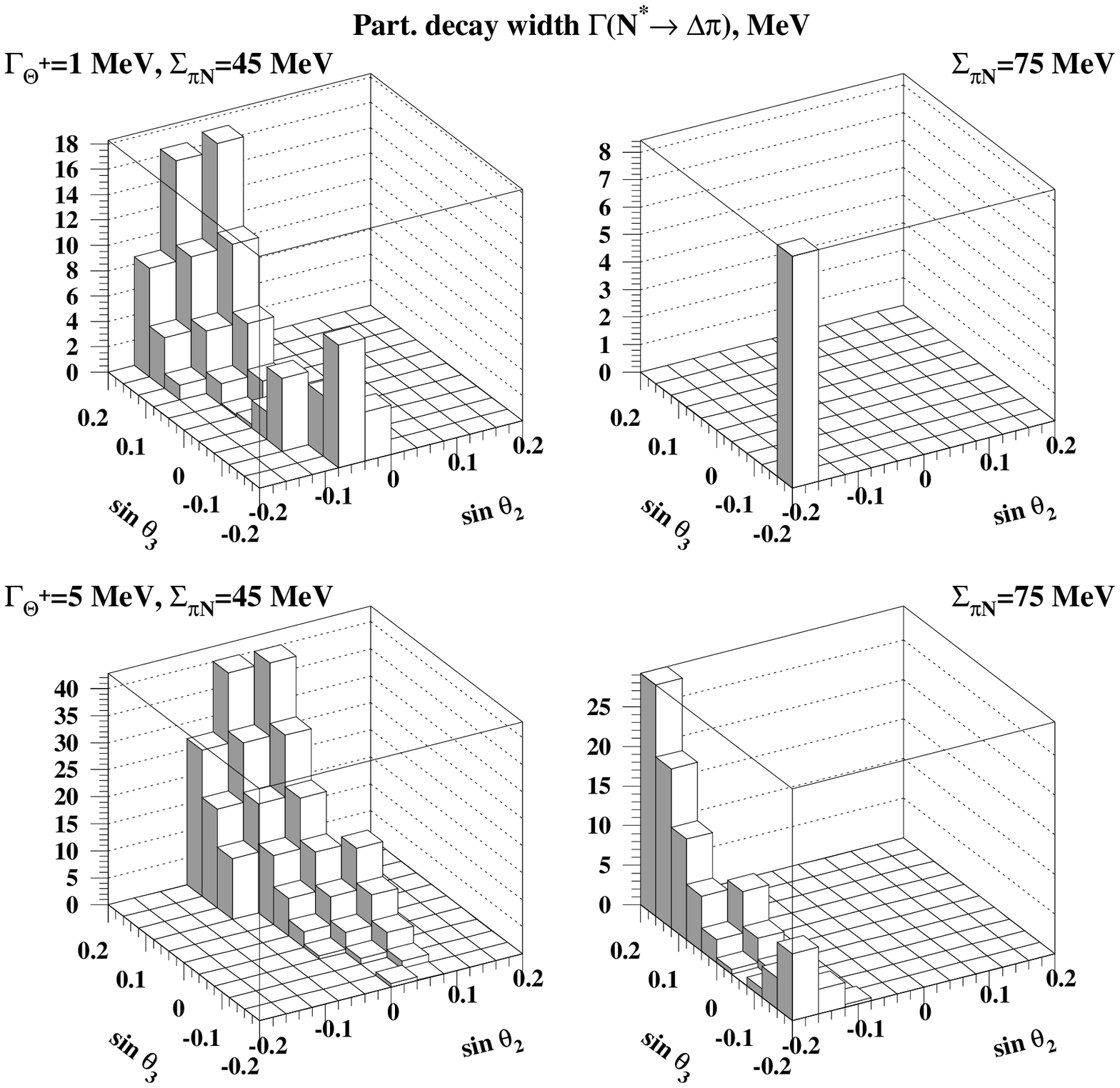}
\caption{$\Gamma_{N_{\at} \to \Lambda \, K}$ and 
$\Gamma_{N_{\at} \to \Delta \, \pi}$ as functions of $\theta_2$
and $\theta_3$.  The decay widths are shown only where 
 $\Gamma_{N_{\at} \to  N\, \pi} \leq 1$ MeV.}
\label{fig:ndpi_cut_paper}
\end{figure}

Figure~\ref{fig:ugli} presents the allowed regions of $\sin \theta_{2,3}$
when the $\Gamma_{N_{\at} \to  N\, \pi} \leq 1$ MeV condition is imposed.
At given $\sin \theta_{2}$, the two solid curves present the maximal and 
minimal values of $\sin \theta_{3}$.

As seen from Figs.~\ref{fig:nneta_cut_paper} and \ref{fig:ndpi_cut_paper}, 
an appropriate choice of the 
$\theta_2$ and $\theta_3$ mixing angles allows to simultaneously suppress
the  $\Gamma_{N_{\at} \to  N\, \pi}$ and $\Gamma_{N_{\at} \to  \Lambda \, K}$
(the latter is much more significantly suppressed at small values of
the total width of $\Theta^+$)
decay widths and to have the unsuppressed $\Gamma_{N_{\at} \to  N\, \eta}$
partial decay widths -- in accord with the present experimental situation
with the $N_{\at}$ decays, if $N_{\at}$ is identified with the 
GRAAL's $N(1670)$.

In addition, imposing the $\Gamma_{N_{\at} \to  N\, \pi} \leq 1$ MeV
constraint, we find that the sum of the considered two-body partial decay
 widths of $N_{\at}$, $\Gamma_{N_{\at}}^{{\rm 2-body}}$,  varies in the 
interval summarized in Table~\ref{table:n_int}.
Note that our analysis predicts the $N_{\at}$ total width which is
  somewhat larger than predicted by the PWA of \cite{Arndt2004}.
\begin{table}
\begin{tabular}{|c|c|c|c|}
\hline
$\Gamma_{\Theta^+}$ (MeV) & $\Sigma_{\pi \, N}$ (MeV)  & $\Gamma_{N_{\at}}^{{\rm 2-body, min}}$ (MeV) &  $\Gamma_{N_{\at}}^{{\rm 2-body, max}}$ (MeV) \\
\hline
1 & 45 & 2.1   & 30 \\
1 & 75 & 8.2   & 18 \\
5 & 45 & 5.2   & 66 \\
5 & 75 & 7.8   & 44 \\
\hline
\end{tabular}
\caption{The range of change of $\Gamma_{N_{\at}}^{{\rm 2-body}}$.}
\label{table:n_int}
\end{table} 

Note that because of the approximate correlation between
 $\Gamma_{N_{\at} \to N \, \pi}$ and
$\Gamma_{N_{\at} \to \Lambda \, K}$ in our analysis,
 it seems unnatural to simultaneously have  sizable 
$\Gamma_{N_{\at} \to \Lambda \, K}$ and suppressed  
$\Gamma_{N_{\at} \to N \, \pi}$. Therefore,
 our analysis disfavors the
identification of the peak at 1734 MeV seen by the STAR collaboration
in the $\Lambda \, K_{S}$ invariant mass \cite{STAR} with $N_{\at}$,
which should have a suppressed partial decay width for the $N\,\pi$ final 
state \cite{Arndt2004}.

\subsection{Decays of $\Sigma_{\at}$}

\begin{figure}
\includegraphics[width=12cm,height=12cm]{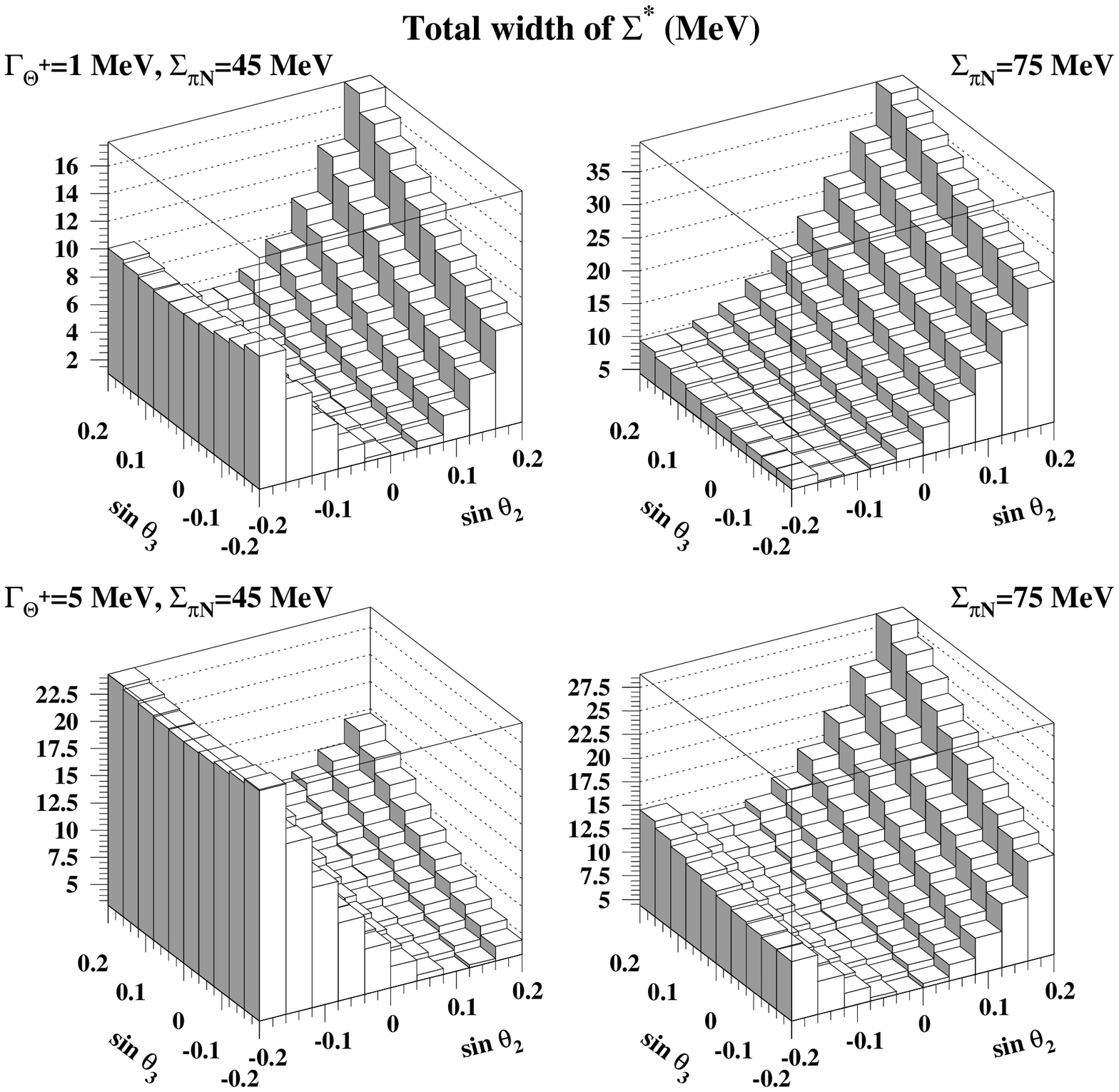}
\caption{The total width of $\Sigma_{\at}$ as a function of $\theta_2$
and $\theta_3$ for $\Gamma_{\Theta^+}=1$ and 5 MeV and for
$\Sigma_{\pi \, N}=45$ and 75 MeV.}
\label{fig:stot}
\end{figure}

Next we turn to the decays of $\Sigma_{\at}$ which we consider analogously
to the decays of  $N_{\at}$. Figure~\ref{fig:stot} depicts the
 dependence of the
total width of the $\Sigma_{\at}$, $\Gamma_{\Sigma_{\at}}$, on the $\theta_2$
and $\theta_3$ mixing angles and on $\Gamma_{\Theta^+}$ and
 $\Sigma_{\pi \, N}$. Unlike the total width of $N_{\at}$,
 $\Gamma_{\Sigma_{\at}}$ depends on both $\Gamma_{\Theta^+}$ and 
$\Sigma_{\pi \, N}$.

\begin{figure}
\includegraphics[width=11cm,height=11cm]{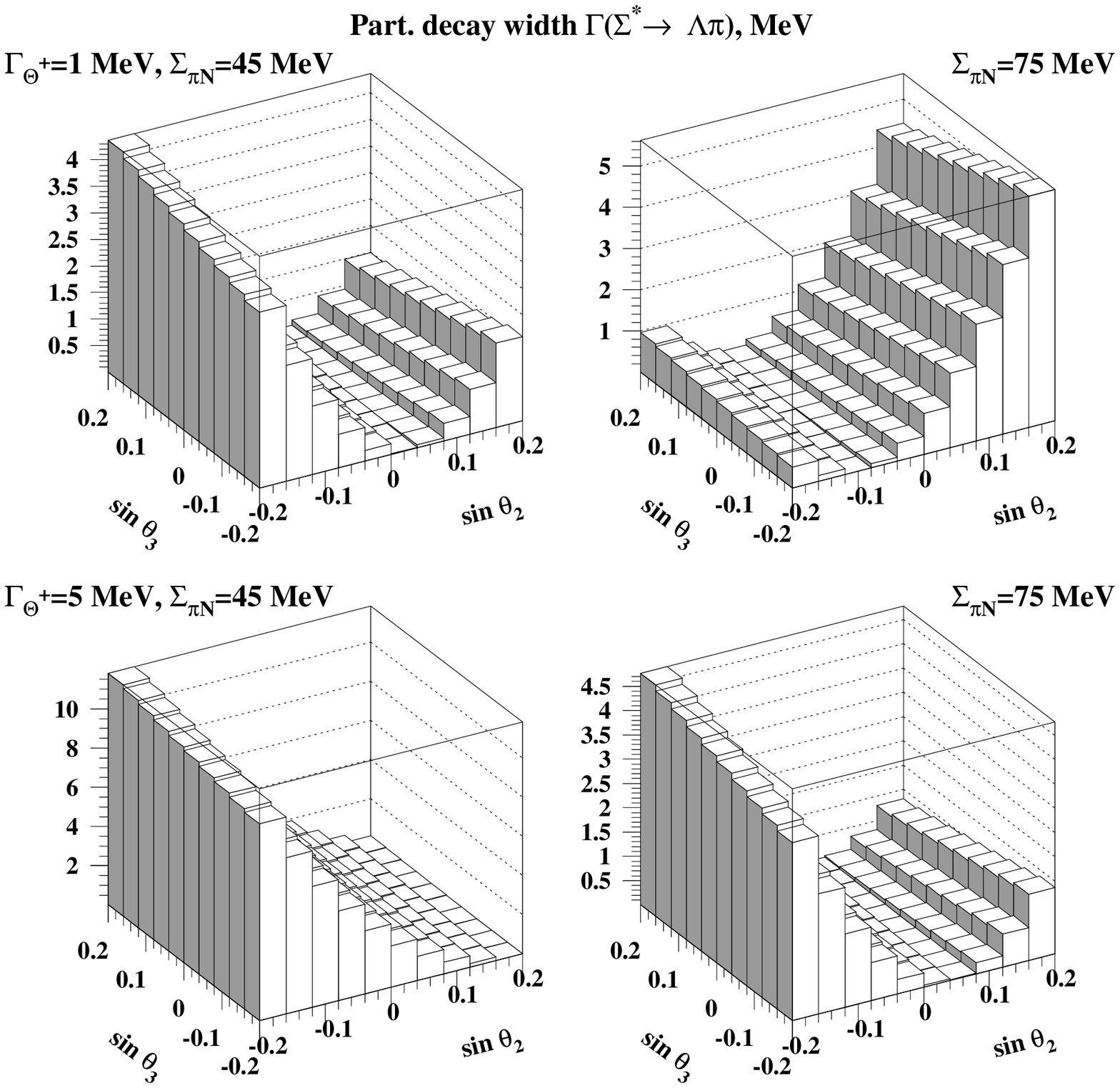}
\includegraphics[width=11cm,height=11cm]{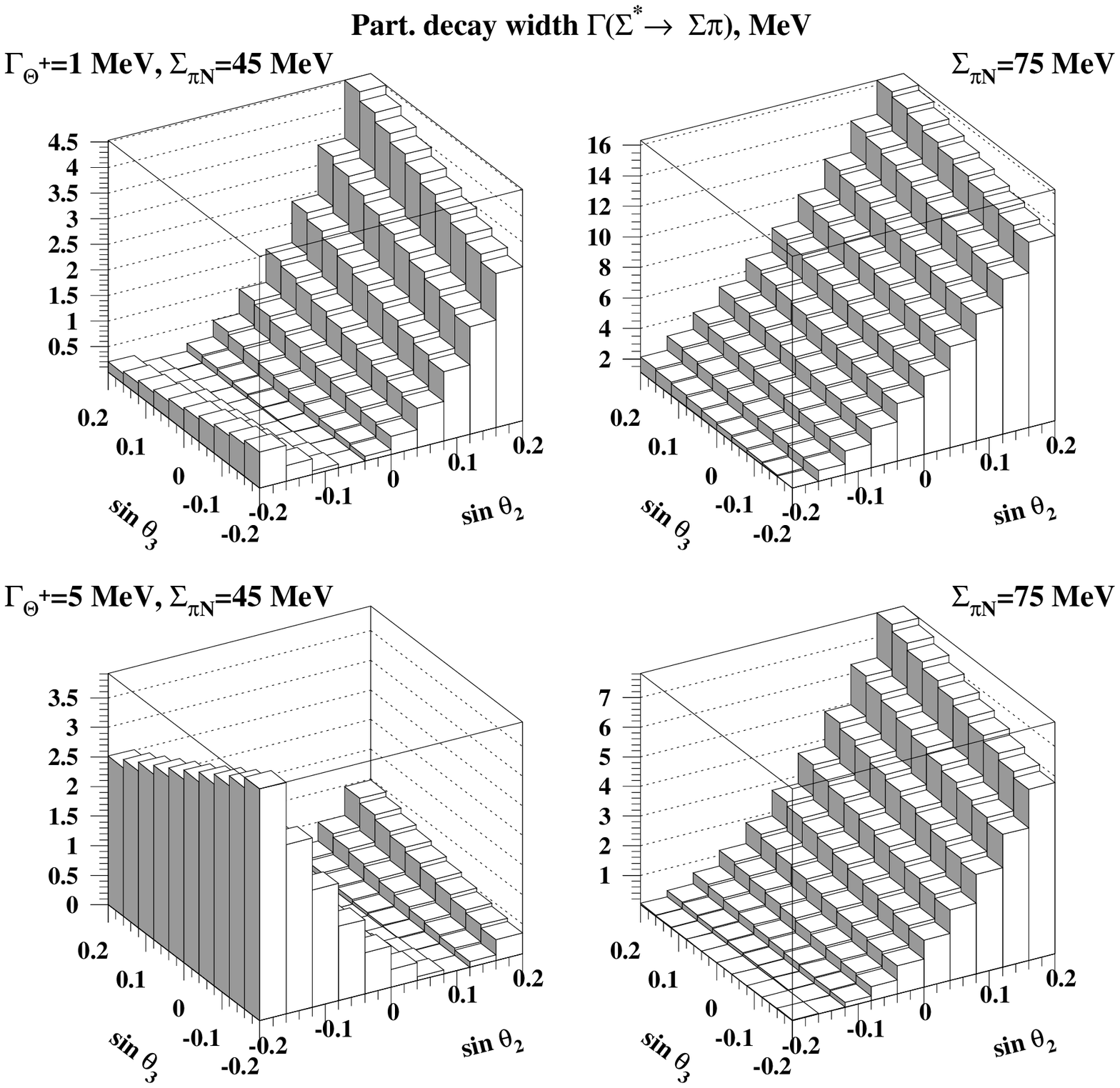}
\caption{$\Gamma_{\Sigma_{\at} \to \Lambda \, \pi}$ and 
$\Gamma_{\Sigma_{\at} \to \Sigma \, \pi}$ as functions
 of $\theta_2$
and $\theta_3$ for $\Gamma_{\Theta^+}=1$ MeV and 5 and 
$\Sigma_{\pi \, N}=45$ and 75 MeV.}
\label{fig:sspi}
\end{figure}

\begin{figure}
\includegraphics[width=11cm,height=11cm]{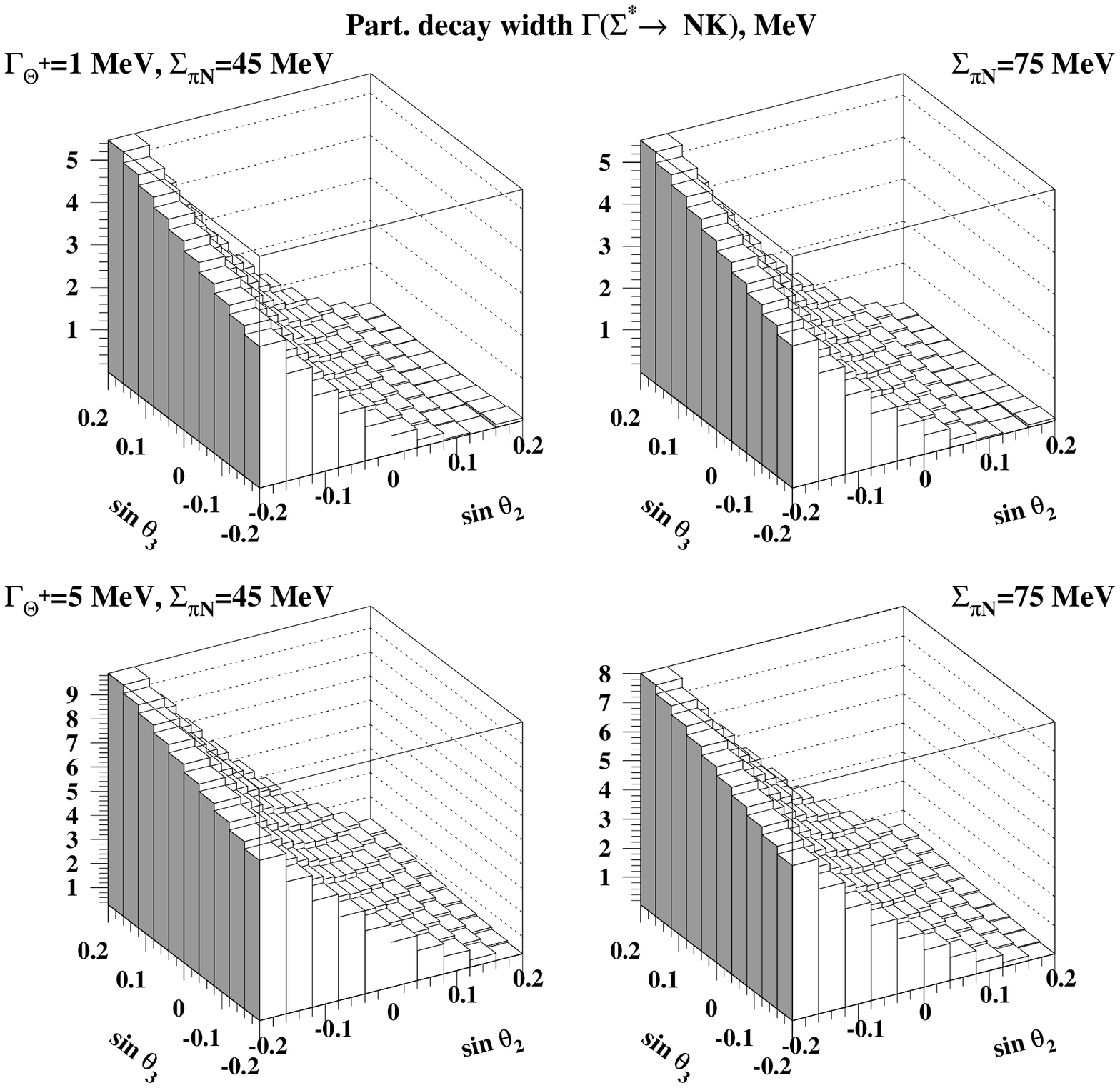}
\includegraphics[width=11cm,height=11cm]{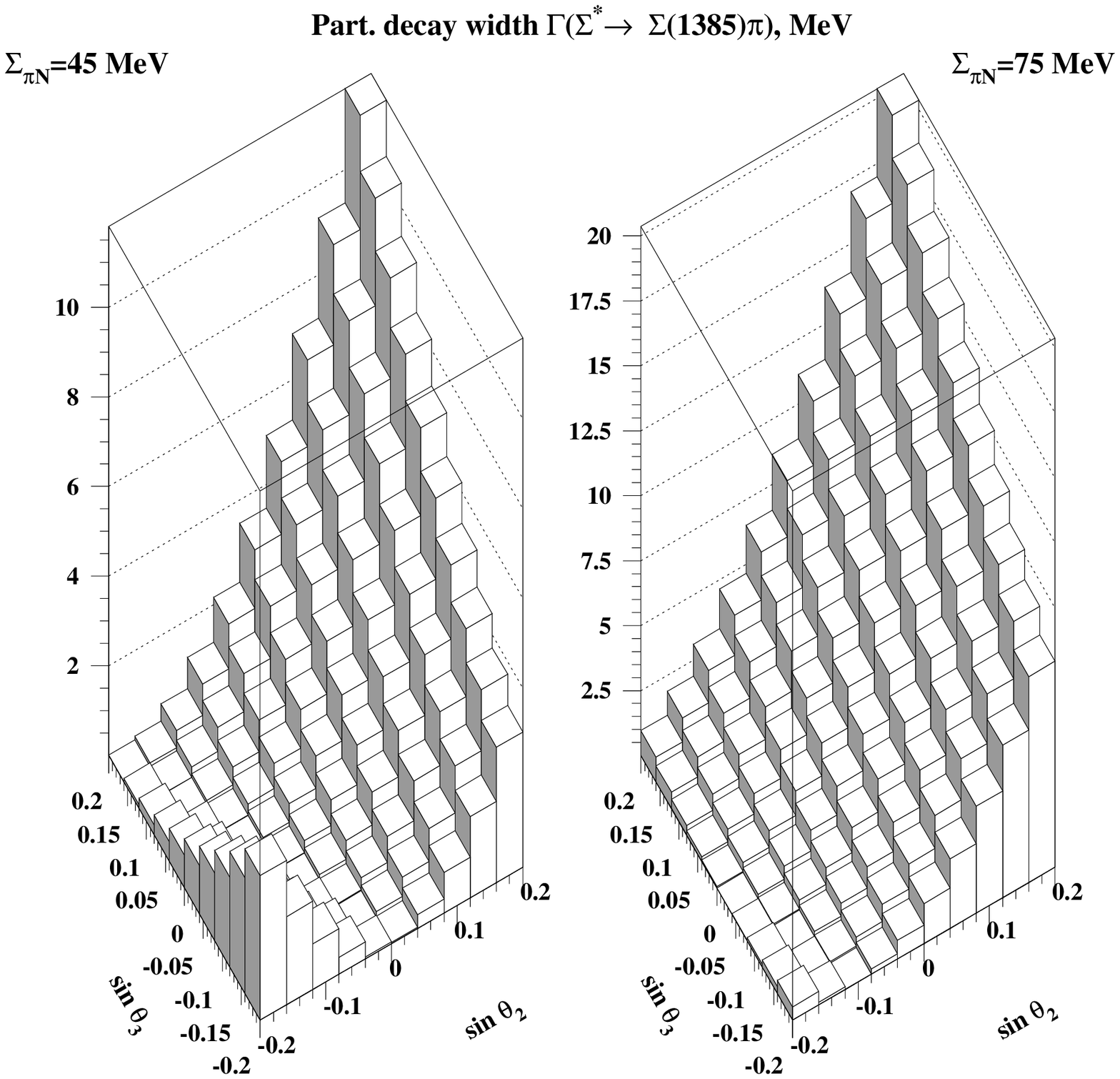}
\caption{$\Gamma_{\Sigma_{\at} \to N \, K}$ and 
$\Gamma_{\Sigma_{\at} \to \Sigma(1385) \, \pi}$ as functions
 of $\theta_2$
and $\theta_3$ for $\Gamma_{\Theta^+}=1$ and 5 MeV and 
$\Sigma_{\pi \, N}=45$ and 75 MeV. $\Gamma_{\Sigma_{\at} \to \Sigma(1385) \, \pi}$ does not depend on $\Gamma_{\Theta^+}$.}
\label{fig:ss10pi}
\end{figure}

Next we examine correlations between partial decay widths of $\Sigma_{\at}$. 
In Figs.~\ref{fig:sspi} and \ref{fig:ss10pi}
 we present $\Gamma_{\Sigma_{\at} \to  \Lambda \, \pi}$,
$\Gamma_{\Sigma_{\at} \to  \Sigma \, \pi}$, $\Gamma_{\Sigma_{\at} \to  N \, \overline{K}}$ and $\Gamma_{\Sigma_{\at} \to  \Sigma_{10} \, \pi}$ partial decay
widths as functions of $\theta_2$ and $\theta_3$ for $\Gamma_{\Theta^+}=1$ and 
5 MeV
and for $\Sigma_{\pi \, N}=45$ and 75 MeV.
The $\Gamma_{\Sigma_{\at} \to  \Sigma \, \eta}$ partial decay width is very
small because the decay takes place very near its threshold -- we do not
show $\Gamma_{\Sigma_{\at} \to  \Sigma \, \eta}$ in this work.

The approximate correlation between the partial decay widths, which is seen
in the $N_{\at}$ case, is much less pronounced in the case of 
$\Sigma_{\at}$. The following, very approximate, trend
can be seen in 
Figs.~\ref{fig:sspi} and \ref{fig:ss10pi}: At 
$\sin \theta_2 \approx -0.2$,
 large $\Gamma_{\Sigma_{\at} \to  \Lambda \, \pi}$  correspond to large
 $\Gamma_{\Sigma_{\at} \to  N \, \overline{K}}$ and to
the suppressed $\Gamma_{\Sigma_{\at} \to  \Sigma \, \pi}$ (except for
$\Gamma_{\Theta^+}=5$ MeV and $\Sigma_{\pi \, N}=45$ MeV); 
suppressed  $\Gamma_{\Sigma_{\at} \to  \Lambda \, \pi}$ correspond to
 suppressed $\Gamma_{\Sigma_{\at} \to  N \, \overline{K}}$
($\Gamma_{\Sigma_{\at} \to  \Lambda \, \pi}$ at $\Gamma_{\Theta^+}=1$ MeV
 and $\Sigma_{\pi \, N}=75$ MeV is an exception); 
$\Gamma_{\Sigma_{\at} \to  \Sigma(1385) \, \pi}$ increases towards 
large and positive $\theta_2$ and $\theta_3$  mixing angles.

These trends can be traced back to  the general expressions for
the $\Sigma_{\at}$ coupling constants, see Eq.~(\ref{eq:decays:s}), along 
with the octet coupling constants of Eq.~(\ref{eq:newcoupling:ns}).
For instance, at $\Gamma_{\Theta^+}=1$ MeV, the increase of
 $\theta_2$ towards its maximal value (as seen from 
Eq.~(\ref{eq:newcoupling:ns}), mixing with octet 4 hardly matters
for the $\Sigma_{\at}$ decays) leads to an increase of 
$g_{\Sigma_{\at} \to \Lambda \, \pi}$ and $g_{\Sigma_{\at} \to \Sigma \, \pi}$
and to a decrease of  $g_{\Sigma_{\at} \to N \, \overline{K}}$.

\begin{figure}
\includegraphics[width=11cm,height=11cm]{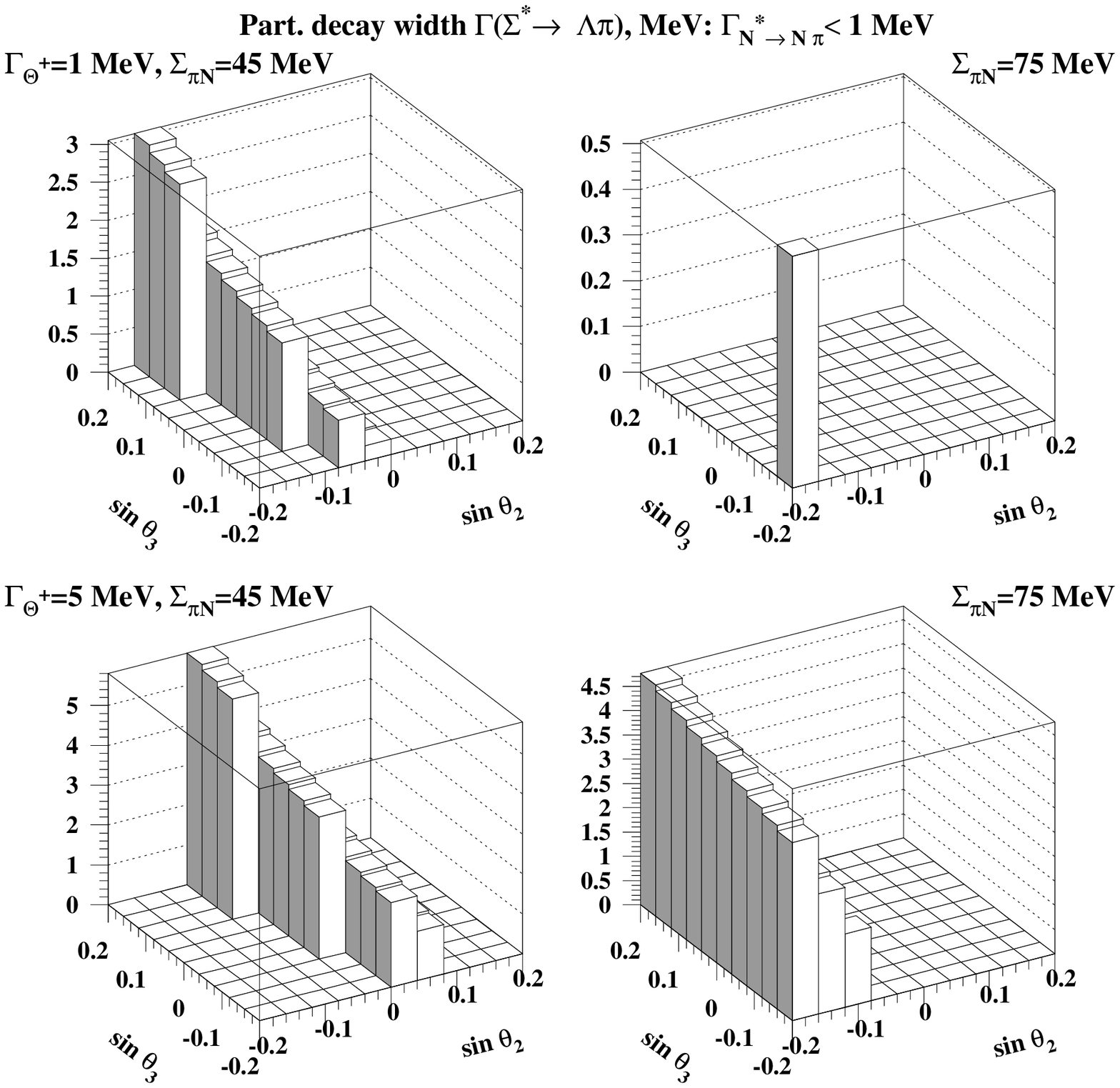}
\includegraphics[width=11cm,height=11cm]{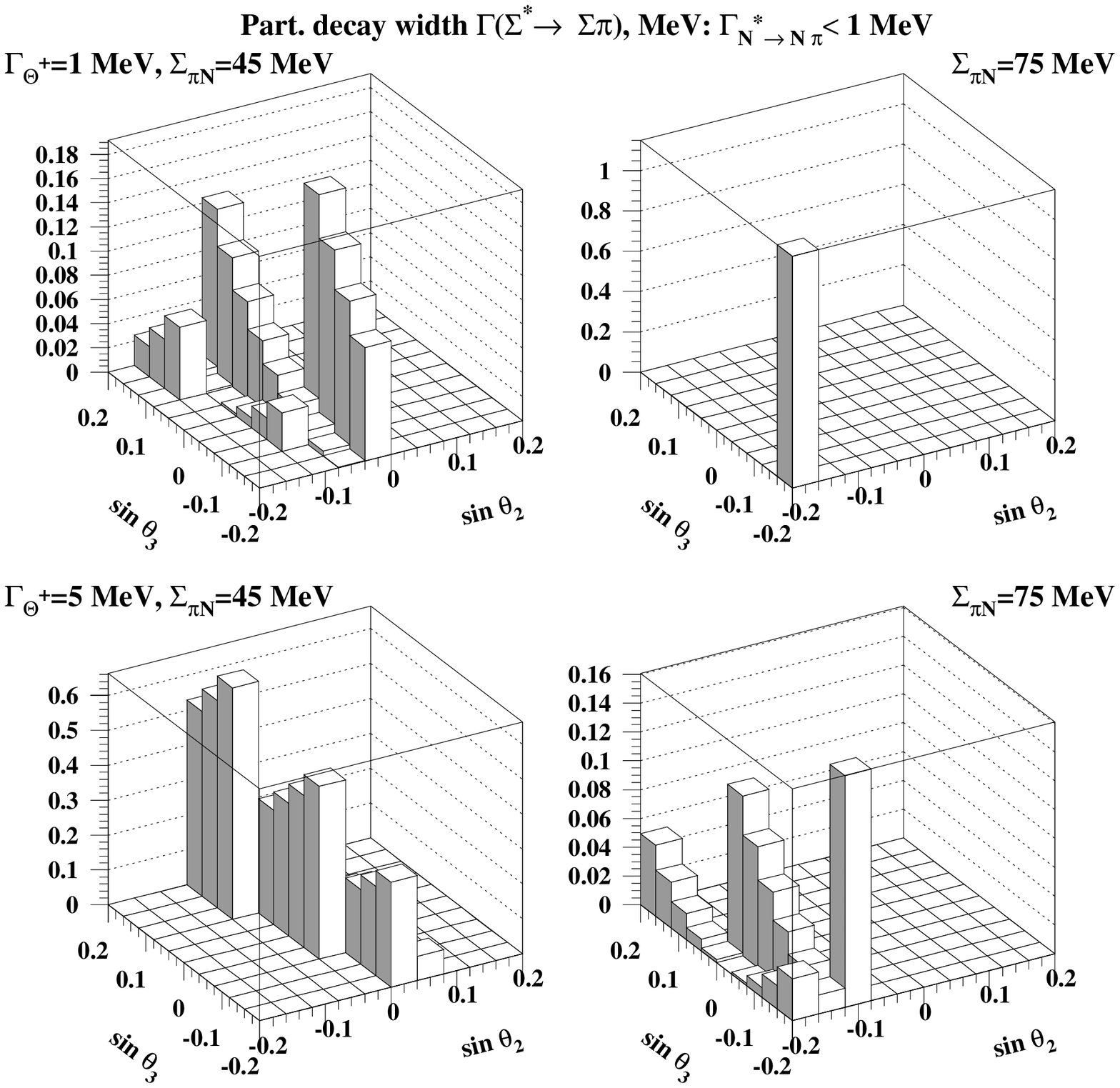}
\caption{$\Gamma_{\Sigma_{\at} \to \Lambda \, \pi}$ and 
$\Gamma_{\Sigma_{\at} \to \Sigma \, \pi}$ as functions of $\theta_2$
and $\theta_3$. The decay widths are shown only where 
 $\Gamma_{N_{\at} \to  N\, \pi} \leq 1$ MeV.}
\label{fig:sspi_cut}
\end{figure}

\begin{figure}
\includegraphics[width=11cm,height=11cm]{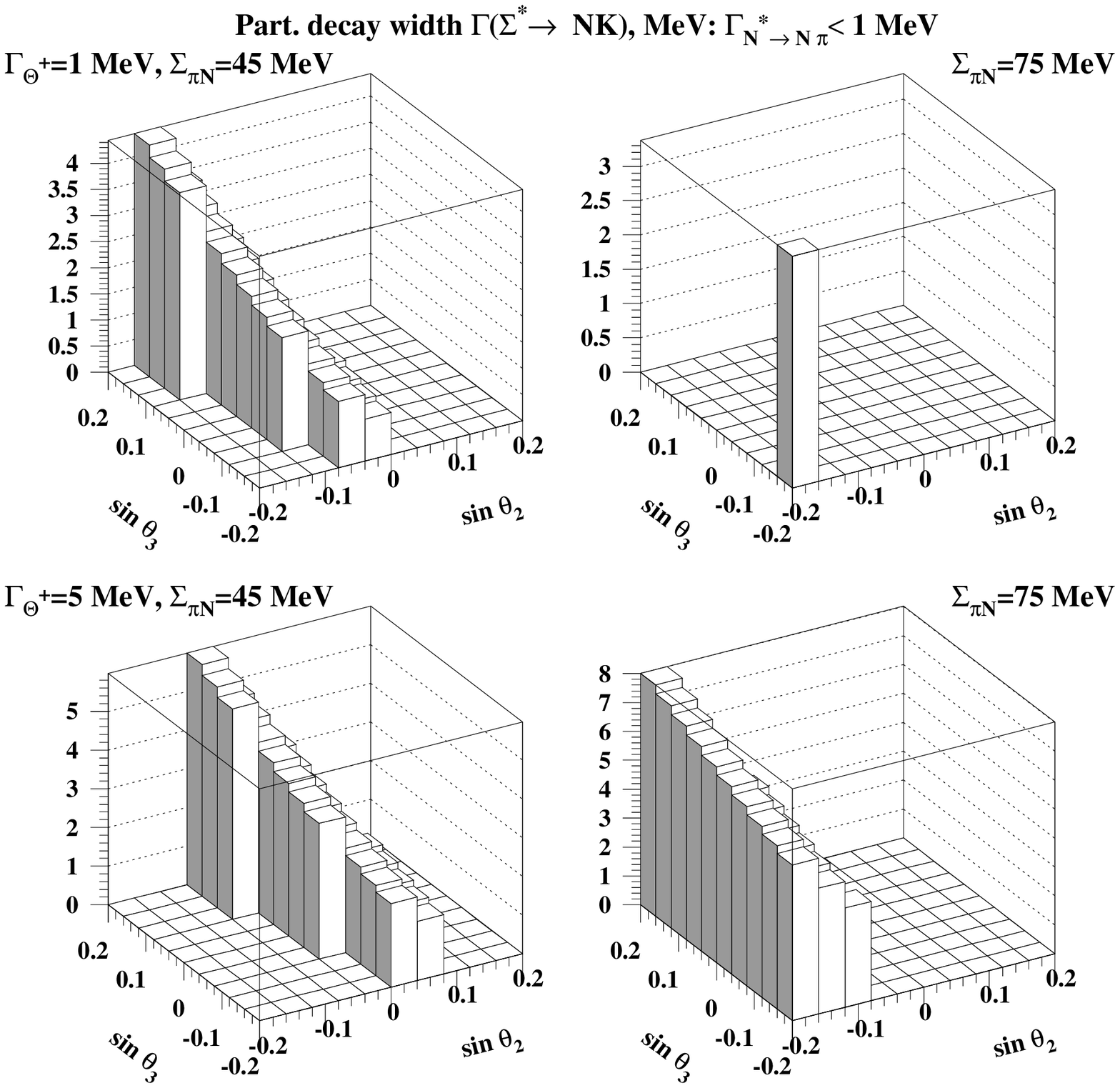}
\includegraphics[width=11cm,height=11cm]{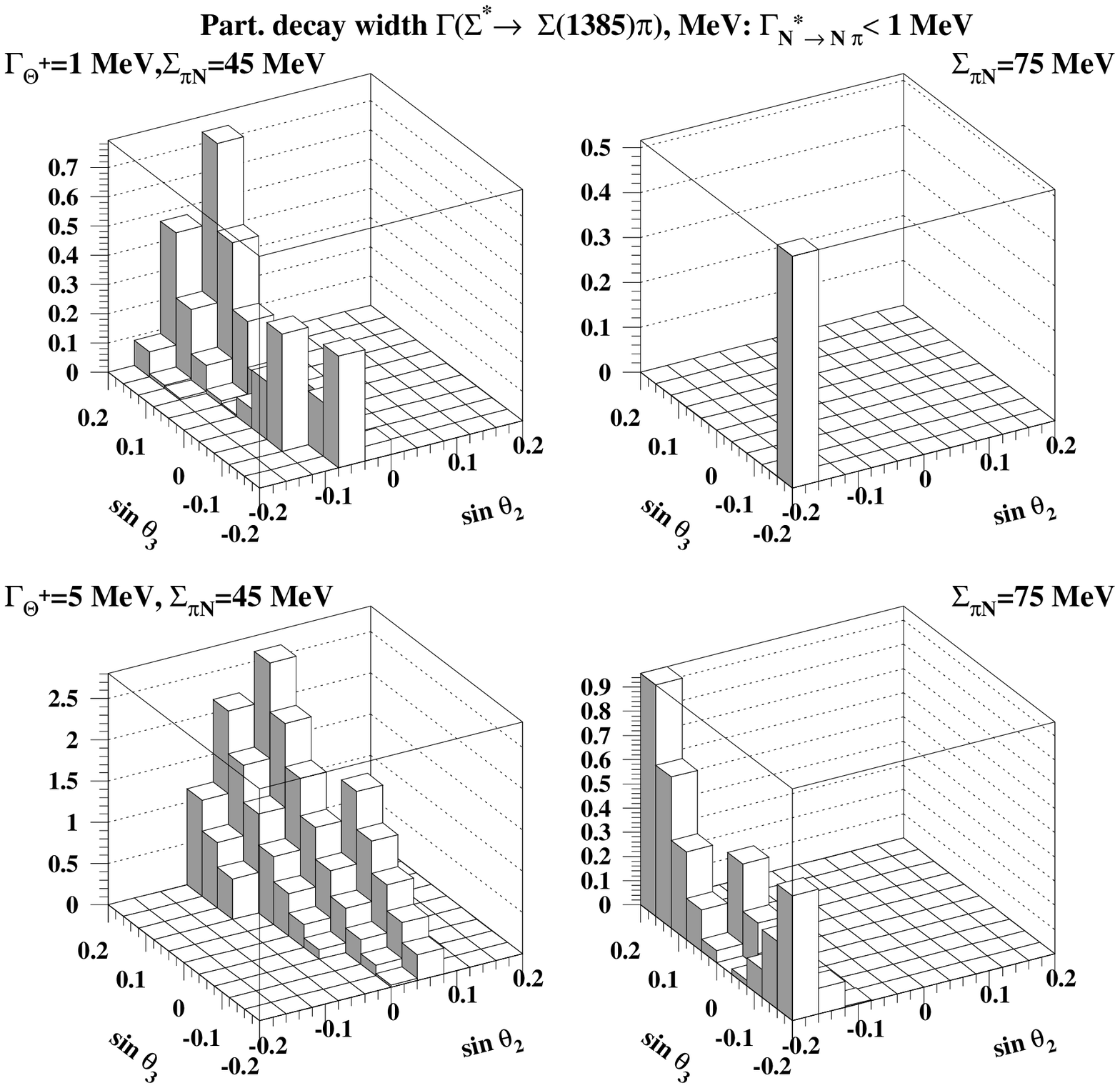}
\caption{$\Gamma_{\Sigma_{\at} \to N \, \overline{K}}$ and 
$\Gamma_{\Sigma_{\at} \to \Sigma(1385) \, \pi}$ as functions of $\theta_2$
and $\theta_3$. The decay widths are shown only where 
 $\Gamma_{N_{\at} \to  N\, \pi} \leq 1$ MeV.}
\label{fig:ss10pi_cut}
\end{figure}

Finally, in Figs.~\ref{fig:sspi_cut}
and \ref{fig:ss10pi_cut}
we show the partial decay widths of $\Sigma_{\at}$  
in the range of $\theta_2$ and $\theta_3$ where 
$\Gamma_{N_{\at} \to  N\, \pi} \leq 1$ MeV. 


Taking the sum of the considered two-body partial decay widths of 
$\Sigma_{\at}$,  $\Gamma_{\Sigma_{\at}}^{{\rm 2-body}}$, 
 we find that in presence of 
 the $\Gamma_{N_{\at} \to  N\, \pi} \leq 1$ MeV
constraint, $\Gamma_{\Sigma_{\at}}^{{\rm 2-body}}$ varies in the 
interval summarized in Table~\ref{table:s_int}.
\begin{table}
\begin{tabular}{|c|c|c|c|}
\hline
$\Gamma_{\Theta^+}$ (MeV) & $\Sigma_{\pi \, N}$ (MeV)  & $\Gamma_{\Sigma_{\at} }^{{\rm 2-body, min}}$ (MeV) &  $\Gamma_{N_{\at}}^{{\rm 2-body, max}}$ (MeV) \\
\hline
1 & 45 & 0.95  & 9.1 \\
1 & 75 & 5.7  & 6.5 \\
5 & 45 & 2.9   & 15 \\
5 & 75 & 5.3   & 15 \\
\hline
\end{tabular}
\caption{The range of change of $\Gamma_{\Sigma_{\at}}^{{\rm 2-body}}$.}
\label{table:s_int}
\end{table} 
In the spirit of the antidecuplet, $\Sigma_{\at}$ appears
 to be much narrower than 
any known $\Sigma$ baryon with mass larger than 1650 MeV \cite{PDG}.

Among the experiments reporting the $\Theta^+$ signal, 
there were four experiments \cite{Neutrino,SVD, HERMES,ZEUS} 
where the $\Theta^+$ was observed as a peak in the $p \, K_S$ invariant mass
and strangeness was not tagged.
Since $\Sigma_{\at}$ decays in the same final state ($N\, \overline{K}$),
the four experiments give direct information on the 
$\Sigma_{\at} \to N\, \overline{K}$ decay -- virtually the only experimental
 piece of information on $\Sigma_{\at}$! Below we shall consider the relevant 
results of the four experiments in some detail.

The analysis of neutrino-nuclear (mostly neon) interaction data \cite{Neutrino}
 clearly reveals  the $\Theta^+$ peak as well as a number of other peaks in
 the $1650 < M_{p \, K_S} < 1850$ MeV mass region, which cannot be 
suppressed by the random-star elimination procedure, see Fig.~3 of \cite{Neutrino}. This is an agreement with the present analysis (Figs.~\ref{fig:ss10pi}
and \ref{fig:ss10pi_cut}), which shows that the branching ratio for the
$\Sigma_{\at} \to N\, \overline{K}$ decay width is essential. 
Obviously, any of the peaks of \cite{Neutrino} in the 1700-1800 MeV mass range
could be a good candidate  for  $\Sigma_{\at}$.

Similar conclusions apply to the SVD collaboration result \cite{SVD}. Before
the cuts aimed to enhance the $\Theta^+$ signal are imposed, 
the $p \, K_S$ invariant
 mass spectrum contains at least two prominent peaks in the 1700-1800 MeV mass
 range (see Fig.~5 of \cite{SVD}), each of which can be interpreted as
$\Sigma_{\at}$. 

The HERMES \cite{HERMES} and ZEUS \cite{ZEUS} $p \, K_S$ invariant 
mass spectra extend only up to
1.7 MeV and, therefore, do not allow to make any conclusions about the 
$\Sigma_{\at}$. 

In addition to the $p \, K_S$ invariant mass spectrum, the HERMES 
collaboration also presents the $\Lambda \, \pi$ invariant mass spectrum
in order to see if the observed peak in the $p \, K_S$ final state is indeed 
generated by the $\Theta^+$ and not by some yet unknown $\Sigma^{\ast}$ 
resonance \cite{Lorenzon}. The  $\Lambda \, \pi$ invariant mass spectrum
has no resonance structures except for the prominent $\Sigma(1385)$ peak. 
According to our analysis, the $\Gamma_{\Sigma_{\at} \to \Lambda\, \pi}$
 partial decay width is in general not large. Moreover, at a small total
width of the $\Theta^+$ and large 
$\Sigma_{\pi \, N}$, $\Gamma_{\Sigma_{\at} \to \Lambda\, \pi}$ is 
dramatically suppressed which seems to be exactly what is needed to 
comply with no-observation of $\Sigma_{\at}$ in the HERMES 
$\Lambda \, \pi$ invariant mass spectrum! 

However, it is difficult to make any quantitative conclusions from the
 HERMES spectrum because of its course scale. Indeed, if one naively
assumes that the HERMES spectrum
 reveals only the well-known $\Sigma(1385)$, several well-established
 $\Sigma$ baryons \cite{PDG}
with noticeable branching ratios to the  $\Lambda \, \pi$ final state are
 missed.

In our opinion, the \cite{Neutrino,SVD} data already contain an indication 
 for a narrow 
$\Sigma_{\at}$ member of the antidecuplet in the 1700-1800 MeV mass range
 and the \cite{HERMES,ZEUS,Lorenzon} data do not rule out its existence.
 Obviously, a dedicated search for the $\Sigma_{\at}$ signal in the
 $p \, K_S$ and $\Lambda \, \pi$ invariant mass spectra is needed in
 order to address several key issues surrounding this least known member
 of the antidecuplet. 

It is interesting that one can offer a candidate $\Sigma_{\at}$ state, 
$\Sigma(1770)$, which has been known for almost 
three decades \cite{PDG,Gopal77}.
Indeed, the one-star $\Sigma(1770)$ has the required $J^P=1/2^+$ and the mass,
 a $14 \pm 4$ \% branching ratio in the $N \, \overline{K}$ final state and 
poorly known but still probably rather small branching ratios into the
$\Lambda \, \pi$ and $\Sigma \, \pi$ final states.
Moreover, our comprehensive SU(3) analysis of baryon multiplets \cite{GPreview}
disfavors that $\Sigma(1770)$ belongs to any octet or decuplet, i.e. it
is very natural to assign $\Sigma(1770)$ to the antidecuplet, see also
\cite{Hosaka}.

However, the $\Sigma(1770)$ with the total width $72 \pm 10$ MeV appears to be 
too wide for the antidecuplet, see Table~\ref{table:s_int}.
 On the other hand, taking the $\Sigma(1770)$ 
branching ratios at their face values \cite{Gopal77}, 
\begin{eqnarray}
 &&Br(N \, \overline{K})=0.14 \pm 0.04 \nonumber\\
&&\sqrt{Br(N \, \overline{K})Br(\Lambda \, \pi)} <  0.04  \nonumber\\
&&\sqrt{Br(N \, \overline{K})Br(\Sigma\, \pi)} < 0.04 \,,
\end{eqnarray}
we observe that the experimental value for the sum of the two-hadron branching 
ratios is less than 20\%, i.e. the two-hadron decays constitute
 less than 1/5 of all  possible (including many-body)  decays.
This indicates that our predictions for the 
$\Gamma_{\Sigma_{\at}}^{{\rm 2-body}}$, which is of the order of $7-15$ MeV,
do not exclude the $\Sigma(1770)$ as candidate for $\Sigma_{\at}$.

In conclusion, our qualitatively reasonable description of the decays of the 
$\Sigma(1770)$ along with its ``correct'' spin, parity and mass makes 
$\Sigma(1770)$ an appealing candidate for $\Sigma_{\at}$. This conjecture 
can be tested only by a dedicated analysis of the $\Sigma$ baryon spectrum in 
the 1700-1800 mass range.

It was argued in \cite{27plet} that the antidecuplet mixing with
 27-plet and 35-plet SU(3) representations
 has a significant impact on the antidecuplet
 decays. Therefore, in order to study how robust our predictions 
for the antidecuplet
 decays  with respect to the additional mixing, in Appendix~\ref{sec:2735}
 we explicitly add the 27-plet and 35-plet contributions to the antidecuplet
 coupling constants and repeat the entire analysis of Sect.~\ref{sec:main}.
We arrive at the following two scenarios corresponding to two possible
solutions for the $G_{\at}$ coupling constant.

When we use the larger and always positive  $G_{\at}$ solution, 
imposing the $\Gamma_{N_{\at} \to N \, \pi} < 1$ MeV condition, we reproduce
the  qualitative picture of the $N_{\at}$ decays presented in 
Sect.~\ref{sec:main}. At the same time, the correlation between the 
$\Sigma_{\at}$ change. For instance, we generally have 
$\Gamma_{\Sigma_{\at} \to \Lambda \, \pi} > \Gamma_{\Sigma_{\at} \to N \,
 \overline{K}}$, which makes it impossible to identify 
$\Sigma_{\at}$ with $\Sigma(1770)$.

Using the other solution for $G_{\at}$, which is mostly negative and becomes
positive and small in magnitude towards larger $\Sigma_{\pi \, N}$, we can 
obtain a picture of the $N_{\at}$ decays, 
which is marginally compatible with the 
one presented in Sect.~\ref{sec:main}, only if $\Gamma_{\Theta^+}=1$ MeV
and $\Sigma_{\pi \, N}=75$ MeV.
At the same time, the correlation between the partial decay widths
 of $\Sigma_{\at}$ reminds the pattern of the $\Sigma(1770)$ decays. 
A characteristic feature of this scenario of the antidecuplet decays is 
rather narrow $N_{\at}$ and $\Sigma_{\at}$.

\section{Conclusions and Discussion}

In this work, we consider mixing of the antidecuplet with three 
$J^P=1/2^+$  octets in the framework of approximate flavor SU(3) symmetry.
These are the ground-state octet, the octet containing $N(1440)$, 
$\Lambda(1600)$, $\Sigma(1660)$
and $\Xi(1690)$ (referred to as octet 3) and the octet containing $N(1710)$,
$\Lambda(1800)$, $\Sigma(1880)$ and $\Xi(1950)$ (referred to as octet 4).
Assuming that SU(3) symmetry is broken only by non-equal masses of hadrons 
within a given unitary multiplet and by small mixing among multiplets
  and that SU(3) symmetry is exact in the
decay vertices, we derived expressions for the partial decay widths
all members of the antidecuplet in the limit of small mixing angles.
The results are expressed in terms of the universal SU(3) coupling constants
and three mixing angles $\theta_i$. For the transition between 
the antidecuplet and the ground-state octet, the coupling constants and the
$\theta_1$ mixing angle are determined by the chiral quark soliton
model. For the transition between the antidecuplet and octets 3 and 4, 
the coupling constants are determined by fitting to the octet decays, while
the $\theta_2$ and $\theta_3$ mixing angles are left as free parameters.
Finally, the total width of the $\Theta^+$ and the pion-nucleon sigma term
are treated as external parameters which are varied in the following
intervals:
$1 \leq \Gamma_{\Theta^+} \leq 5$ MeV; $45 \leq \Sigma_{\pi \, N} \leq 75$ MeV.
The $\theta_2$ and $\theta_3$ mixing angles are varied in the $-0.2 \leq \sin \theta_{2,3} \leq 0.2$ interval.

In this analysis, the $\theta_2$ and $\theta_3$ mixing angles were
 constrained by identifying
the $N_{\at}$ state with the peak around 1670 MeV observed by the 
GRAAL experiment in the $\gamma \, n \to n \, \eta$ reaction \cite{GRAAL2}.
The fact that $N_{\at}$ might have mass around 1670 was earlier 
predicted in \cite{DP2004,Arndt2004}.
In general, the nowadays experimental information on the $N_{\at}$
 decays can be
qualitatively  summarized
as follows: $\Gamma_{N_{\at} \to N \, \eta}$ is sizable \cite{GRAAL2};
 $\Gamma_{N_{\at} \to N \, \pi}$ is small \cite{Arndt2004}; 
$\Gamma_{N_{\at} \to \Lambda \, K}$ is possibly suppressed in order to
 comply with the STAR result \cite{STAR}; the total width of $N_{\at}$
is of the order of 10-20 MeV \cite{Arndt2004}.
We find that all these conditions can be met by a suitable choice of
 $\theta_2$ and $\theta_3$, see Figs.~\ref{fig:ugli}, \ref{fig:nneta_cut_paper} 
and \ref{fig:ndpi_cut_paper}.

Our approach based on the mixing of the antidecuplet with three octets
 appears to be an improvement over the scenario of \cite{Arndt2004},
 which implied that the antidecuplet mixes only with
 the ground-state octet, 
  because we are able to demonstrate that a narrow $\Theta^+$ and small
 $\Gamma_{N_{\at} \to N \, \pi}$ become consistent due to the mixing with 
several multiplet (this was only hypothesized in \cite{Arndt2004}).

After the mixing angles are constrained by the $N_{\at}$ decays, we
make definite predictions for the $\Sigma_{\at}$ decays, see 
Figs.~\ref{fig:sspi_cut} and  \ref{fig:ss10pi_cut}. In particular, at 
$\Gamma_{\Theta^+}=1$ MeV and $\Sigma_{\pi \, N}=75$ MeV, we predict that 
$\Gamma_{\Sigma_{\at} \to N \, \overline{K}}$ is significantly 
enhanced compared to $\Gamma_{\Sigma_{\at} \to \Lambda \, \pi}$
and $\Gamma_{\Sigma_{\at} \to \Sigma\, \pi}$. This correctly reproduces the 
trend of the branching ratios of $\Sigma(1770)$, a known
 one-star $\Sigma$ baryon
with $J^P=1/2^+$ \cite{PDG}. However, the sum of the predicted
 two-body partial 
decay widths is much smaller than the experimental
value for the total width of $\Sigma(1770)$, $\Gamma_{\Sigma(1770)}=72 \pm 10$
MeV.
In any case, we believe that our conjecture that $\Sigma_{\at}$ could be 
identified with 
$\Sigma(1770)$ deserves further experimental and phenomenological analyses.

We discuss that a narrow $\Sigma_{\at}$ state with mass near 
1770 MeV could be searched for in the $K_S \, p$ 
invariant mass spectrum using the available data which already 
revealed the $\Theta^+$ signal \cite{Neutrino,SVD}.

In order to access a possible theoretical uncertainty of our 
predictions,  we examine how our predictions for the antidecuplet decays
 change when we introduce an additional mixing of the antidecuplet with
 a 27-plet \cite{27plet}. 
We observe the following two scenarios, which correspond to two possible
solutions for the $G_{\at}$ coupling constant.
Using the larger  $G_{\at}$ solution and 
imposing the $\Gamma_{N_{\at} \to N \, \pi}$ constraint, we reproduce
the  qualitative picture of the $N_{\at}$ decays presented in 
Sect.~\ref{sec:main}. At the same time, the correlation between the 
$\Sigma_{\at}$ change, which makes it impossible to identify 
$\Sigma_{\at}$ with $\Sigma(1770)$.

Using the smaller $G_{\at}$ solution, which is mostly negative and becomes
positive at small $\Gamma_{\Theta^+}$ and large $\Sigma_{\pi \, N}$, we can 
still obtain a picture of the $N_{\at}$ decays, 
which is marginally compatible with the 
one presented in Sect.~\ref{sec:main}.
At the same time, the correlation between the partial decay widths
 of $\Sigma_{\at}$ is similar to that of $\Sigma(1770)$. 
In addition, this scenario predicts rather narrow  $N_{\at}$ and 
$\Sigma_{\at}$ states.

Any further progress in our understanding of the properties of the
antidecuplet should come from experiments. One of the main purposes 
of this work was to show that it is possible to bring order to a multitude
 of direct and indirect experimental information on the antidecuplet decays
 using the very fundamental and successful principle of approximate flavor SU(3) symmetry and to ignite interest among experimentalists
 in studying all members of the antidecuplet. 

\begin{acknowledgments}

This work is supported by the Sofia Kovalevskaya Program of the Alexander
von Humboldt Foundation and by FNRS and IISN (Belgium).
We thank Ya.I. Azimov and I. Strakovsky
 for carefully reading the manuscript and making 
numerous useful comments.

\end{acknowledgments}

\appendix
\section{Determination of SU(3) coupling constants of octets 3 and 4}
\label{sec:34}

In this appendix, we derive the SU(3) coupling constants of octets 3 and 4,
which are summarized in  Eq.~(\ref{eq:newcoupling:ns}.
Octet 3 consists of $N(1440)$, $\Lambda(1600)$, $\Sigma(1660)$ and
$\Xi(1690)$; octet 4 consists of
$N(1710)$, $\Lambda(1810)$, $\Sigma(1880)$ and
$\Xi(1950)$.

In general, the SU(3) coupling constants of a given unitary multiplet can be 
determined by performing a $\chi^2$ fit to the experimentally measured
partial decay widths, see e.g. \cite{Samios74,GPreview}.
In some cases, the fit is unsuccessful, which indicates that the considered
multiplet is most likely mixed with some other multiplet(s).
Our analysis shows that while approximate flavor SU(3) symmetry can account 
for the known decays of octet 4, SU(3) fails for octet 4 because 
SU(3) incorrectly predicts  the sign of the $\Sigma(1880) \to \Sigma \, \pi$
amplitude.
A possible solution, which remedies the problem, 
 is to introduce a mixing between octets 3 and 4.

The mixing of two octets can be parameterized in terms of four mixing
angles: $\Theta_N$, $\Theta_{\Lambda}$, $\Theta_{\Sigma}$
and $\Theta_{\Xi}$. However,
only two mixing angles are independent. In our analysis, we take
 $\Theta_N$ and $\Theta_{\Sigma}$ as the independent mixing angles.
Then, the physical decay coupling constants of octet 3 and 4,
which we generically call $G_2$ and $G_3$, are expressed in terms of
 the unmixed coupling constants, $G_2^{0}$ and $G_3^{0}$,
and the mixing angle $\theta$ 
\begin{eqnarray}
&&G_2=\cos \theta\, G_2^0+\sin \theta \,G_3^0 \nonumber\\
&&G_3=-\sin \theta\, G_2^0+\cos \theta\, G_3^0 \,.
\end{eqnarray}

In particular,
the SU(3) coupling constants of $N(1440)$ from octet 3, which are proportional
to the relevant coupling constants entering
Eq.~(\ref{eq:decays:n}), have the following form
\begin{eqnarray}
&&G_{N_2  N \,\pi}=\frac{1}{2\sqrt{5}}g_{N_2  N \,\pi}=\frac{9}{20} 
\left(G_8^{(2)}(1+\alpha^{(2)}) \cos \phi_N+ G_8^{(3)}(1+\alpha^{(3)})\sin \phi_N  \right) \nonumber\\
&&G_{N_2  N \,\eta}=\frac{1}{2\sqrt{5}}g_{N_2  N \,\eta}=-\frac{3}{20} 
\left(G_8^{(2)}(1-3\,\alpha^{(2)}) \cos \phi_N+ G_8^{(3)}(1-3\,\alpha^{(3)})\sin \phi_N \right) \nonumber\\
&&G_{N_2  \Lambda \,K}=\frac{1}{2\sqrt{5}}g_{N_2  \Lambda \,K}=-\frac{3 }{20} 
\left(G_8^{(2)}(1+3\,\alpha^{(2)}) \cos \phi_N+ G_8^{(3)}(1+3\,\alpha^{(3)})\sin \phi_N \right) \nonumber\\
&&G_{N_2  \Delta \,\pi}=\frac{2}{\sqrt{5}}g_{N_2  \Delta \,\pi}=-\frac{2}{\sqrt{5}}\left(G_{10}^{(2)} \cos \phi_N+G_{10}^{(3)} \sin \phi_N \right) \,.
\label{eq:n1440}
\end{eqnarray} 
For the $N(1710)$ from octet 4, the relevant $g_{N_3}$ coupling constants
 are obtained from Eq.~(\ref{eq:n1440}) after the replacement
 $\cos \phi_N \to -\sin \phi_N$
and $\sin \phi_N \to \cos \phi_N$. 

The parameters $G_8^{(2,3)}$, $\alpha^{(2,3)}$ and $G_{10}^{(2,3)}$ and the
 mixing angles $\theta_{N,\Sigma}$ are determined from the $\chi^2$ fit to
the combined set of experimentally measured partial decay widths of octets 3 and 4.

The SU(3) coupling constants of $\Sigma(1660)$, which enter
Eq.~(\ref{eq:decays:s}), have the following structure
\begin{eqnarray}
&&G_{\Sigma_2  \Lambda \,\pi}=\frac{1}{2\sqrt{5}}g_{\Sigma_2  \Lambda \,\pi}=\frac{3}{10} 
\left(G_8^{(2)} \cos \phi_{\Sigma}+ G_8^{(3)}\sin \phi_{\Sigma}  \right)
 \nonumber\\
&&G_{\Sigma_2  \Sigma \,\eta}=\frac{1}{2\sqrt{5}}g_{\Sigma_2  \Sigma \,\eta}=\frac{3}{10}
\left(G_8^{(2)}\cos \phi_{\Sigma}+ G_8^{(3)}\sin \phi_{\Sigma} \right)
 \nonumber\\
&&G_{\Sigma_2  \Sigma \,\pi}=\frac{1}{\sqrt{30}}g_{\Sigma_2  \Sigma \,\pi}=\frac{3 \sqrt{6}}{10}
\left(G_8^{(2)}\,\alpha^{(2)} \cos \phi_{\Sigma}+
 G_8^{(3)}\,\alpha^{(3)} \sin \phi_{\Sigma} \right) \nonumber\\
&&G_{\Sigma_2  N \,\overline{K}}=\frac{1}{\sqrt{30}}g_{\Sigma_2  N \,\overline{K}}=-\frac{3 \sqrt{3}}{10\sqrt{2}}
\left(G_8^{(2)}(1-\alpha^{(2)}) \cos \phi_{\Sigma}+
 G_8^{(3)}(1-\alpha^{(3)} )\sin \phi_{\Sigma} \right) \nonumber\\
&&G_{\Sigma_2  \Sigma_{10} \,\pi}=\frac{\sqrt{30}}{15}g_{\Sigma_2  \Sigma_{10} \,\pi}=-\frac{\sqrt{30}}{15}\left(G_{10}^{(2)} \cos \phi_{\Sigma}+
G_{10}^{(3)} \sin \phi_{\Sigma} \right)\,.
\label{eq:s1660}
\end{eqnarray}
The corresponding $\Sigma(1880)$ coupling constants
 are obtained from Eq.~(\ref{eq:s1660}) after the replacement
 $\cos \phi_{\Sigma} \to -\sin \phi_{\Sigma}$
and $\sin \phi_{\Sigma} \to \cos \phi_{\Sigma}$.

In addition, for the $\chi^2$ fit we need two coupling constants of the 
$\Lambda(1600)$ state
\begin{eqnarray}
&&G_{\Lambda_2  N \,\overline{K}}=\frac{3}{10 \sqrt{2}} 
\left(G_8^{(2)}(1+3\,\alpha^{(2)}) \cos \phi_{\Lambda}+ G_8^{(3)}(1+3\,\alpha^{(3)})\sin \phi_{\Lambda}  \right) \nonumber\\
&&G_{\Lambda_2  \Sigma \,\pi}=-\frac{3 \sqrt{3}}{10} 
\left(G_8^{(2)} \cos \phi_{\Lambda}+ G_8^{(3)}\sin \phi_{\Lambda}  \right) \,.
\label{eq:l1600}
\end{eqnarray}
The $\phi_{\Lambda}$ mixing angle can be expressed in terms of $\phi_{N}$
 and $\phi_{\Sigma}$ and, with good accuracy, $\phi_{\Lambda} \approx -\phi_{\Sigma}$. 
Naturally, the relevant $\Lambda(1810)$ coupling constants are obtained
from Eq.~(\ref{eq:l1600}) after the replacement
$\cos \phi_{\Lambda} \to -\sin \phi_{\Lambda}$
and $\sin \phi_{\Lambda} \to \cos \phi_{\Lambda}$. 

The SU(3) predictions for the partial decay widths are formed by squaring the
coupling constants of Eqs.~(\ref{eq:n1440}), (\ref{eq:s1660}) and 
(\ref{eq:l1600}) and multiplying them by the phase space factors of 
Eqs.~(\ref{eq:ps}) and (\ref{eq:ps2}).

We performed an eight-parameter $\chi^2$ fit to the combined set of
twelve observables of octets 3 and 4. The following observables of octet 3 
were used 
\begin{itemize}
{\item $\Gamma(N \to N \,\pi)$}
{\item $Br(N \to N \,\pi)/Br(N \to \Delta \, \pi)$}
{\item $\Gamma(\Lambda \to N \overline{K})$}
{\item $Br(\Lambda \to N \overline{K})/\sqrt{Br(\Lambda \to N \overline{K}) Br(\Lambda \to \Sigma \pi)}$}
{\item $\Gamma(\Sigma \to N \overline{K})$}
{\item $Br(\Sigma \to N \overline{K})/\sqrt{Br(\Sigma \to N \overline{K})Br(\Sigma \to \Sigma \pi)}$.}
\end{itemize}
Note that we use the ratios of branching ratios
 as our fitted observables  in order to rid of error correlations.

From octet 4, we took the following observables
\begin{itemize}
{\item $\Gamma(N \to N \,\pi)$}
{\item $Br(N \to N \,\pi)/Br(N \to \Delta \, \pi)$}
{\item $\Gamma(\Lambda \to N \overline{K})$}
{\item $Br(\Lambda \to N \overline{K})/\sqrt{Br(\Lambda \to N \overline{K}) Br(\Lambda \to \Sigma \pi)}$}
{\item $Br(\Sigma \to N \overline{K})/\sqrt{Br(\Sigma \to N \overline{K})Br(\Sigma \to \Sigma \pi)}$}
{\item $Br(\Sigma \to N \overline{K})/\sqrt{Br(\Sigma \to N \overline{K})Br(\Sigma \to \Lambda \pi)}$.}
\end{itemize}

The results of the $\chi^2$ fit are summarized in Eq.~(\ref{eq:m34:chi2})
\begin{eqnarray}
&& G_8^{(2)}=12.6 \pm 0.2 \quad \quad G_8^{(3)}=2.20 \pm 1.88   \nonumber\\
&&\alpha^{(2)}=0.37 \pm 0.08 \quad \quad \alpha^{(3)}=0.93 \pm 0.83 \nonumber\\
&&G_{10}^{(2)}=16.4 \pm 2.1 \quad \quad G_{10}^{(3)}=14.4 \pm 4.5 \nonumber\\
&& \phi_N=(1.4 \pm 6.8)^0 \quad \quad \phi_{\Sigma}=(14.9 \pm 5.2)^0 \nonumber\\
&& \chi^2/d.o.f=6.17/4 \,.
\label{eq:m34:chi2}
\end{eqnarray}
The acceptably good  value of $\chi^2$ per degree of freedom is a result
of the fact that all the fitted decay amplitudes, including the
$\Sigma(1880) \to \Sigma \, \pi$ amplitude, are described rather well.

Finally, substituting the values of the coupling constants and mixing
angles from Eq.~(\ref{eq:m34:chi2}) into Eqs.~(\ref{eq:n1440}) and
(\ref{eq:s1660}),one readily obtains the values of the coupling constants
entering our predictions for the antidecuplet decays, which are
summarized in Eq.~(\ref{eq:newcoupling:ns}).

\section{Additional mixing with 27-plet}
\label{sec:2735}

It was argued in \cite{27plet} that the mixing of the antidecuplet with
 (yet fictitious) 
27-plet and 35-plet SU(3) representations has a significant impact on 
the antidecuplet
 decays. Therefore, in order to study how robust our predictions 
for the antidecuplet
 decays in Sect.~\ref{sec:888} with respect to the additional mixing,
 we explicitly add the 27-plet and 35-plet contributions to the antidecuplet
 coupling constants
(\ref{eq:decays:theta}), (\ref{eq:decays:n}), (\ref{eq:decays:s}) and (\ref{eq:decays:xi}). 
In doing this, we borrow the required coupling constants,
$H_{27}$ and  $H_{27}^{\prime}$, and mixing angles, which are 
proportional to $d_{27}$ and $c_{27}$,
 from \cite{27plet}. 
Equation~(\ref{eq:27}) summarizes which replacements of the previously used  
coupling constants one has to make in order to include the 27-plet  
contributions (note that mixing with a 35-plet does not
enter the expressions for $\bf{\at} \to \bf{8} + \bf{8}$ and 
$\bf{\at} \to \bf{10} + \bf{8}$ decays) \cite{27plet}
\begin{eqnarray}
&&g_{\Theta^+ N\,K} \to g_{\Theta^+ N\,K} -\frac{1}{\sqrt{5}} \frac{7}{4} c_{27} \, H_{27}^{\prime} \,, \nonumber\\
&& g_{N_{\at}  N\, \pi} \to g_{N_{\at}  N\, \pi}+ \frac{1}{2 \sqrt{5}}\left(\frac{49}{12}
c_{27} \, H_{27}^{\prime}+\frac{1}{15}d_{27} \, H_{27} \right ) \,, \nonumber\\
&& g_{N_{\at}  N\, \eta} \to g_{N_{\at}  N\, \eta} + \frac{1}{2 \sqrt{5}}\left(
\frac{7}{4} c_{27} \, H_{27}^{\prime}+\frac{1}{5}d_{27} \, H_{27} \right)\,,  \nonumber\\
&& g_{N_{\at}  \Lambda\, K} \to g_{N_{\at}  \Lambda\, K}+
  \frac{1}{2 \sqrt{5}}\left(-\frac{7}{2} c_{27} \, H_{27}^{\prime}+\frac{1}{5}d_{27} \, H_{27} \right)\,,  \nonumber\\
&& g_{N_{\at}  \Sigma\, K} \to g_{N_{\at}  \Sigma\, K}-
 \frac{1}{2 \sqrt{5}}\left(\frac{7}{3} c_{27} \, H_{27}^{\prime}+\frac{1}{15}d_{27} \, H_{27} \right) \,, \nonumber\\
&& g_{N_{\at}  \Delta\, \pi} \to g_{N_{\at}  \Delta\, \pi} \,, \nonumber\\
&& g_{\Sigma_{\at}  \Lambda\, \pi} \to g_{\Sigma_{\at}  \Lambda\, \pi}+
 \frac{1}{2 \sqrt{5}}\left(\frac{7}{2} c_{27} \, H_{27}^{\prime}
+\frac{4}{15}d_{27} \, H_{27} \right) \,, \nonumber\\
&& g_{\Sigma_{\at}  \Sigma\, \eta} \to g_{\Sigma_{\at}  \Sigma\, \eta}
 +\frac{1}{2 \sqrt{5}}\left(\frac{7}{3} c_{27} \, H_{27}^{\prime}
+\frac{4}{15}d_{27} \, H_{27}  \right) \,, \nonumber\\
&& g_{\Sigma_{\at}  \Sigma\, \pi} \to g_{\Sigma_{\at}  \Sigma\, \pi}
+\frac{1}{\sqrt{30}}\left(\frac{7}{2} c_{27} \, H_{27}^{\prime}\right) \,, \nonumber\\
&& g_{\Sigma_{\at}  \Xi\, K} \to g_{\Sigma_{\at}  \Xi\, K} 
+ \frac{1}{\sqrt{30}}\left(-\frac{14}{3} c_{27} \, H_{27}^{\prime}
+\frac{4}{15}d_{27} \, H_{27} \right) \,, \nonumber \\
&& g_{\Sigma_{\at}  N\, \overline{K}} \to g_{\Sigma_{\at}  N\, \overline{K}} 
+ \frac{1}{\sqrt{30}}\left(-\frac{7}{6} c_{27} \, H_{27}^{\prime}
+\frac{4}{15}d_{27} \, H_{27} \right) \,, \nonumber \\
&& g_{\Sigma_{\at}  \Sigma_{10}\, \pi} \to g_{\Sigma_{\at}  \Sigma_{10}\, \pi} \,, \nonumber \\
&&g_{\Xi_{\at} \Sigma \, \overline{K}} \to g_{\Xi_{\at} \Sigma \, \overline{K}}
+\frac{1}{\sqrt{10}} \left(-\frac{7}{12} c_{27} \, H_{27}^{\prime}
+\frac{1}{3}d_{27} \, H_{27}\right) \,, \nonumber \\
&&g_{\Xi_{\at} \Xi \, \pi} \to g_{\Xi_{\at} \Xi \, \pi}+
\frac{1}{\sqrt{10}} \left(\frac{7}{6} c_{27} \, H_{27}^{\prime}+\frac{1}{3}d_{27} \, H_{27}\right) \,.
\label{eq:27}
\end{eqnarray}  
We neglect the contribution of the 27-plet to the $\at \to 10 + 8$ decays
because the corresponding coupling constant is extremely small, see the first
of Refs.~\cite{27plet}.

\subsection{The $G_{\at}$ coupling constant}

\begin{figure}
\includegraphics[width=12cm,height=12cm]{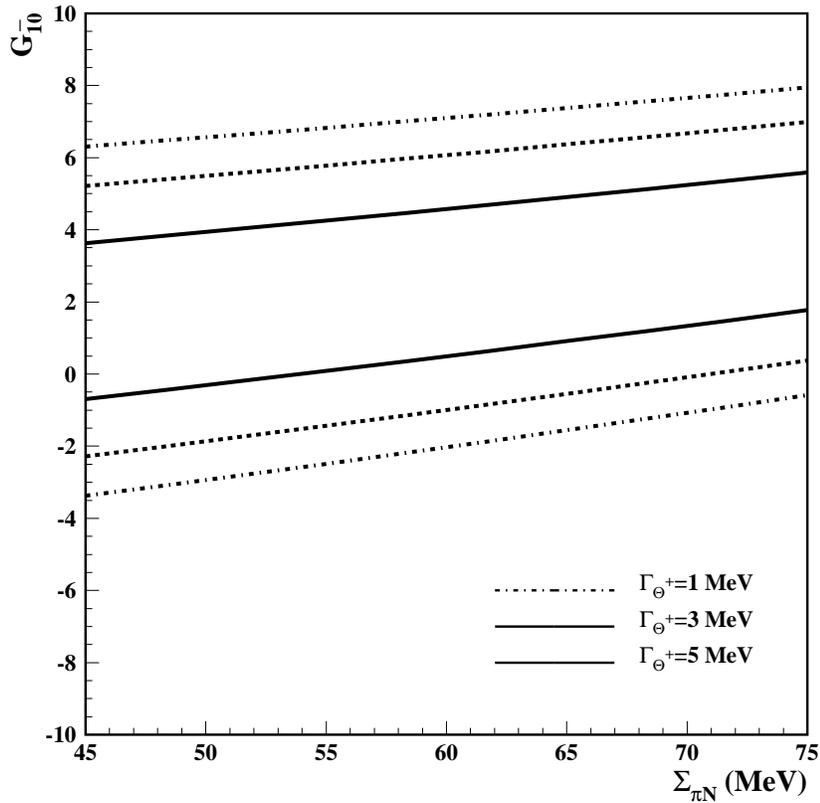}
\caption{The $G_{\at}$ coupling constant as a function of
$\Sigma_{\pi \, N}$ and at different $\Gamma_{\Theta^+}$. 
There are two solution of 
$G_{\at}$: Positive and mostly negative.
}
\label{fig:g10_27}
\end{figure}

The 27-plet contribution to the total width of $\Theta^+$ affects the 
values of the $G_{\at}$ coupling constant which we extract from 
$\Gamma_{\Theta^+}$.
 Figure~\ref{fig:g10_27}
presents the resulting $G_{\at}$ as a function of
$\Sigma_{\pi \, N}$ and at different $\Gamma_{\Theta^+}$. 
A comparison of Figs.~\ref{fig:g10} and \ref{fig:g10_27} shows that while
previously there was one positive and one negative solution for $G_{\at}$, now 
there is one positive solution and one solution, which changes sign:
$G_{\at}$ at $\Gamma_{\Theta^+}=1$ and 3 MeV changes sign and becomes
positive at large $\Sigma_{\pi \, N}$. 
Since the positive sign of  $G_{\at}$ is essential in order to obtain 
a qualitatively correct picture of the $N_{\at}$ 
decays, we present our predictions for the antidecuplet decays
using the both solutions for the $G_{\at}$. The two solutions for 
$G_{\at}$ will be referred to as ``positive'' and ``mostly negative'' 
solutions.

\subsection{Decays of $\Xi_{\at}$}

\begin{figure}
\includegraphics[width=12cm,height=12cm]{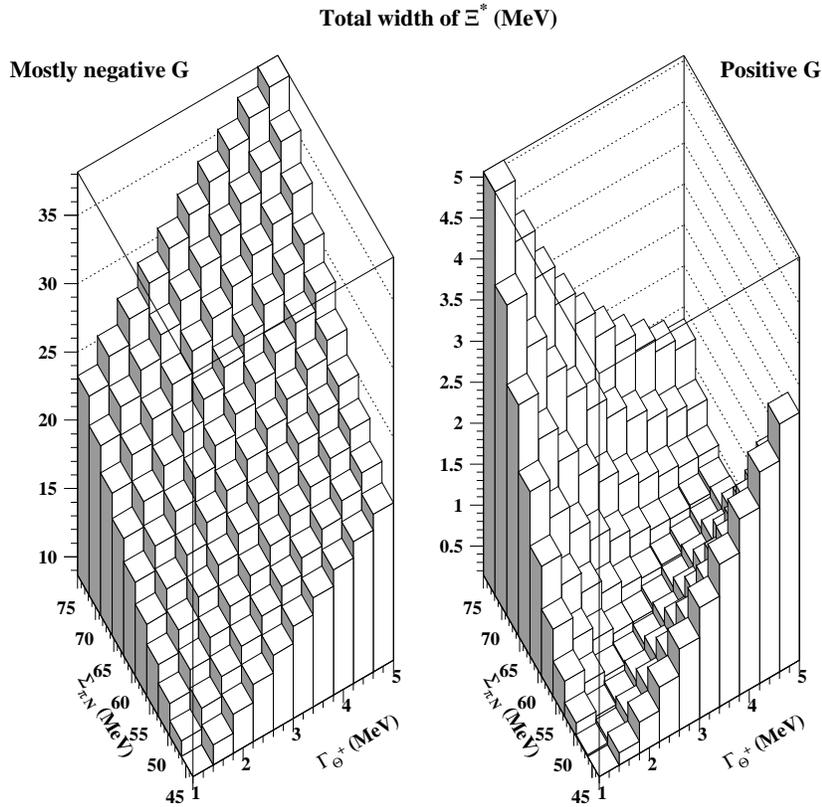}
\caption{The total width of $\Xi_{\at}$ as a function of $\Gamma_{\Theta^+}$  and $\Sigma_{\pi \, N}$ in the presence of the mixing with the
27-plet. The two plots correspond to the positive  and mostly negative
$G_{\at}$ solutions.
}
\label{fig:xi_27}
\end{figure}

In what follows, we repeat the analysis of the antidecuplet decays including 
the 27-plet contribution. We start with the total width of 
$\Xi_{\at}$. Figure~\ref{fig:xi_27} presents $\Gamma_{\Xi_{\at}}$
as a function of $\Gamma_{\Theta^+}$ and $\Sigma_{\pi \, N}$ for the 
two possible
solutions for $G_{\at}$. 
In agreement with the analysis of \cite{27plet}, the mixing with the 
27-plet is rather important for the decays of the $\Xi_{\at}$.

Note that the positive $G_{\at}$ solution at large $\Gamma_{\Theta^+}$ and 
$\Sigma_{\pi \, N}$ results in the values of $\Gamma_{\Xi_{\at}}$ which are 
higher than
the present upper limit on the total width of $\Xi_{\at}$. However, until
the existence of $\Xi_{\at}$ and its properties receive firmer experimental 
support \cite{BABAR,NA49:critique}, one should not make any quantitative
 statements about which values
$\Gamma_{\Theta^+}$, $\Sigma_{\pi \, N}$, $\theta_2$ and $\theta_3$ 
are appropriate for the sufficiently narrow $\Gamma_{\Xi_{\at}}$.

\subsection{Decays of $N_{\at}$}

We found that the mixing with the 27-plet insignificantly lowers the total
 width of  $N_{\at}$ and, hence, Fig.~\ref{fig:ntot} changes only little 
when the 27-plet admixture is included.
 The change is small because, in general, $\Gamma_{N_{\at}}$ receives a 
dominant contribution from 
$\Gamma_{N_{\at} \to \Delta \, \pi}$, which is insensitive to the 27-plet 
contribution, see Eqs.~\ref{eq:27}.

\begin{figure}
\includegraphics[width=15cm,height=15cm]{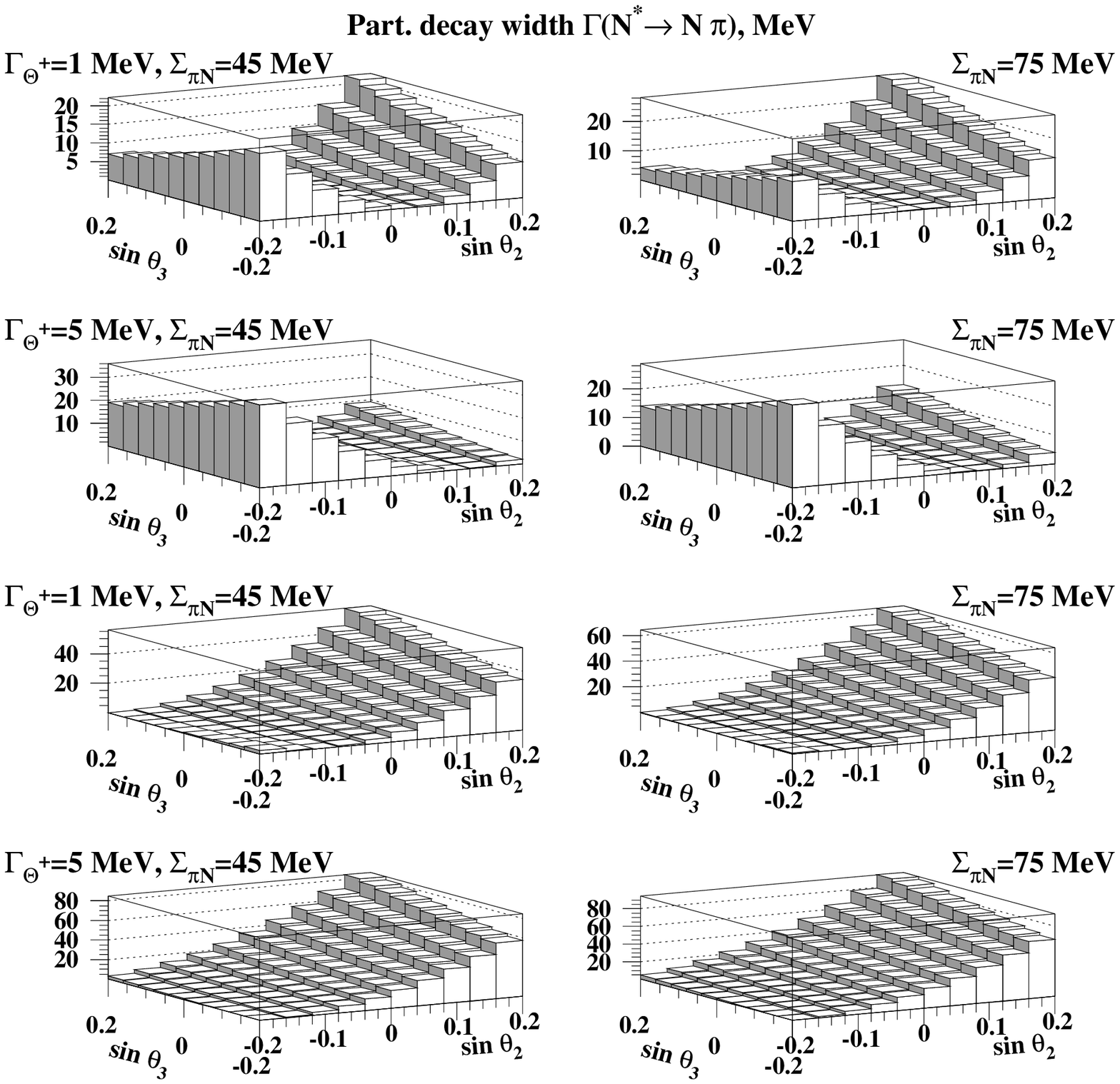}
\caption{$\Gamma_{N_{\at} \to N \, \pi}$ as a function of $\theta_2$
and $\theta_3$ in the presence of the mixing with the
27-plet. The four upper panels correspond to the positive $G_{\at}$ solution;
the four lower panels correspond to the mostly negative $G_{\at}$ solution.
}
\label{fig:nnpi_27}
\end{figure}

\begin{figure}
\includegraphics[width=15cm,height=15cm]{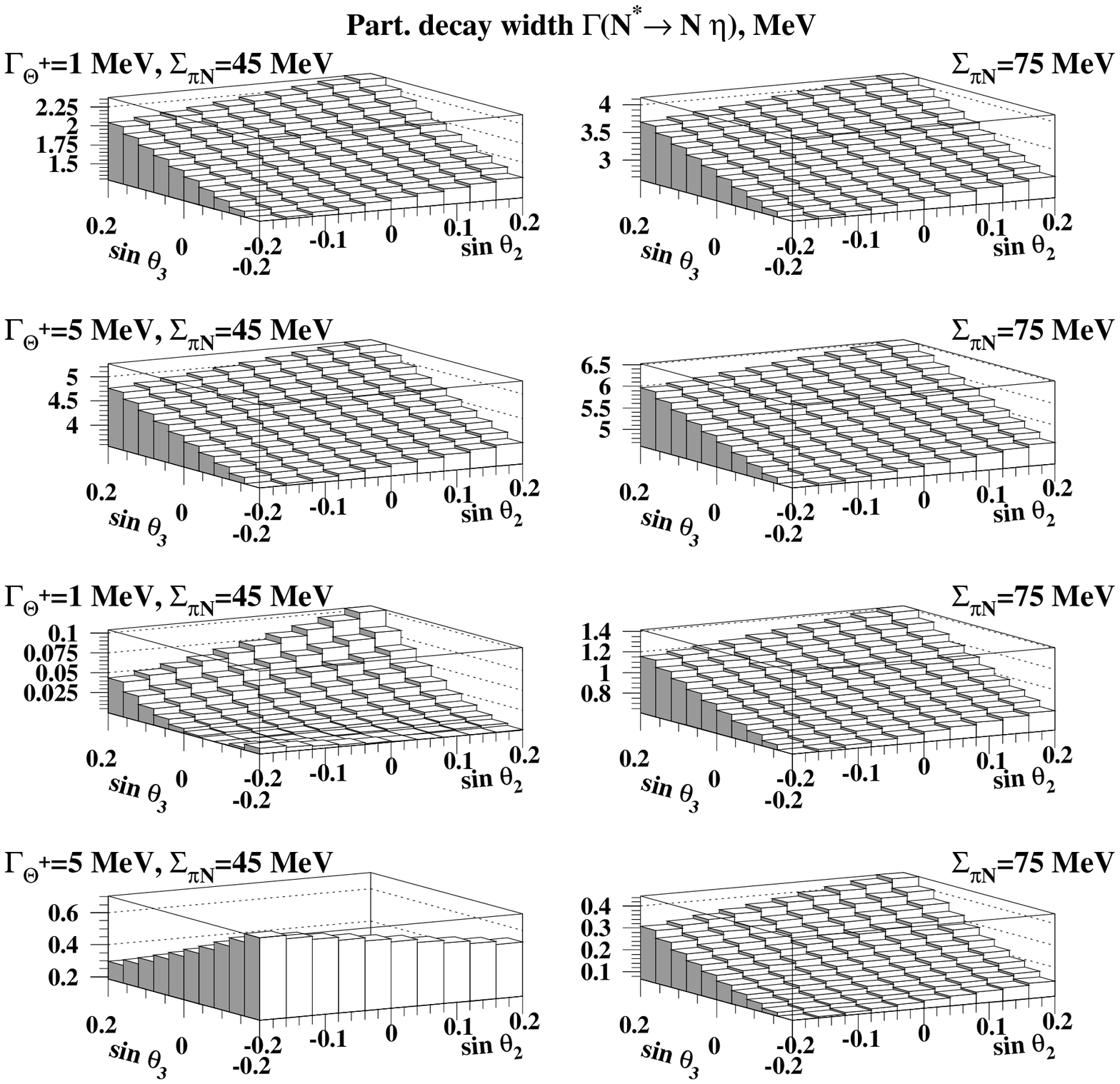}
\caption{$\Gamma_{N_{\at} \to N \, \eta}$ as a function of $\theta_2$
and $\theta_3$ in the presence of the mixing with the
27-plet. 
The four upper panels correspond to the positive $G_{\at}$ solution;
the four lower panels correspond to the mostly negative $G_{\at}$ solution.
}
\label{fig:nneta_27}
\end{figure}

\begin{figure}
\includegraphics[width=15cm,height=15cm]{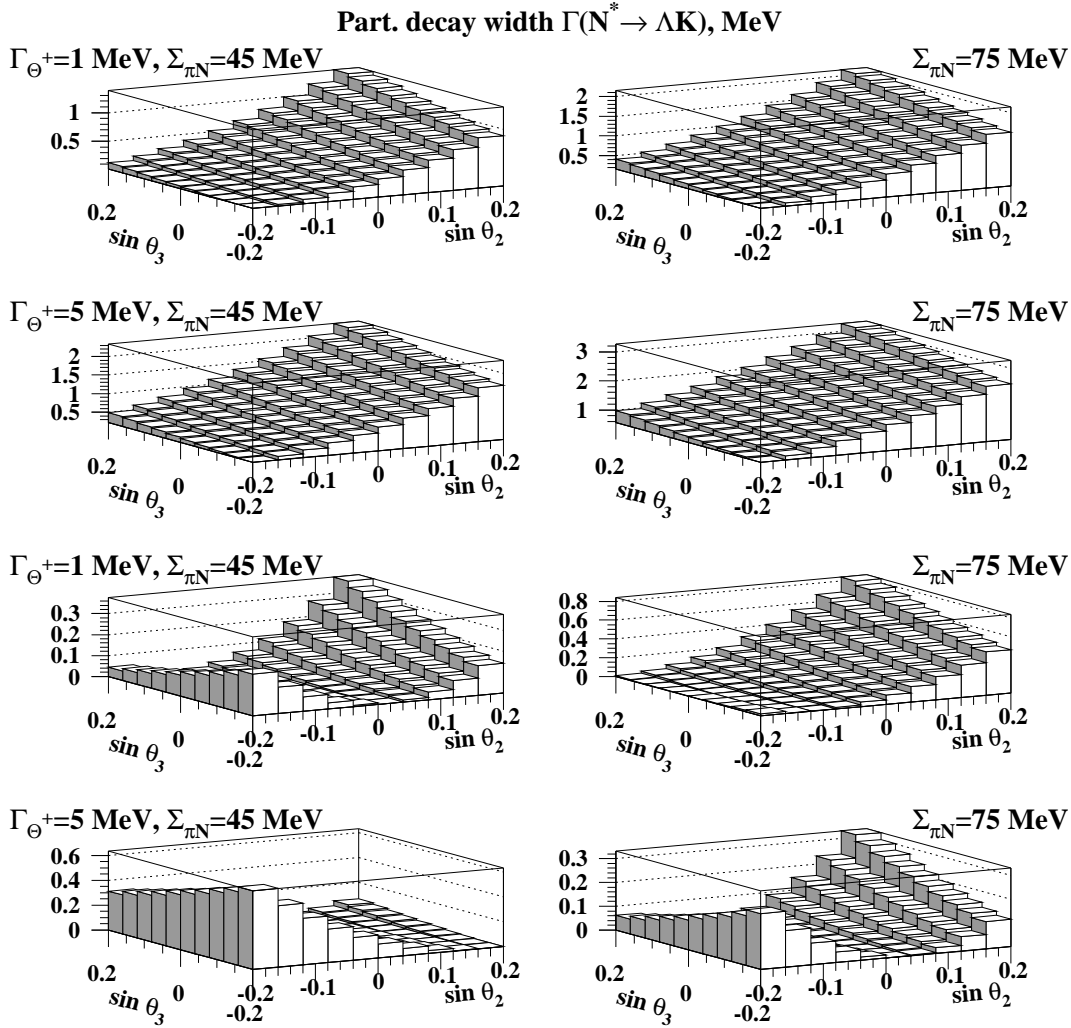}
\caption{$\Gamma_{N_{\at} \to \Lambda \, K}$ as a function of $\theta_2$
and $\theta_3$ in the presence of the mixing with the
27-plet.
 The four upper panels correspond to the positive $G_{\at}$ solution;
the four lower panels correspond to the mostly negative $G_{\at}$ solution.}
\label{fig:nlk_27}
\end{figure}

Figures~\ref{fig:nnpi_27}, \ref{fig:nneta_27} and \ref{fig:nlk_27}
present the $\Gamma_{N_{\at} \to N \, \pi}$, $\Gamma_{N_{\at} \to N \, \eta}$
and $\Gamma_{N_{\at} \to \Lambda \, K}$ partial decay widths in the presence
of the mixing with the 27-plet as functions of the $\theta_2$
and $\theta_3$ mixing angles at $\Gamma_{\Theta^+}=1$ and 5 MeV and at
$\Sigma_{\pi \, N}=45$ and 75 MeV.
In each plot, the four upper panels correspond to the positive $G_{\at}$
 solution; the four lower panels correspond to the mostly negative $G_{\at}$ solution.
Note that the $\Gamma_{N_{\at} \to \Delta \, \pi}$ is presented in 
Fig.~\ref{fig:ndpi}.

An examination of Figs.~\ref{fig:nnpi_27}, \ref{fig:nneta_27} and
 \ref{fig:nlk_27} shows that the difference between the predicted partial 
widths using the two solutions for $G_{\at}$ is dramatic: The use of the 
mostly negative $G_{\at}$ instead of the positive $G_{\at}$
increases $\Gamma_{N_{\at} \to N \, \pi}$ and reduces 
$\Gamma_{N_{\at} \to N \, \eta}$ and $\Gamma_{N_{\at} \to \Lambda \, K}$. 
However, we shall still be able to find regions of the $\theta_2$ and
$\theta_3$ mixing angles where 
$\Gamma_{N_{\at} \to N \, \eta} > \Gamma_{N_{\at} \to N \, \pi}$.



\clearpage

\clearpage

\begin{figure}
\includegraphics[width=15cm,height=15cm]{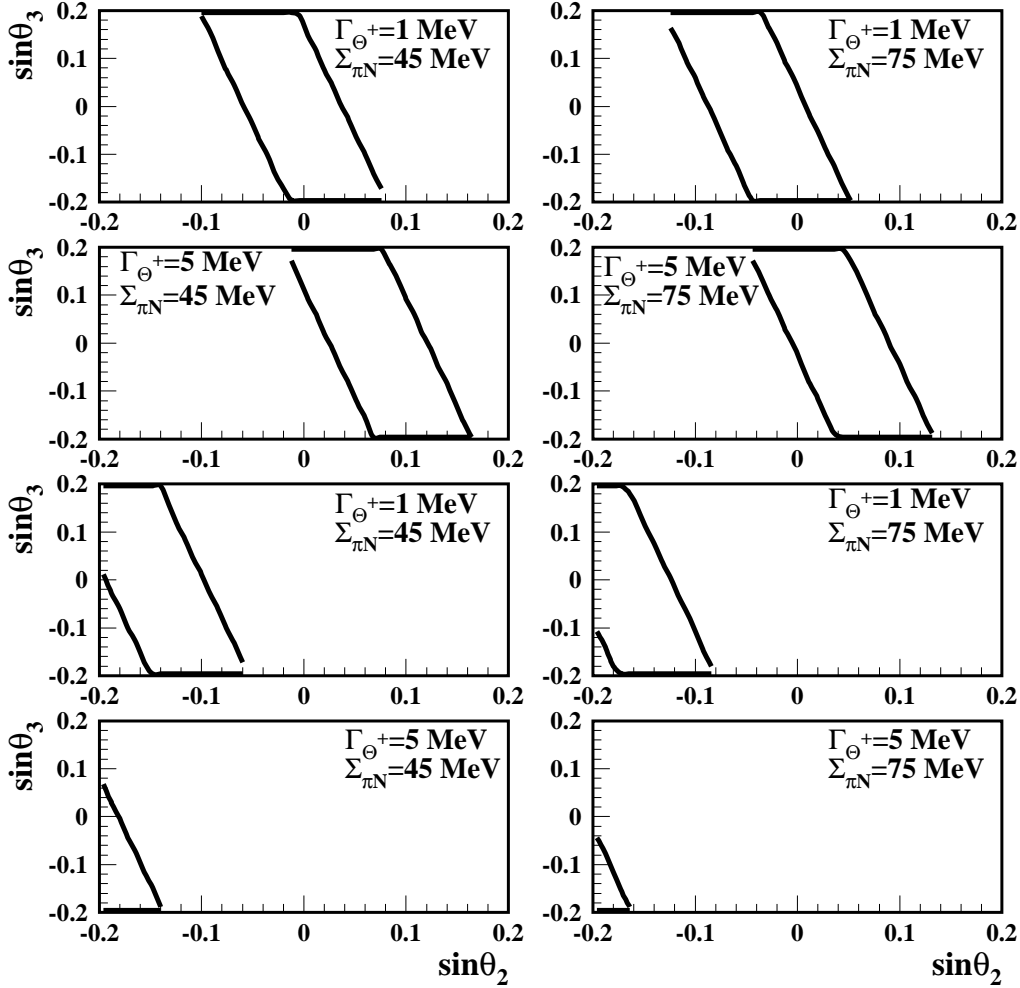}
\caption{The regions of the $\theta_2$ and $\theta_2$
mixing angles allowed by the  
$\Gamma_{N_{\at} \to  N\, \pi} \leq 1$ MeV condition 
in  the presence of the mixing with the
27-plet. The four upper panels correspond to the positive $G_{\at}$ solution;
the four lower panels correspond to the mostly negative $G_{\at}$ solution.
}
\label{fig:ugli_27}
\end{figure}

We are interested in the values of the $\theta_{2,3}$ mixing angles 
which correspond to $\Gamma_{N_{\at} \to N \, \pi} \leq 1$ MeV.
Figure~\ref{fig:ugli_27} presents the allowed regions of $\sin \theta_{2,3}$
in the presence of the $\Gamma_{N_{\at} \to  N\, \pi} \leq 1$ MeV condition.
At given $\sin \theta_{2}$, the two solid curves present the maximal and 
minimal values of $\sin \theta_{3}$.
The four upper panels correspond to the positive $G_{\at}$ solution;
the four lower panels correspond to the mostly negative $G_{\at}$ solution.

\begin{figure}
\includegraphics[width=15cm,height=15cm]{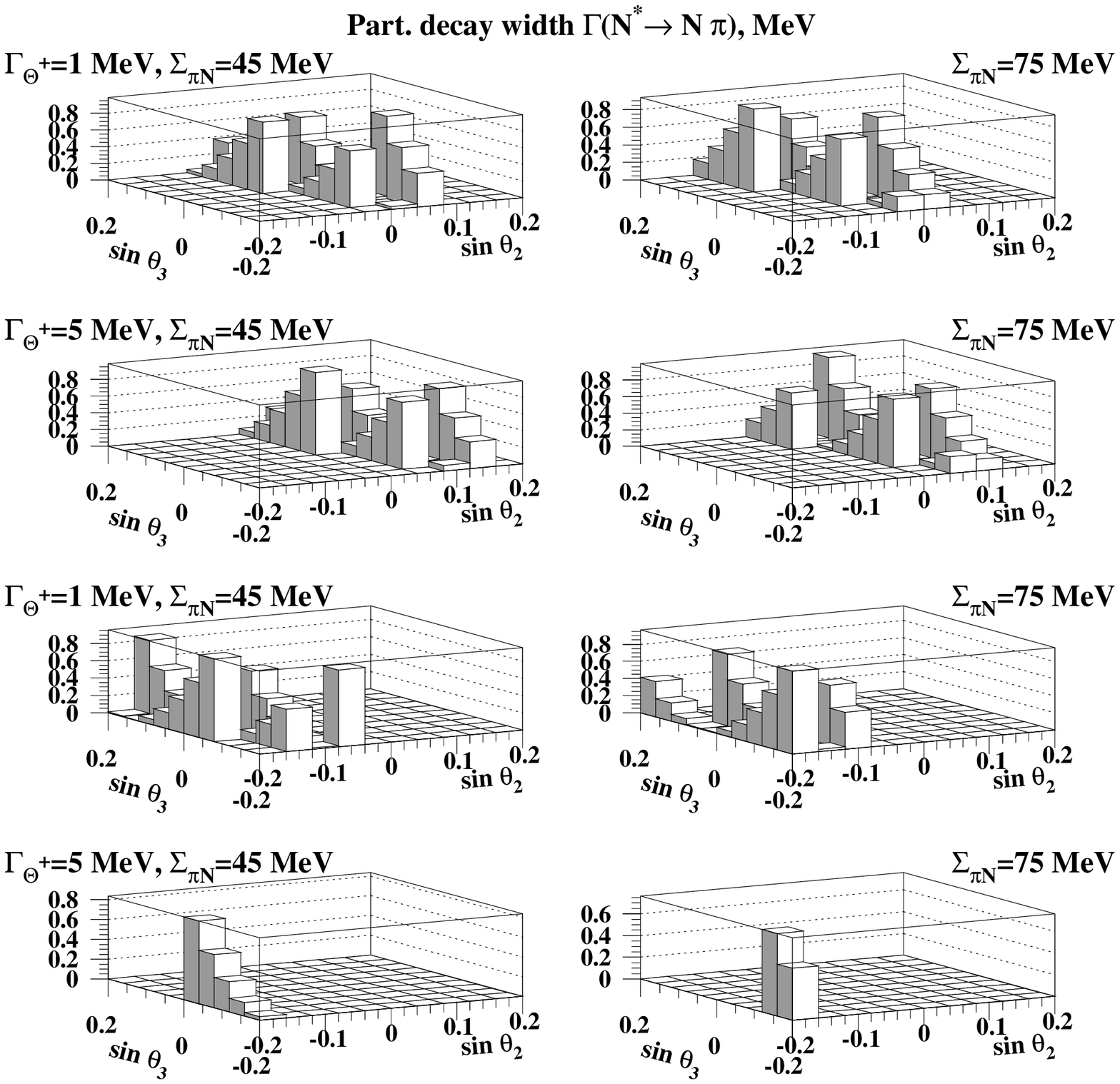}
\caption{$\Gamma_{N_{\at} \to N \, \pi}$ as a function of $\theta_2$
and $\theta_3$ in  the presence of the mixing with the
27-plet. The decay width is shown only where 
 $\Gamma_{N_{\at} \to  N\, \pi} \leq 1$ MeV.
The four upper panels correspond to the positive $G_{\at}$ solution;
the four lower panels correspond to the mostly negative $G_{\at}$ solution.
}
\label{fig:nnpi_cut_27_paper}
\end{figure}

\begin{figure}
\includegraphics[width=15cm,height=15cm]{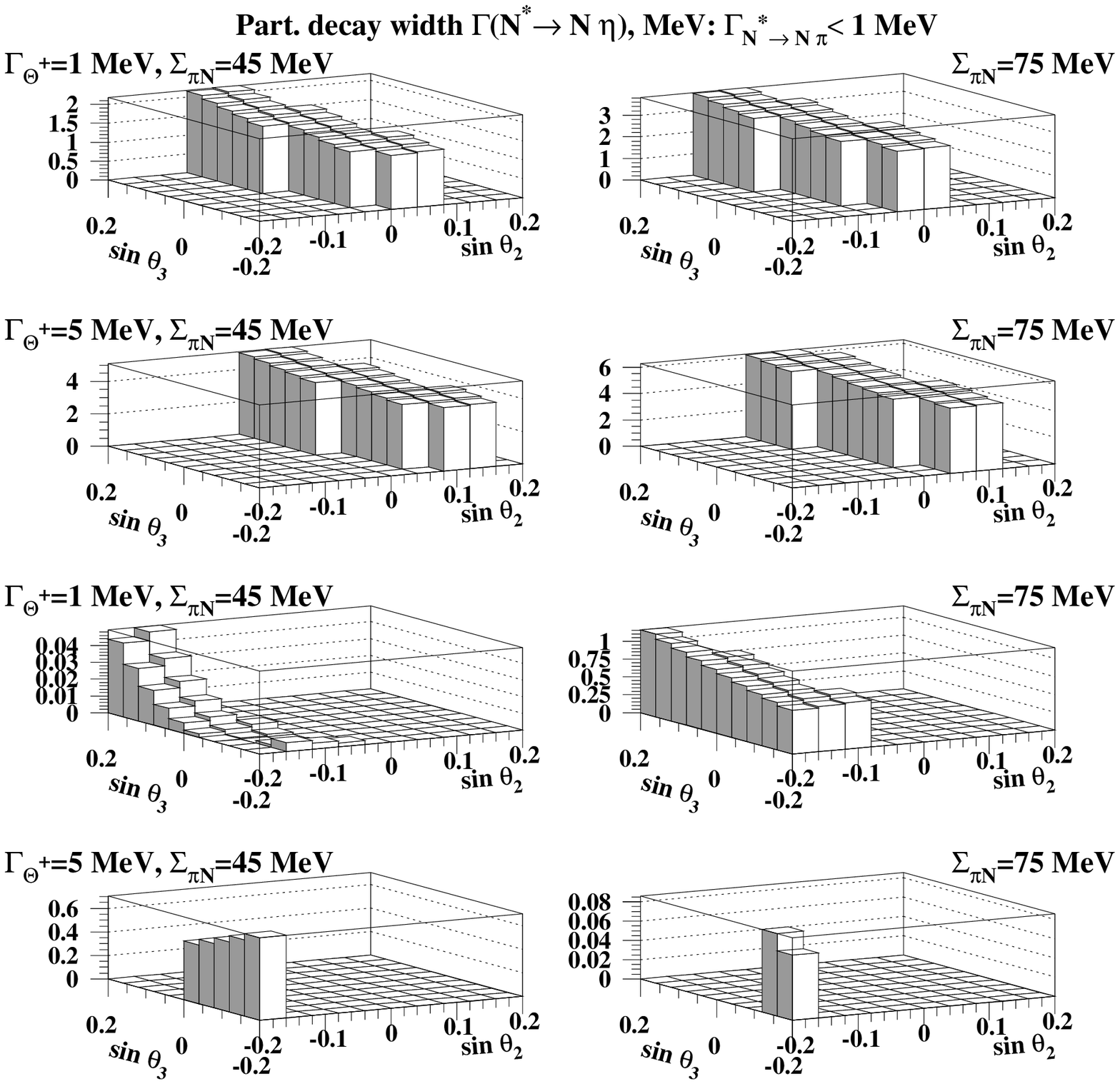}
\caption{$\Gamma_{N_{\at} \to N \, \eta}$ as a function of $\theta_2$
and $\theta_3$ in  the presence of the mixing with the
27-plet.  The decay width is shown only where 
 $\Gamma_{N_{\at} \to  N\, \pi} \leq 1$ MeV.
The four upper panels correspond to the positive $G_{\at}$ solution;
the four lower panels correspond to the mostly negative $G_{\at}$ solution.
}
\label{fig:nneta_cut_27_paper}
\end{figure}

\begin{figure}
\includegraphics[width=15cm,height=15cm]{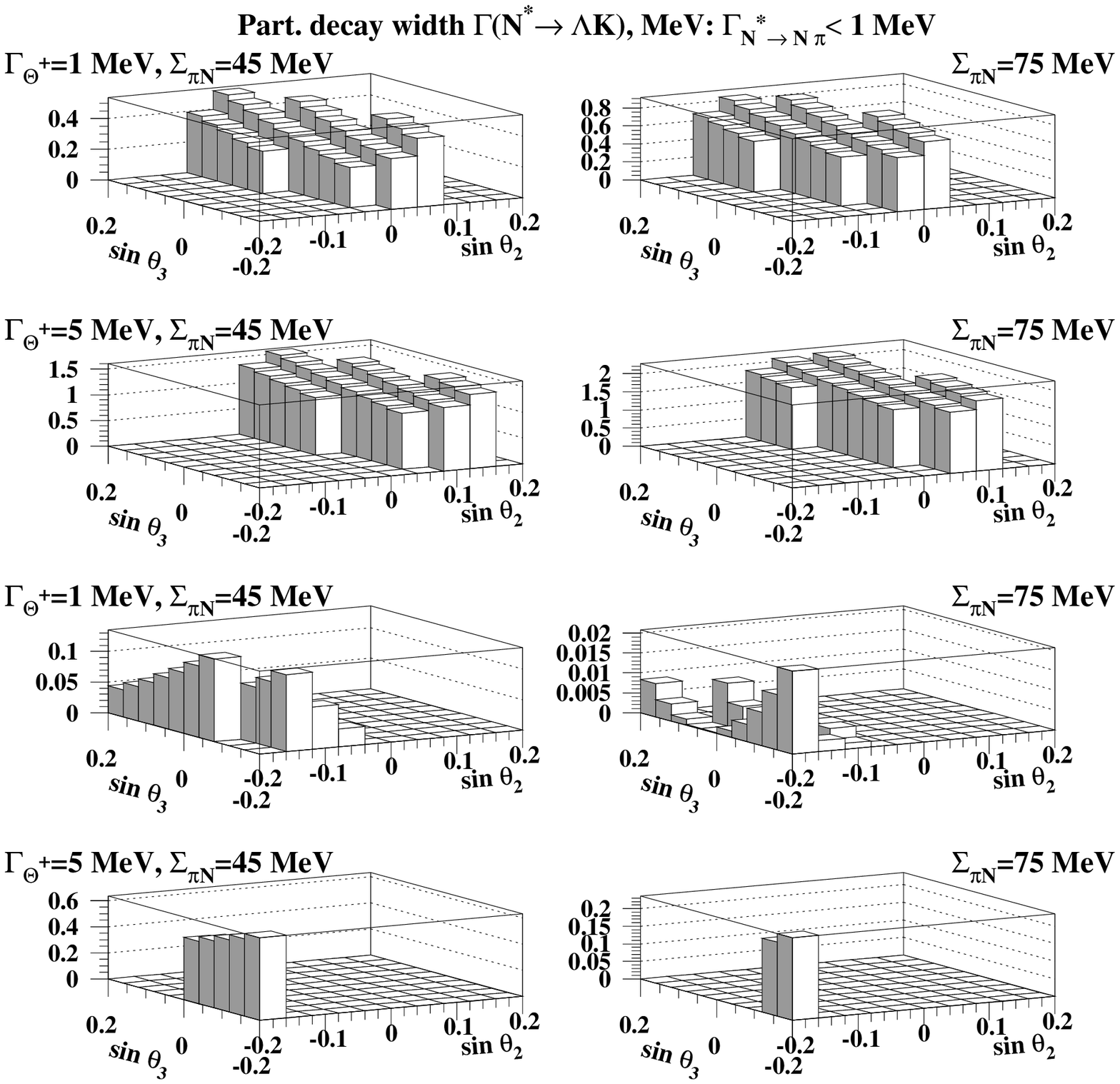}
\caption{$\Gamma_{N_{\at} \to \Lambda \, K}$ as a function of $\theta_2$
and $\theta_3$ in  the presence of the mixing with the
27-plet.  The decay width is shown only where 
 $\Gamma_{N_{\at} \to  N\, \pi} \leq 1$ MeV.
The four upper panels correspond to the positive $G_{\at}$ solution;
the four lower panels correspond to the mostly negative $G_{\at}$ solution.
}
\label{fig:nlk_cut_27}
\end{figure}

\begin{figure}
\includegraphics[width=15cm,height=15cm]{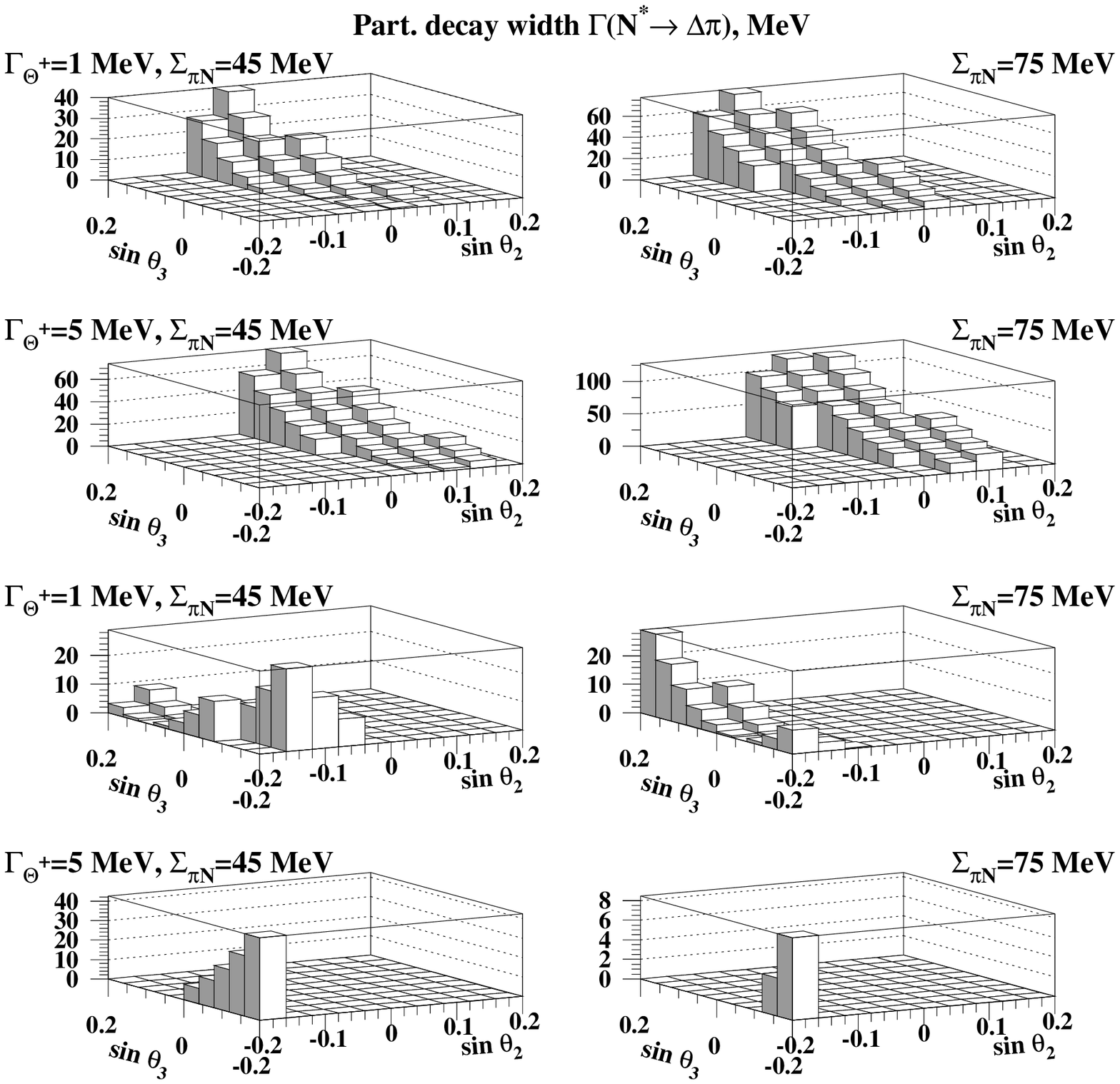}
\caption{$\Gamma_{N_{\at} \to \Delta \, \pi}$ as a function of $\theta_2$
and $\theta_3$ in  the presence of the mixing with the
27-plet.  The decay width is shown only where 
 $\Gamma_{N_{\at} \to  N\, \pi} \leq 1$ MeV.
The four upper panels correspond to the positive $G_{\at}$ solution;
the four lower panels correspond to the mostly negative $G_{\at}$ solution.
}
\label{fig:ndpi_cut_27_paper}
\end{figure}

Figures~\ref{fig:nnpi_cut_27_paper}, \ref{fig:nneta_cut_27_paper},
\ref{fig:nlk_cut_27} and \ref{fig:ndpi_cut_27_paper}
present the $\Gamma_{N_{\at} \to N \, \pi}$, $\Gamma_{N_{\at} \to N \, \eta}$,
$\Gamma_{N_{\at} \to \Lambda \, K}$ and $\Gamma_{N_{\at} \to \Delta \, \pi}$
partial decay widths in the region of $\theta_2$ and $\theta_3$ where
$\Gamma_{N_{\at} \to N \, \pi} \leq 1$ MeV.

For the positive $G_{\at}$ solution, 
 the 27-plet contribution lowers $\Gamma_{N_{\at} \to N \, \pi}$. As 
a result, the kinematic region where 
$\Gamma_{N_{\at} \to N \, \pi} \leq 1$ MeV (four upper 
panels in Fig.~\ref{fig:nnpi_cut_27_paper})
is somewhat wider than
in Fig.~\ref{fig:nneta_cut_paper}. In addition, the region is shifted 
towards positive $\sin \theta_2$.
As to the mostly negative  $G_{\at}$ solution, the 27-plet contribution 
increases $\Gamma_{N_{\at} \to N \, \pi}$ and, thus, makes the region
$\Gamma_{N_{\at} \to N \, \pi} \leq 1$ rather narrow. The allowed
region corresponds to large and negative $\sin \theta_2$, see four lower 
panels in Fig.~\ref{fig:nnpi_cut_27_paper}.

Turning to the $\Gamma_{N_{\at} \to N \, \eta}$ partial decay width 
(Fig.~\ref{fig:nneta_cut_27_paper}), we notice that the positive 
$G_{\at}$ solution corresponds to the $\Gamma_{N_{\at} \to N \, \eta}$, 
which is of the order of several MeV. 
On the other hand, among the cases corresponding to the mostly negative 
$G_{\at}$ solution (four lower panels of Fig.~\ref{fig:nneta_cut_27_paper}),
 only the  
$\Gamma_{\Theta^+}=1$ MeV and $\Sigma_{\pi \, N}=75$ MeV case
fits our qualitative picture of the $N_{\at}$ decays, which
assumes that while the $\Gamma_{N_{\at} \to N \, \pi}$ is suppressed and 
$\Gamma_{N_{\at} \to \Lambda \, K}$ is possibly suppressed, 
$\Gamma_{N_{\at} \to N \, \eta}$ is sizable.

We explained in Sect.~\ref{sec:main}  that the STAR result 
on the $\Lambda \, K_S$ invariant mass spectrum \cite{STAR} can be 
interpreted as an indication that $\Gamma_{N_{\at} \to \Lambda \, K}$ is 
possibly suppressed. Therefore, all cases considered in 
Fig.~\ref{fig:nlk_cut_27} (except maybe for the 
$\Gamma_{\Theta^+}=5$ MeV and positive $G_{\at}$ case) fit well the 
hypothesis of the suppressed $\Gamma_{N_{\at} \to \Lambda \, K}$.

The $\Gamma_{N_{\at} \to \Delta \, \pi}$ is a 
steeply rising function of $\theta_2$. For the positive $G_{\at}$
solution, the 27-plet admixture shifts the range of $\sin \theta_{2,3}$
allowed by the $\Gamma_{N_{\at} \to  N\, \pi} \leq 1$ MeV condition
towards positive $\sin \theta_{2}$. As a results, the values of 
$\Gamma_{N_{\at} \to \Delta \, \pi}$ in four upper panels of 
Fig.~\ref{fig:ndpi_cut_27_paper} are significantly higher than in
Fig.~\ref{fig:ndpi_cut_paper}.
This should be contrasted with the predicted much lower 
$\Gamma_{N_{\at} \to \Delta \, \pi}$, which corresponds to the
mostly negative  $G_{\at}$ solution, see the lower four panels of 
Fig.~\ref{fig:ndpi_cut_27_paper}.

The sum of the considered 
two-hadron partial decays widths of $N_{\at}$, in the presence of the 
$\Gamma_{N_{\at} \to  N\, \pi} \leq 1$ MeV condition and the mixing with the
27-plet, 
varies in the interval summarized in Table~\ref{table:n_int_27}.
The first value corresponds to the positive $G_{\at}$ solution;
 the value in the parenthesis corresponds to the mostly negative $G_{\at}$ 
solution. 

\begin{table}
\begin{tabular}{|c|c|c|c|}
\hline
$\Gamma_{\Theta^+}$ (MeV) & $\Sigma_{\pi \, N}$ (MeV)  & $\Gamma_{N_{\at}}^{{\rm 2-body, min}}$ (MeV) &  $\Gamma_{N_{\at}}^{{\rm 2-body, max}}$ (MeV) \\
\hline
1 & 45 & 1.9 (0.1)   & 52 (37)\\
1 & 75 & 4.3 (0.8)  & 103 (38) \\
5 & 45 & 5.7 (3.4)  & 97 (55) \\
5 & 75 & 17 (1.3)   & 157 (14) \\
\hline
\end{tabular}
\caption{The range of change of $\Gamma_{N_{\at}}^{{\rm 2-body}}$.
The first value corresponds to the positive $G_{\at}$ solution;
 the value in the parenthesis corresponds to the mostly negative $G_{\at}$ 
solution.}
\label{table:n_int_27}
\end{table} 

In summary, the additional mixing with the 27-plet does not change our
qualitative picture of the $N_{\at}$ decays when we use the positive 
solution for the $G_{\at}$ coupling constant.
The picture consists in the following observations: 
 $\Gamma_{N_{\at} \to  N\, \pi}$ is suppressed; 
$\Gamma_{N_{\at} \to  \Lambda\, K}$ is possibly suppressed; 
$\Gamma_{N_{\at} \to  N\, \eta}$ is sizable;
$\Gamma_{N_{\at} \to  \Delta\, \pi}$ is not too large such that 
$\Gamma_{N_{\at}}$ is of the order of 10-20 MeV.

Using the mostly negative solution for the $G_{\at}$ coupling constant, the 
above mentioned picture emerges only at 
 $\Gamma_{\Theta^+}=1$ MeV and $\Sigma_{\pi \, N}=75$ MeV.
A particular feature of this scenario is a possibly very narrow
$N_{\at}$ with a vanishingly small $\Gamma_{N_{\at} \to  \Lambda\, K}$.

\subsection{Decays of $\Sigma_{\at}$}

\begin{figure}
\includegraphics[width=15cm,height=15cm]{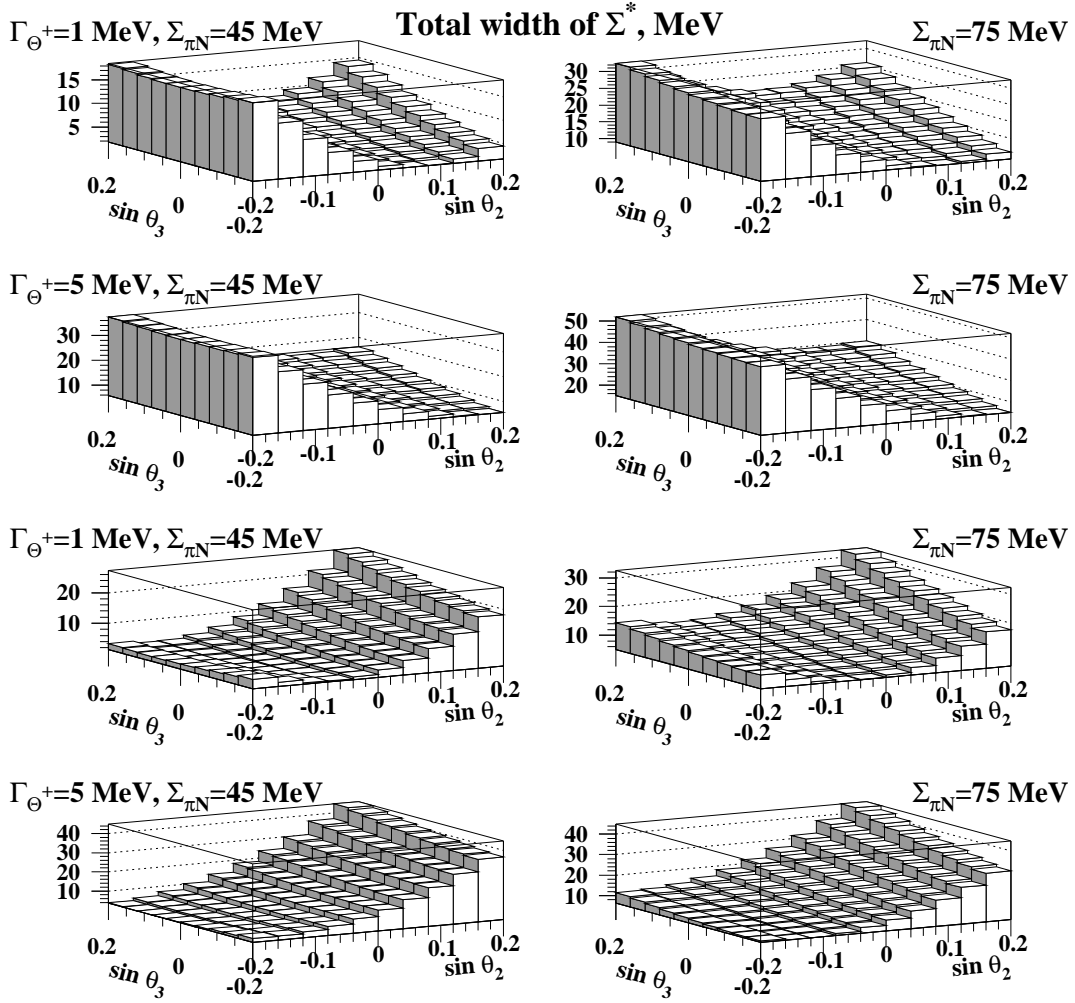}
\caption{The total width of $\Sigma_{\at}$ as a function of $\theta_2$
and $\theta_3$ in  the presence of the mixing with the
27-plet.
The four upper panels correspond to the positive $G_{\at}$ solution;
the four lower panels correspond to the mostly negative $G_{\at}$ solution.
}
\label{fig:stot_27}
\end{figure}

The total width of $\Sigma_{\at}$ is presented in Fig.~\ref{fig:stot_27}. 
A comparison with Fig.~\ref{fig:stot} reveals that using the positive
$G_{\at}$ solution, the 27-plet makes an 
insignificant contribution at small values of $\Gamma_{\Theta^+}$.
 On the other hand, at $\Sigma_{\pi \, N}=75$
MeV, the 27-plet contribution alters the pattern of the $\theta_2$ and 
$\theta_3$ dependence and noticeably changes the size of
 $\Gamma_{\Sigma_{\at}}$.
In addition, our predictions for $\Gamma_{\Sigma_{\at}}$ are very different
when we use the positive and mostly native solutions for $G_{\at}$.

\begin{figure}
\includegraphics[width=15cm,height=15cm]{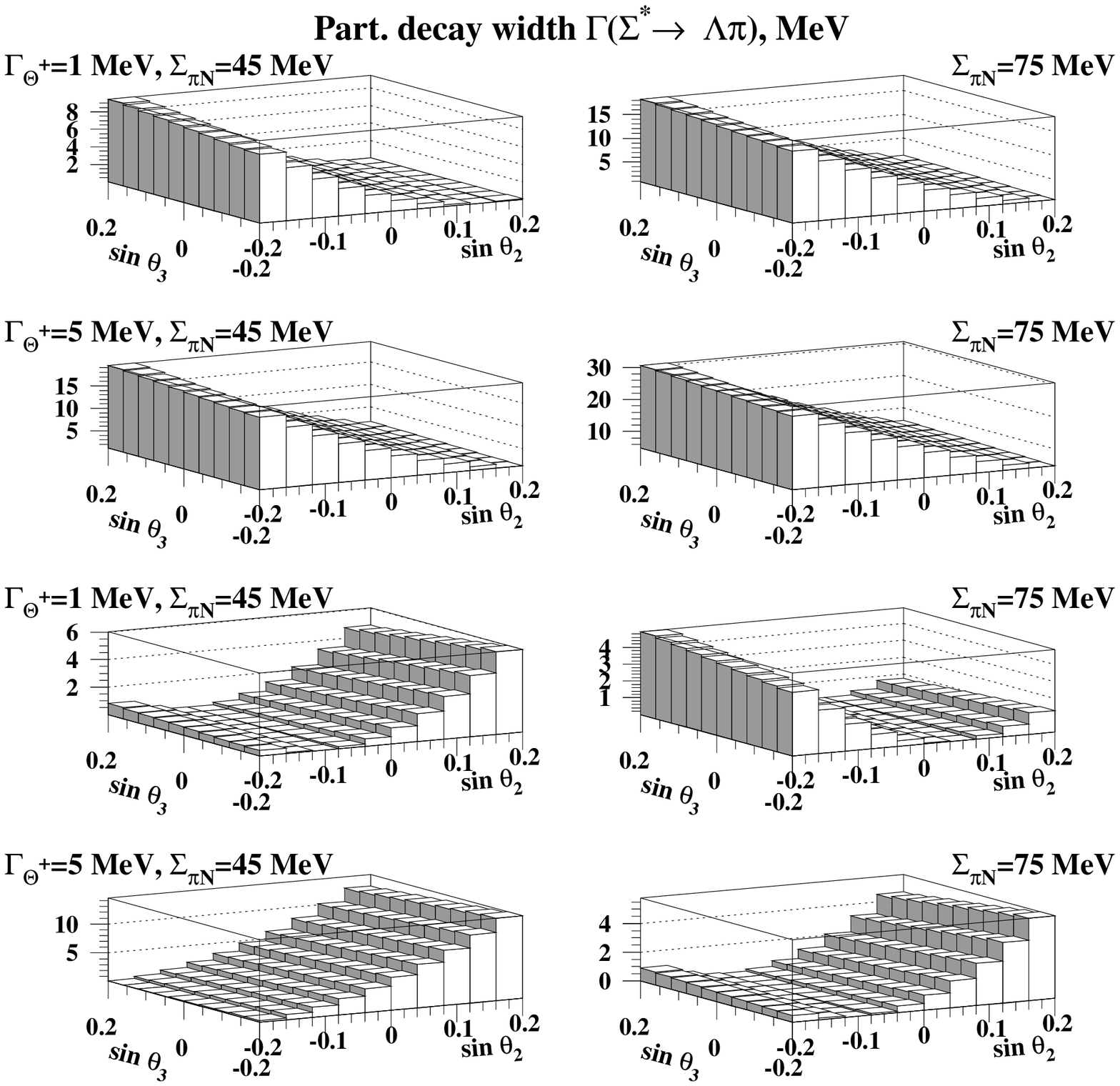}
\caption{$\Gamma_{\Sigma_{\at} \to \Lambda \, \pi}$ as a function
 of $\theta_2$ and $\theta_3$ in  the presence of the mixing with the
27-plet.
The four upper panels correspond to the positive $G_{\at}$ solution;
the four lower panels correspond to the mostly negative $G_{\at}$ solution.
}
\label{fig:slpi_27}
\end{figure}

\begin{figure}
\includegraphics[width=15cm,height=15cm]{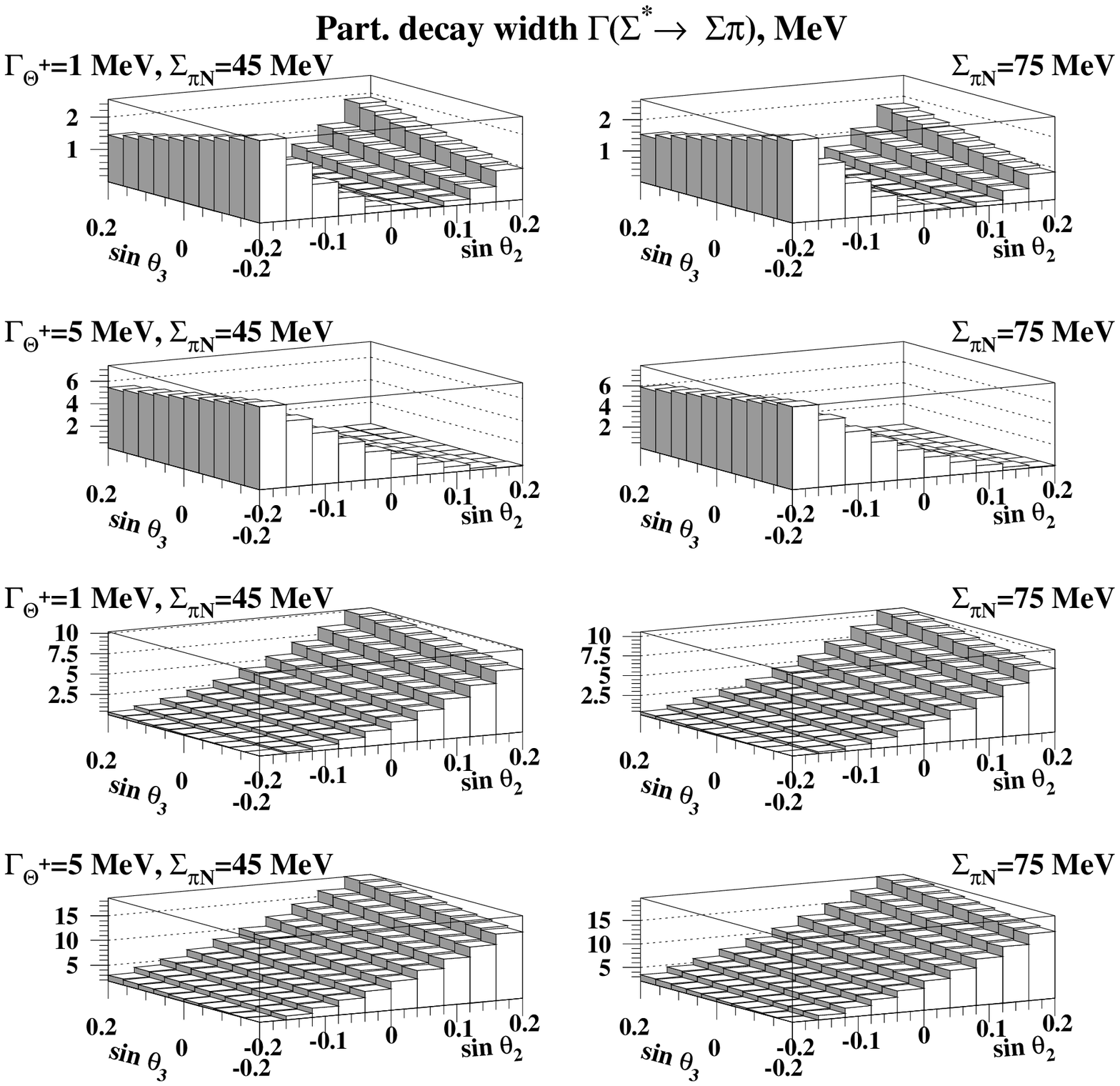}
\caption{$\Gamma_{\Sigma_{\at} \to \Sigma \, \pi}$ as a function
 of $\theta_2$
and $\theta_3$ in  the presence of the mixing with the
27-plet.
The four upper panels correspond to the positive $G_{\at}$ solution;
the four lower panels correspond to the mostly negative $G_{\at}$ solution.
}
\label{fig:sspi_27}
\end{figure}

\begin{figure}
\includegraphics[width=15cm,height=15cm]{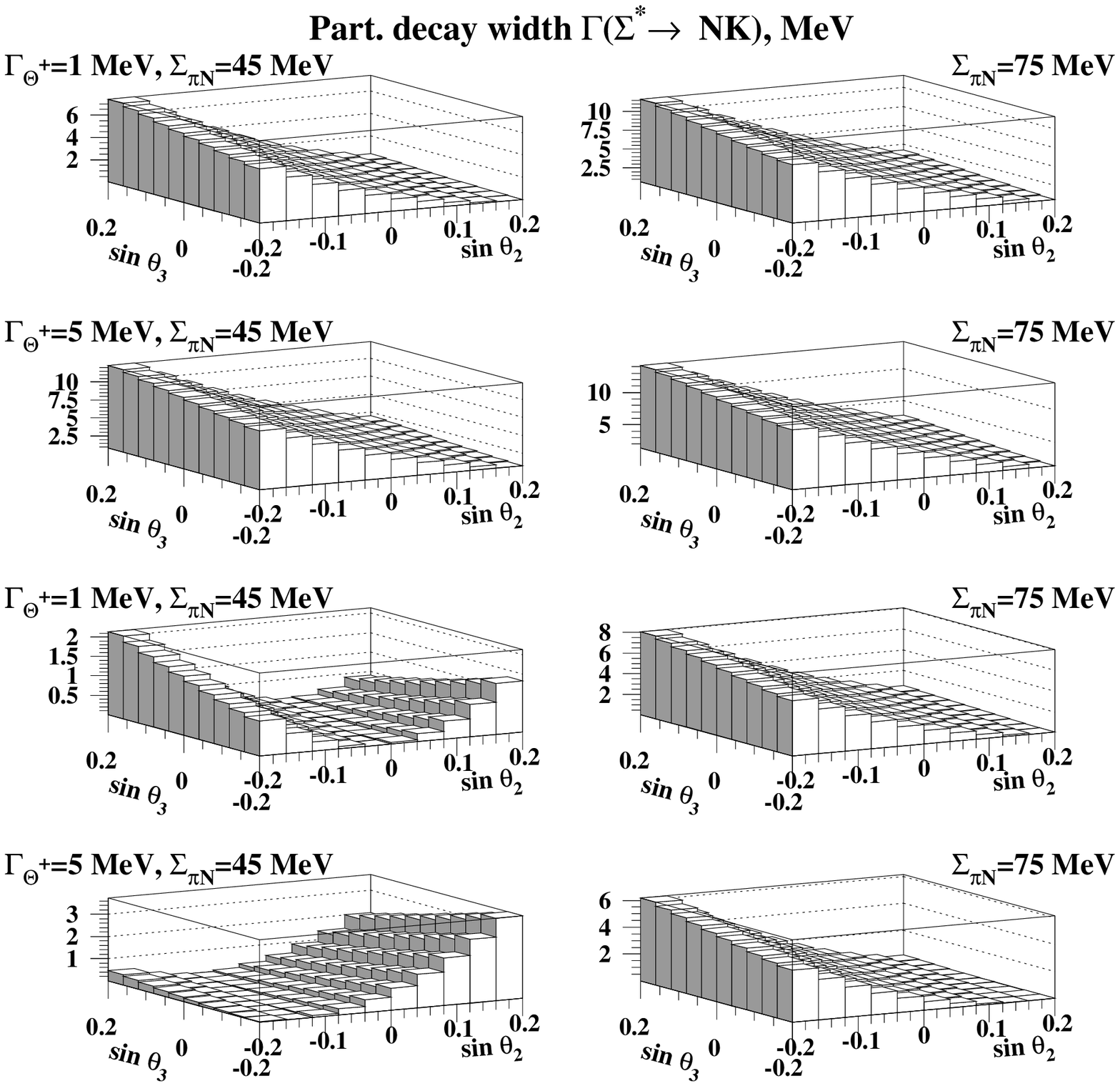}
\caption{$\Gamma_{\Sigma_{\at} \to N \, K}$ as a function
 of $\theta_2$
and $\theta_3$ in  the presence of the mixing with the
27-plet.
The four upper panels correspond to the positive $G_{\at}$ solution;
the four lower panels correspond to the mostly negative $G_{\at}$ solution.
}
\label{fig:snk_27}
\end{figure}

The $\Gamma_{\Sigma_{\at} \to \Lambda \, \pi}$, $\Gamma_{\Sigma_{\at} \to \Sigma \, \pi}$ and $\Gamma_{\Sigma_{\at} \to N \, K}$ partial decay widths in the
presence of the mixing with the 27-plet are presented in 
Figs.~\ref{fig:slpi_27}, \ref{fig:sspi_27} and \ref{fig:snk_27}.
As can be readily seen from a comparison with Figs.~\ref{fig:sspi} and 
\ref{fig:ss10pi}, the influence of the 27-plet is dramatic: Both the 
patterns and the absolute values of the predicted partial decay widths are
different.

\clearpage

\clearpage

\clearpage

\begin{figure}
\includegraphics[width=15cm,height=15cm]{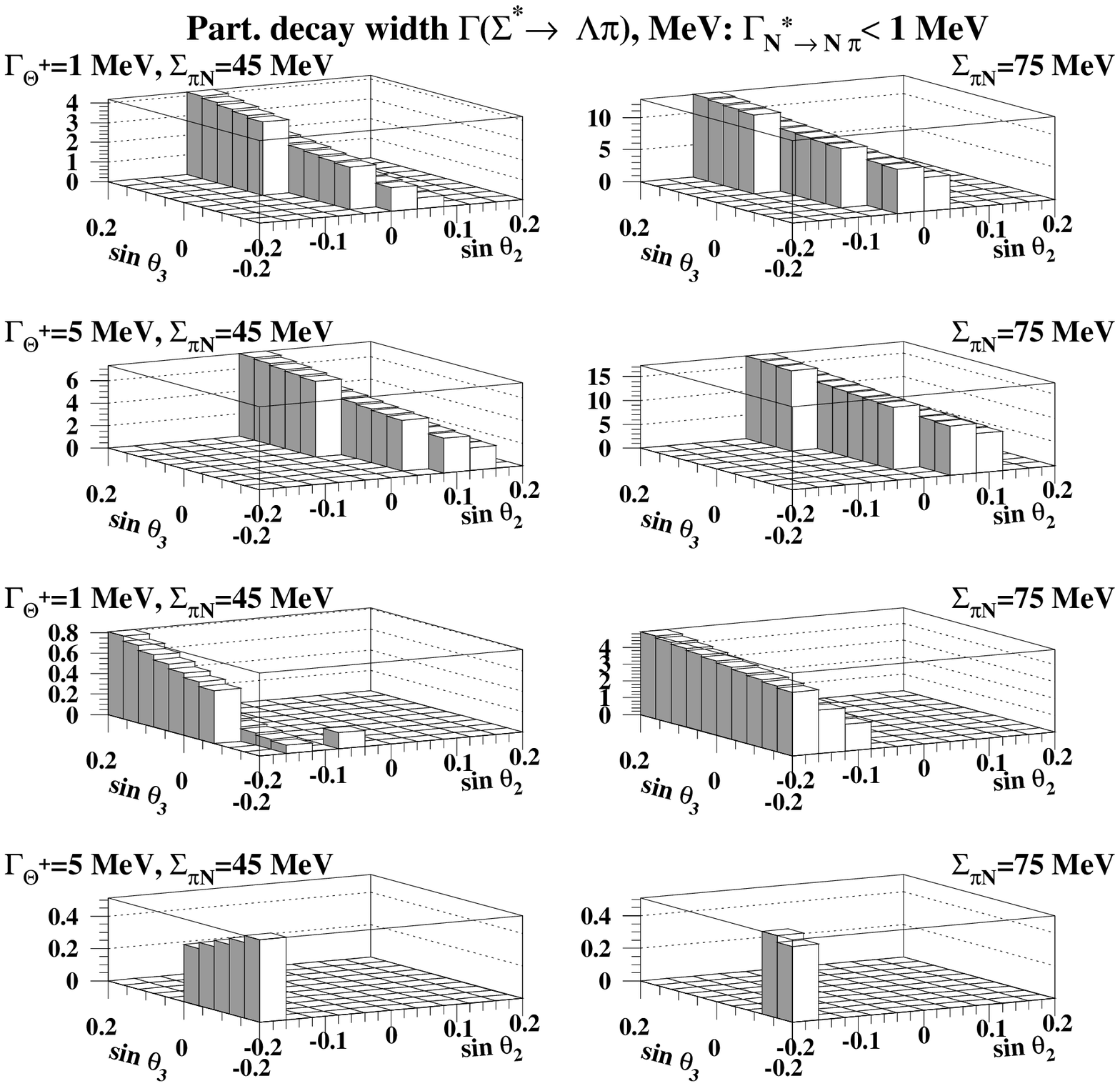}
\caption{$\Gamma_{\Sigma_{\at} \to \Lambda \, \pi}$ as a function of $\theta_2$
and $\theta_3$ in the presence of the mixing with the
27-plet. The decay width is shown only where 
 $\Gamma_{N_{\at} \to  N\, \pi} \leq 1$ MeV.
The four upper panels correspond to the positive $G_{\at}$ solution;
the four lower panels correspond to the mostly negative $G_{\at}$ solution.
}

\label{fig:slpi_cut_27}
\end{figure}

\begin{figure}
\includegraphics[width=15cm,height=15cm]{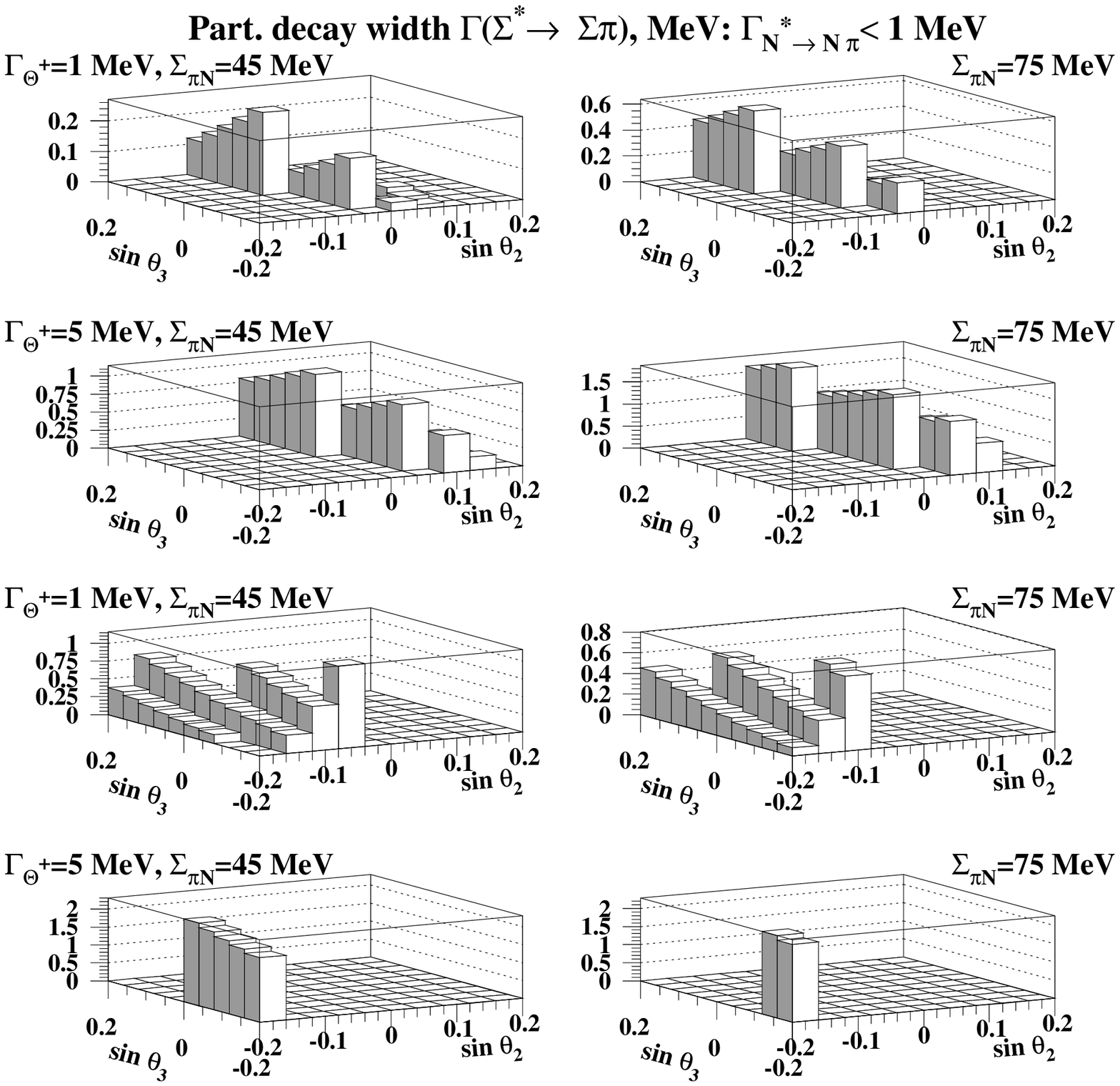}
\caption{$\Gamma_{\Sigma_{\at} \to \Sigma \, \pi}$ as a function of $\theta_2$
and $\theta_3$ in the presence of the mixing with the
27-plet. The decay width is shown only where 
 $\Gamma_{N_{\at} \to  N\, \pi} \leq 1$ MeV.
The four upper panels correspond to the positive $G_{\at}$ solution;
the four lower panels correspond to the mostly negative $G_{\at}$ solution.
}
\label{fig:sspi_cut_27}
\end{figure}

\begin{figure}
\includegraphics[width=15cm,height=15cm]{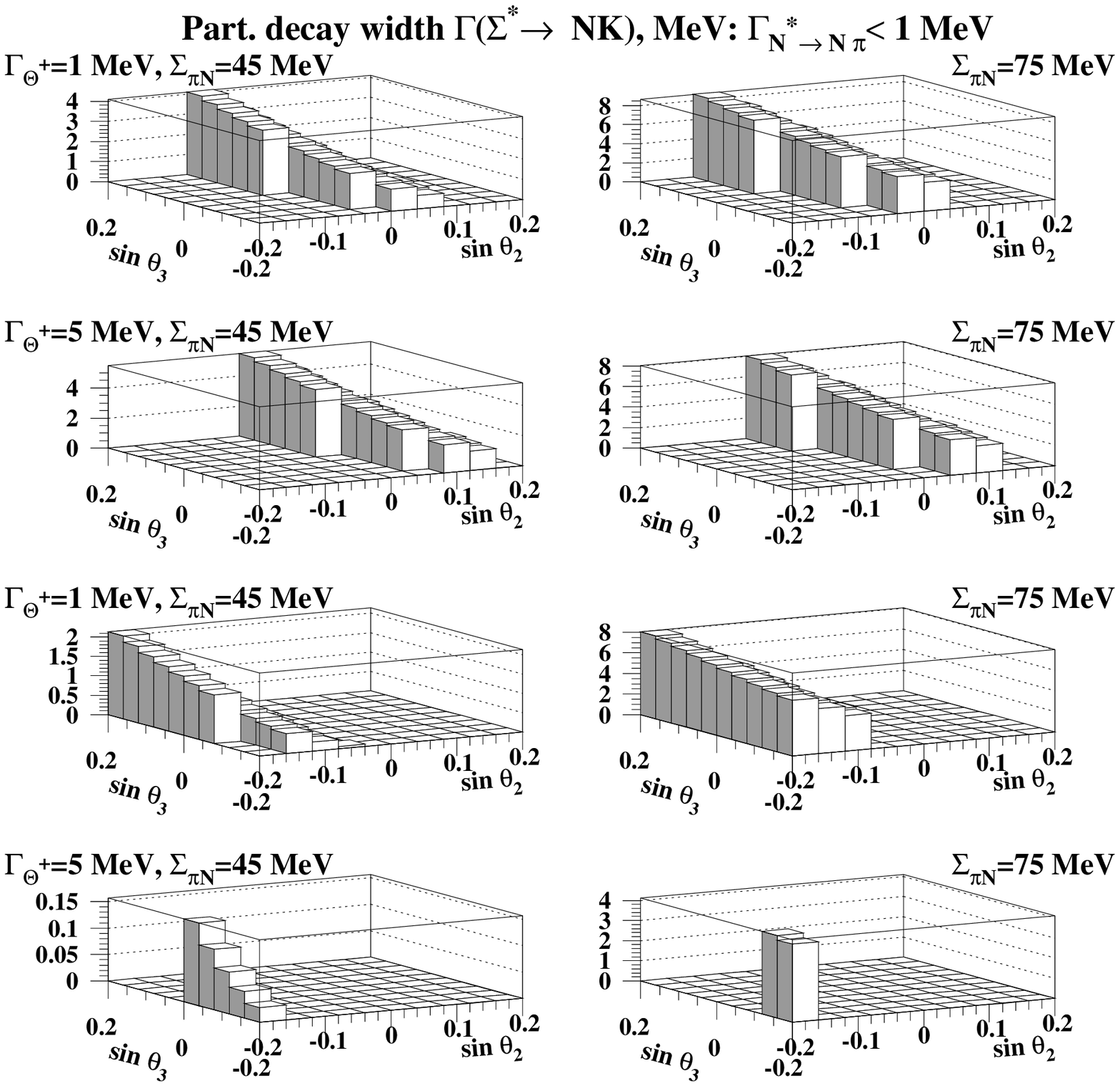}
\caption{$\Gamma_{\Sigma_{\at} \to N \, \overline{K}}$ as a function of $\theta_2$
and $\theta_3$ in the presence of the mixing with the
27-plet. The decay width is shown only where 
 $\Gamma_{N_{\at} \to  N\, \pi} \leq 1$ MeV.
The four upper panels correspond to the positive $G_{\at}$ solution;
the four lower panels correspond to the mostly negative $G_{\at}$ solution.
}
\label{fig:snk_cut_27}
\end{figure}

\begin{figure}
\includegraphics[width=15cm,height=15cm]{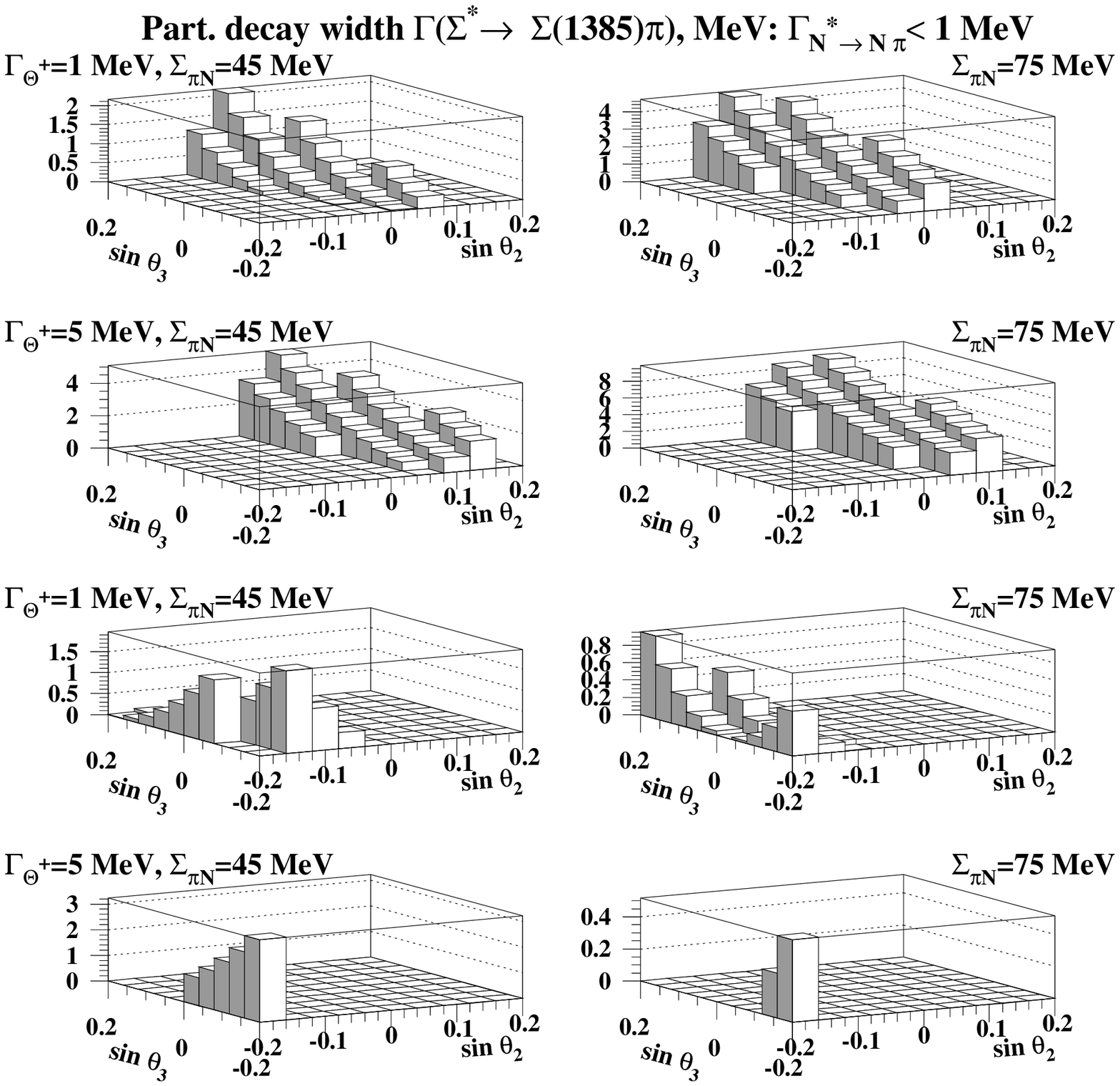}
\caption{$\Gamma_{\Sigma_{\at} \to \Sigma(1385) \, \pi}$ as a function of $\theta_2$
and $\theta_3$ in the presence of the mixing with the
27-plet. The decay width is shown only where 
 $\Gamma_{N_{\at} \to  N\, \pi} \leq 1$ MeV.
The four upper panels correspond to the positive $G_{\at}$ solution;
the four lower panels correspond to the mostly negative $G_{\at}$ solution.
}
\label{fig:ss10pi_cut_27}
\end{figure}

Next we examine the considered partial decay widths of $\Sigma_{\at}$ 
 in the domain of the $\theta_2$ and $\theta_3$ mixing angles where
 $\Gamma_{N_{\at} \to  N\, \pi} \leq 1$ MeV. The results are presented in
Figs.~\ref{fig:slpi_cut_27}, \ref{fig:sspi_cut_27}, \ref{fig:snk_cut_27}
and \ref{fig:ss10pi_cut_27}.

First we consider the case of the positive $G_{\at}$ coupling constant.
As seen from a comparison of Figs.~\ref{fig:slpi_cut_27} and
\ref{fig:snk_cut_27},  
$\Gamma_{\Sigma_{\at} \to \Lambda \, \pi} > \Gamma_{\Sigma_{\at} \to N\, K}$,
which makes it difficult or even impossible to
identify $\Sigma_{\at}$ with 
$\Sigma(1770)$ because the data suggests that $Br(\Sigma(1770) \to N \, \overline{K})$
 is several times larger than $Br(\Sigma(1770) \to \Lambda \, \pi)$.

Turning to the mostly negative  $G_{\at}$ solution, we see that the 
emerging pattern of the $\Sigma_{\at}$ decays reminds that of 
$\Sigma(1770)$: The $\Gamma_{\Sigma_{\at} \to N\, K}$ partial decay widths
is several times larger than $\Gamma_{\Sigma_{\at} \to \Lambda \, \pi}$ and
$\Gamma_{\Sigma_{\at} \to \Sigma \, \pi}$.

Taking the sum of the considered two-hadron partial decay widths of 
$\Sigma_{\at}$, 
 we find that in presence of 
 the $\Gamma_{N_{\at} \to  N\, \pi} \leq 1$ MeV
constraint and mixing with the 27-plet,
 $\Gamma_{\Sigma_{\at}}^{{\rm 2-body}}$ varies in the interval summarized
in Table~\ref{table:s_int_27}.
\begin{table}
\begin{tabular}{|c|c|c|c|}
\hline
$\Gamma_{\Theta^+}$ (MeV) & $\Sigma_{\pi \, N}$ (MeV)  & $\Gamma_{\Sigma_{\at}}^{{\rm 2-body, min}}$ (MeV) &  $\Gamma_{\Sigma_{\at}}^{{\rm 2-body, max}}$ (MeV) \\
\hline
1 & 45 & 1.2 (1.5)  & 13 (4.2)\\
1 & 75 & 7.8  (5.3) & 30 (17)\\
5 & 45 & 3.8  (3.1) & 21 (6.8) \\
5 & 75 & 13 (6.9)   & 40 (8.2)\\
\hline
\end{tabular}
\caption{The range of change of $\Gamma_{\Sigma_{\at}}^{{\rm 2-body}}$.
The first value corresponds to the positive $G_{\at}$ solution;
 the value in the parenthesis corresponds to the mostly negative $G_{\at}$ 
solution.}
\label{table:s_int_27}
\end{table} 

In summary, the antidecuplet mixing with a 27-plet significantly affects
the $\Sigma_{\at}$ decays.
Using the positive $G_{\at}$ solution, we predict that 
$\Gamma_{\Sigma_{\at} \to \Lambda \, \pi} > \Gamma_{\Sigma_{\at} \to N\, K}$
and that the both partial widths of the order of 5-15 MeV.
With the mostly negative $G_{\at}$ solution, we obtain a rather narrow 
$\Sigma_{\at}$ with decays properties qualitatively reminding those of
$\Sigma(1770)$. Of course, our $\Sigma_{\at}$ is much narrower than 
$\Sigma(1770)$.


\clearpage

\clearpage

\clearpage

\end{document}